\documentclass[prd,aps,twocolumn,preprintnumbers, showpacs, nofootinbib,superscriptaddress,notitlepage]{revtex4-1}
\usepackage{mathrsfs}
\usepackage{amsfonts}
\usepackage{amsmath}
\usepackage{slashed}
\usepackage{array}
\usepackage{verbatim}
\usepackage{epsfig}
\usepackage{graphicx}
\usepackage{color}
\usepackage[dvipsnames]{xcolor}
\usepackage[colorlinks,linkcolor=red,anchorcolor=green,citecolor=blue,CJKbookmarks=True]{hyperref}
\definecolor{red}{rgb}{1,0,0}

\usepackage{multirow}
\usepackage{float}

\newcommand{\beq}{\begin{eqnarray}}
\newcommand{\eeq}{\end{eqnarray}}

\def\be{\begin{equation}}
\def\ee{\end{equation}}
\def\bea{\begin{eqnarray}}
\def\eea{\end{eqnarray}}
\def\bal#1\eal{\begin{align}#1\end{align}}

\begin{document}

\title{Precision determination of nucleon iso-vector scalar and tensor charges at the physical point}

\collaboration{\bf{CLQCD Collaboration}}

\author{
\includegraphics[scale=0.30]{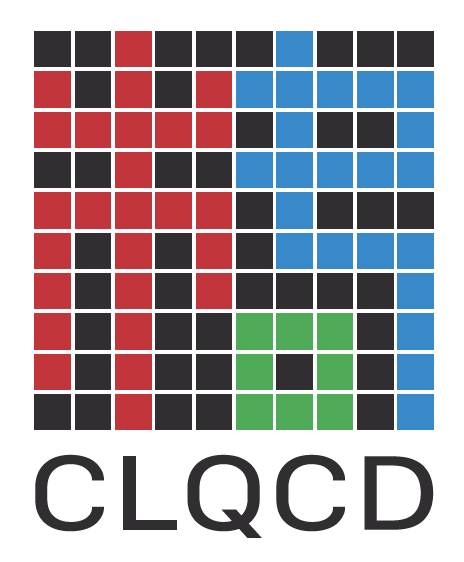}\\
Ji-Hao Wang}
\affiliation{Institute of Theoretical Physics, Chinese Academy of Sciences, Beijing 100190, China}
\affiliation{University of Chinese Academy of Sciences, School of Physical Sciences, Beijing 100049, China}

\author{Zhi-Cheng Hu}
\affiliation{Institute of Modern Physics, Chinese Academy of Sciences, Lanzhou, 730000, China}
\affiliation{University of Chinese Academy of Sciences, School of Physical Sciences, Beijing 100049, China}

\author{Xiangdong Ji}
\affiliation{Department of Physics, University of Maryland, College Park, MD 20850, USA.}

\author{Xiangyu Jiang}
\affiliation{CAS Key Laboratory of Theoretical Physics, Institute of Theoretical Physics, Chinese Academy of Sciences, Beijing 100190, China}

\author{Yushan Su}
\email[Corresponding author: ]{ysu12345@umd.edu}
\affiliation{Department of Physics, University of Maryland, College Park, MD 20850, USA.}

\author{Peng Sun}
\affiliation{Institute of Modern Physics, Chinese Academy of Sciences, Lanzhou, 730000, China}

\author{Yi-Bo Yang}
\email[Corresponding author: ]{ybyang@itp.ac.cn}
\affiliation{CAS Key Laboratory of Theoretical Physics, Institute of Theoretical Physics, Chinese Academy of Sciences, Beijing 100190, China}
\affiliation{University of Chinese Academy of Sciences, School of Physical Sciences, Beijing 100049, China}
\affiliation{School of Fundamental Physics and Mathematical Sciences, Hangzhou Institute for Advanced Study, UCAS, Hangzhou 310024, China}
\affiliation{International Centre for Theoretical Physics Asia-Pacific, Beijing/Hangzhou, China}

\date{\today}

\begin{abstract}
We report a high precision calculation of the isospin vector charge $g_{S,T}$ of the nucleon using recently proposed ``blending" method which provides a high-precision stochastic estimate of the all-to-all fermion propagator. {Through multiplying the current operator by the traditional nucleon interpolator, we create a new operator that captures the major excited state contaminations. The linear combination of this new operator and traditional nucleon interpolator reduces these excited states and improves the robustness of the multi-state fit.} Using 15 $N_f=2+1$ lattice ensembles which cover 5 lattice spacing, 5 combinations with the same quark masses and lattice spacing but multiple volumes, including three at the physical pion mass, we report so far most precise lattice QCD prediction $g_T^{\rm QCD} = 1.0264[77]_{\rm tot}(53)_{\rm stat}   (13)_{a}  (46)_{\rm FV}  (01)_\chi (28)_{\rm ex} (04)_{\rm re}$ and $g_S^{\rm QCD} = 1.106[43]_{\rm tot}(31)_{\rm stat}   (03)_{a}  (28)_{\rm FV}  (01)_\chi (08)_{\rm ex} (08)_{\rm re}$ at $\overline{\mathrm{MS}}$ 2~GeV, with the systematic uncertainties from continuum, infinite volume, chiral extrapolations, excited state contamination and also renormalization. 
\end{abstract}

\maketitle

\textit{Introduction}: A wide range of experiments use nucleons and nuclei as targets to probe the limits of the Standard Model and to search for new physics. Interpreting the results from these experiments requires precise knowledge of the matrix elements that describe how a target interacts with a probe (like a current or an effective operator). Obtaining those matrix elements with percent-level uncertainty is essential for the precision test of the standard model and can only be done by the first-principle lattice QCD calculations.

The nucleon iso-vector charges $\bar{u}\Gamma_X u-\bar{d}\Gamma_X d$—specifically for the axial-vector ($\Gamma_A=\gamma_5\gamma_{i}$), scalar ($\Gamma_S=\mathbf{1}$), and tensor ($\Gamma_T=\sigma_{ij}$) currents—are among the simplest matrix elements to compute. This simplicity arises because the disconnected quark diagrams cancel out, up to isospin-breaking (ISB) effects. While the axial current can be tightly constrained from the neutron weak decay, modulo ISB and QED corrections~\cite{ParticleDataGroup:2024cfk}, the scalar and tensor charges are in higher demand. They are particularly crucial for probing TeV-scale new physics~\cite{Bhattacharya:2011qm,Bhattacharya:2016zcn}, enabling high-precision calculations of the proton-neutron mass difference~\cite{BMW:2014pzb}, and providing constraints on parton distribution functions~\cite{Lin:2017stx,LatticeParton:2022xsd}.

However, precise predictions of these charges remain highly challenging. This is primarily due to the competing difficulties of significant excited-state contamination (ESC) at small source-sink separations ($t_f$) and exponentially growing statistical noise at large $t_f$. Consequently, although sophisticated methods (e.g., multi-state fits, such as \cite{Harris:2019bih,Bali:2023sdi,Hasan:2019noy,Alexandrou:2019brg,Ottnad:2020qbw}, multi-channel analyses through generalized eigen-value problem (GEVP) with $\pi N$~\cite{Alexandrou:2024tps, Wang:2023omf} and/or $\sigma N$ states~\cite{Barca:2024hrl}, {Summation~\cite{Fucito:1982ff,Capitani:2012gj, deDivitiis:2012vs} and} Feynman-Hellman inspired method~\cite{Bouchard:2016heu,Chang:2018uxx, He:2021yvm}) exist, the lattice community has not yet converged on a definitive criterion for ensuring ESC is sufficiently suppressed in realistic calculations~\cite{FlavourLatticeAveragingGroupFLAG:2024oxs}.

At the same time, a critical limitation of all existing high-precision lattice QCD results for $g_{A,S,T}$~\cite{Capitani:2012gj, Bali:2014nma, Bhattacharya:2016zcn, Capitani:2017qpc, Chang:2018uxx,Harris:2019bih,Bali:2023sdi,Mainz24,Jang:2023zts} is {the pion mass dependence of the finite-volume effects (FVE). One would argue that the dominance interaction at the long distance through the pion can impose $m_{\pi}^2$ suppression of FVE at physical quark masses~\cite{Beane:2004rf}, while it is not sufficiently constrained by the lattice data around physical pion mass which has sizable statistical uncertainty and/or specific volume. This ambiguity} therefore represents an uncontrolled source of systematic error in the field.

Using the blending method~\cite{Hu:2025vhd} and the idea of current-involved interpolation field~\cite{Barca:2024hrl,Barca:2025det} proposed recently, our study delivers a new calculation of the nucleon iso-vector charges $g_S$ and $g_T$ with better control over systematic errors—notably from ESC, chiral extrapolation, and FVE—compared to all existing lattice determinations. This advancement yields results for $g_{S,T}$ with twice the statistical precision of previous determinations. The $g_S$ result also disfavors the HB$\chi$PT {FVE} ansatz, and a more conservative estimate of FVE is essential to obtain a reliable prediction for both $g_S$ and $g_T$.
The corresponding $g_A$ result is reserved for a separate publication~\cite{gA_work} {which requires additional QED corrections (e.g., Ref.~\cite{Cirigliano:2022hob,Tomalak:2025jtn}) and more conservative treatment on FVE (See Ref.~\cite{Hall:2025ytt} for recent investigation) to compare with experiment}.

\begin{table}[ht!]
  \centering
  \caption{Lattice spacing $a$, lattice volume $n_L^3\times n_T$, pion mass $m_{\pi}$ with $m_{\pi}L$,  {and also the fit values of $g_S$ and $g_T$.}}
  \resizebox{1.00\columnwidth}{!}{
    \begin{tabular}{c|cccccc}
    \hline
    \hline
          & $a$(fm) & $n_L^3\times n_T$ & $m_\pi$(MeV) & $m_\pi L$ & {$g_S$} & {$g_T$} \\
    \hline
    C24P34 & \multirow{8}[2]{*}{0.1053} & 24$\times$64 & 341.1(1.8)  & 4.38  & 1.038(23)  & 1.0167(67)\\
    C24P29 &       &                   24$\times$72    & 292.7(1.2) & 3.75   & 1.027(37)& 0.9904(45) \\
    C32P29 &       &                    32$\times$64   & 292.4(1.1) & 5.01  & 1.091(13)  &  1.0007(55) \\
    C24P23 &       &                   24$\times$64    & 229.5(3.0) & 2.93  & 0.942(33) & 0.9670(39) \\
    C32P23 &       &                   32$\times$64    & 228.0(1.2) & 3.91  & 1.039(19) & 0.9796(38) \\
    C48P23 &       &                   48$\times$96    & 225.6(0.9)  & 5.79  & 1.107(19) & 0.9881(59) \\
    C48P14 &       &                  48$\times$96     & 135.5(1.6)  & {3.47}  & 1.050(34)  & 0.9743(81)
    \\
   C64P14 &       &                  64$\times$128     & 134.5(1.6)  & 4.63  & 1.111(27) & 0.9885(75) \\
    \hline 
        E32P29 & 0.08973 & 32$\times$64 & 286.7(1.8)      &   4.19     & 1.012(29) & 1.0103(64)  \\
    \hline
    F32P30 & \multirow{5}[2]{*}{0.07753} &32$\times$96 & 303.2(1.3)  & 3.56  & 1.004(40)  & 1.0059(37) \\
    F48P30 &       & 48$\times$96 & 303.4(0.9)  & 5.72  & 1.091(17) &  1.0217(35)  \\
    F32P21 &       & 32$\times$64 & 210.9(2.2) & 2.67  & 0.910(42) & 0.9685(38)   \\
    F48P21 &       & 48$\times$96 & 207.2(1.1)  & 3.91  & 1.049(26)  & 0.9897(33) \\
    F64P13 &       & 64$\times$128& 134.1(1.5)   & 3.37 & 0.997(29) & 0.9982(39)  \\
    \hline
    G36P29 & 0.06887 &36$\times$108 & 297.2(0.9) & 3.73 & 0.956(35) & 1.0120(61) \\
    \hline
    H48P32 & 0.05199 &48$\times$144 & 317.2(0.9) &   4.00   & 0.982(36)  & 1.0260(53) \\
    \hline
    \hline
    \end{tabular}%
    }
  \label{tab:info}%
\end{table}%

\textit{Simulation Setup}:
The results in this work are based on the 2+1 flavor (degenerate up and down quarks plus the strange quark)
ensembles~\cite{CLQCD:2023sdb,CLQCD:2024yyn} from the CLQCD collaboration using the tadpole improved tree level Symanzik (TITLS) gauge action and the tadpole improved tree level Clover (TITLC) fermion action. The information of the ensembles used in this work is summarized in Table~\ref{tab:info}.

The nucleon matrix elements $g_{X=S,T}$ we consider are defined through the relation,
\begin{align}
    \langle N(p,s) | {\cal O}_{X} |N(p,s) \rangle = g_X\bar{u}(p,s) \Gamma_{X}u(p,s), 
\end{align}
where the nucleon states $N(p,s)$ and spinors $u(p,s)$ have given momentum $p$ and spin $s$, ${\cal O}_X =\bar{u} \Gamma_X u-\bar{d}\Gamma_X d$ is the iso-vector singlet operator with $\Gamma_S=\mathbf{1}$ and $ \Gamma_T=\sigma_{\mu\nu}$.

Within the framework of quantum field theory, the nucleon matrix element is extracted from the ratio of three-point (3pt) to two-point (2pt) correlation functions:
\begin{align}\label{eq:Ratio}
{\cal R}_{X}(t_f,t;{\cal N})
& = \frac{\int \text{d}^3x\text{d}^3y\text{d}^3z\langle {\cal N}(\vec{x},t_f) {\cal O}_{X}(\vec{y},t) {\cal N}^{\dagger}(\vec{z},0)\rangle}{\int \text{d}^3x\text{d}^3z \langle {\cal N}(\vec{x}, t_f) {\cal N}^{\dagger}(\vec{z}, 0) \rangle} \nonumber\\
=
&\langle {\cal O}_X \rangle_N + \mathcal{O}(e^{-\delta m t}, e^{-\delta m(t_f-t)}, e^{-\delta m t_f}),
\end{align}
where ${\cal N}(\vec{x},t)\equiv \epsilon_{abc}(u_a^{T}C\gamma_5d_b)u_c$ denotes the nucleon interpolating operator and $\delta m$ is the energy gap between the ground state and the first excited state. A significant challenge is that this ratio can suffer from substantial ESC at finite temporal separations $t$ and $t_f$. 

\begin{figure}
    \centering
    \includegraphics[width=1.0\linewidth]{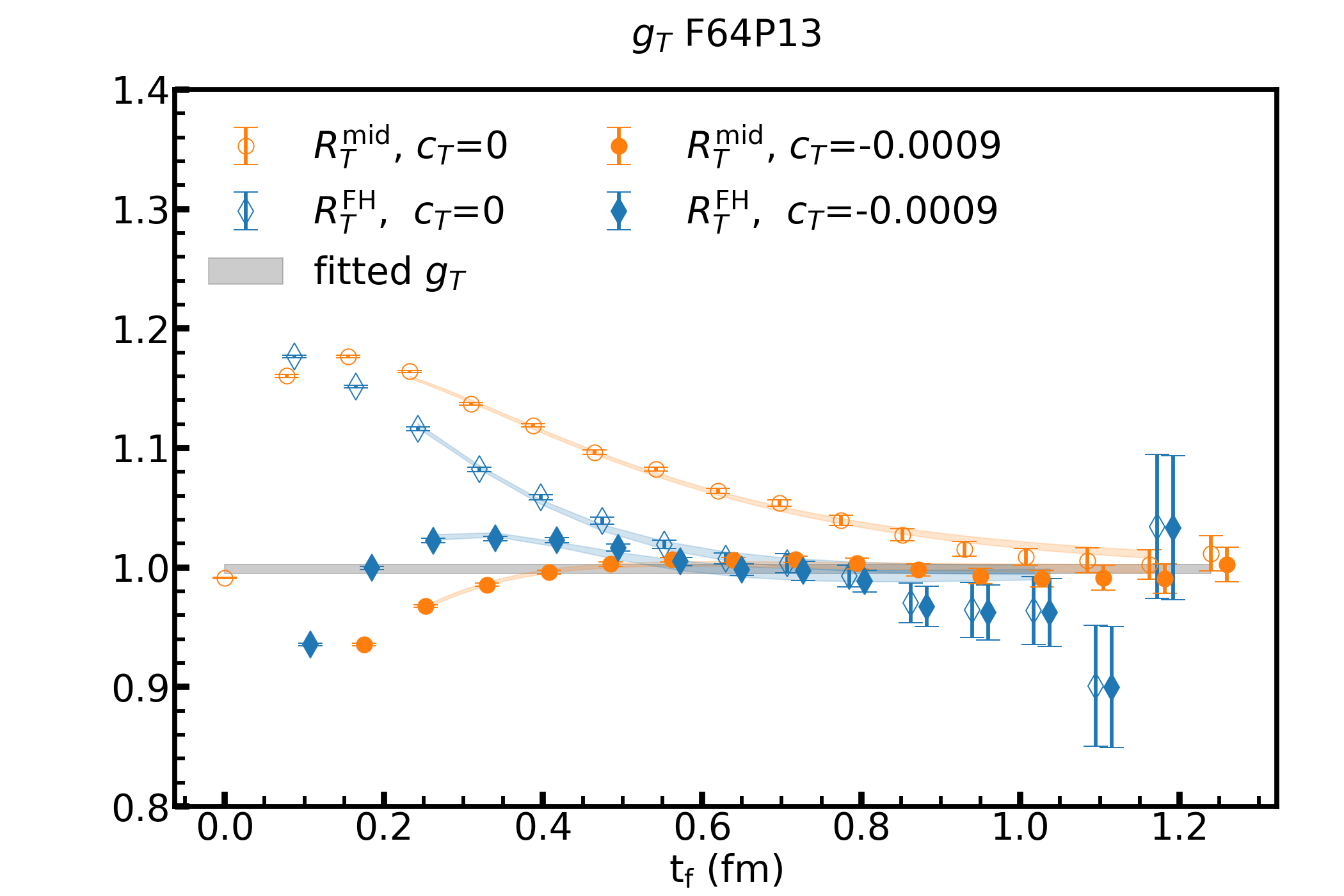}
    \includegraphics[width=1.0\linewidth]{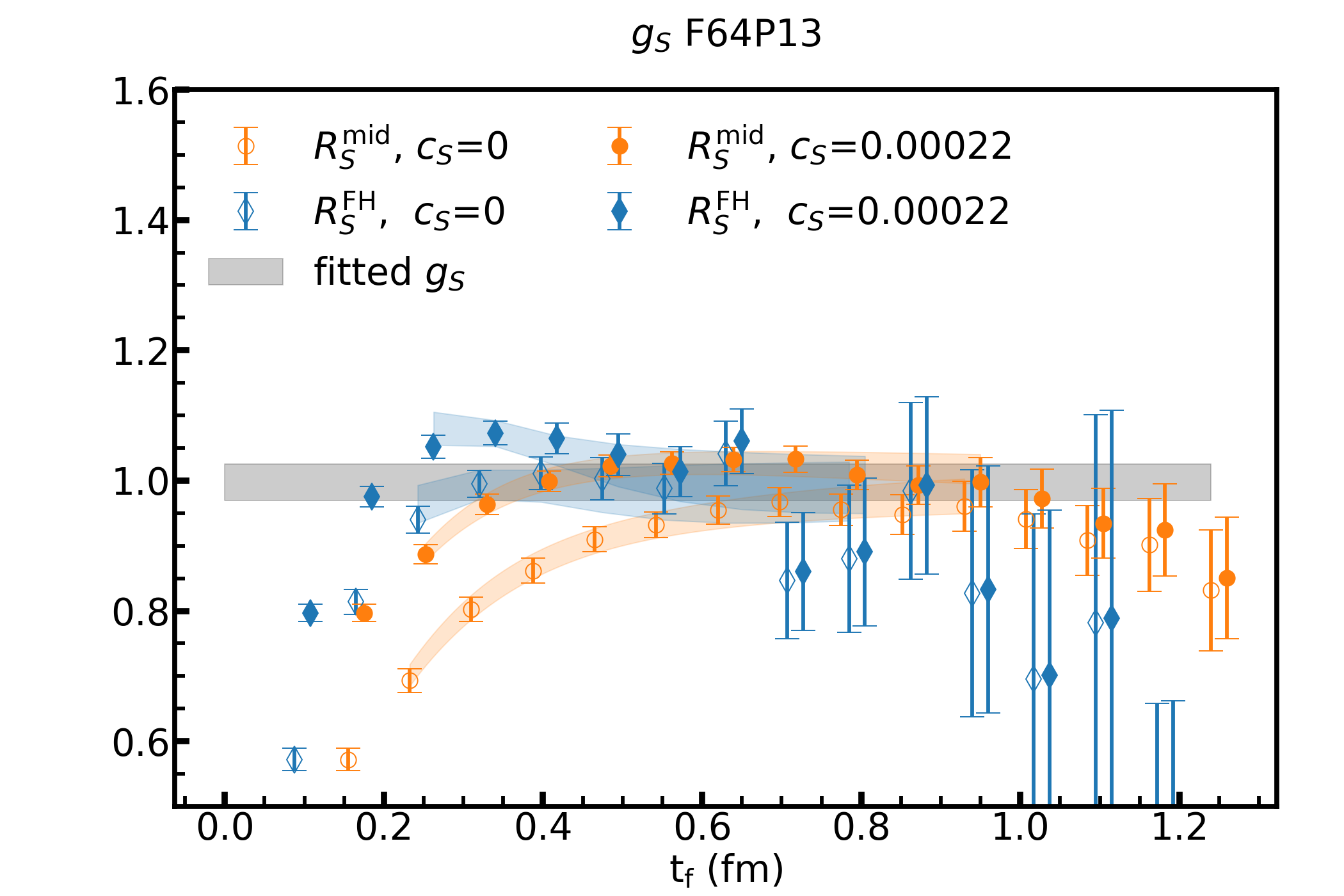}
    \caption{
        The dependence of the source-sink separation \( t_f \) for the ratios \( \mathcal{R}_X^{\rm mid}(t_f)\equiv \mathcal{R}_X(t_f, t = t_f/2) \) and \( \mathcal{R}_X^{\text{FH}}(t_f) \) on the F64P13 ensemble, for the tensor $X=T$ (upper panel) and scalar $X=S$ (lower panel) operator cases. The gray band represents the result from the joint three-state fit. The minimum source-sink separation for fitting is \( t_f^{\text{min}}=3a \sim 0.24~\text{fm} \).
    }\label{fig:gevp_main}
\end{figure}

The signal-to-noise ratio of ${\cal R}_X$ decays exponentially with $t_f$, making robust control of excited-state contamination (ESC) contingent upon two factors: high-precision statistical signals at sufficiently large $t_f$, and the use of a proper interpolating field that suppresses contributions from excited states. The recently developed blending method~\cite{Hu:2025vhd,supplemental} extends the distillation method~\cite{HadronSpectrum:2009krc,Morningstar:2011ka}--which projects the full quark propagators onto their low-energy modes to efficiently suppress ESC--to a stochastic estimation of the full quark propagator itself. The unbiased nature of this method has been rigorously established through both mathematical proof and explicit numerical verification using 2-,3-,4-point correlation functions~\cite{Hu:2025vhd}, and also the nucleon iso-vector $g_V=0.9998(12)$~\cite{supplemental}. This extension enables the precise calculation of $R_{X}(t_f,t)$ for arbitrary combinations of $t$ and $t_f$ values without additional cost for quark propagator production, thereby generating the high-quality data necessary to investigate possible good choices of the interpolation operators.

As detailed in the supplemental materials~\cite{supplemental}, our numerical tests indicate that the dominant ESC can be eliminated using a linear combination of matrix elements of ${\cal O}_X$ within a basis with two of interpolation fields, $H = \{ {\cal N}, {\cal N}_X\equiv {\cal N} O_X \}$ where the second one is ``current-involved" {which couples strongly to excited states enhanced by the current \( \mathcal{O}_X \) itself}~\cite{Barca:2024hrl,Barca:2025det}. The nucleon projection is chosen such that the matrix elements $\langle H_i | O_X | H^{\dagger}_j \rangle$ are non-zero for all $i,j=1,2$.

Based on the studies using the heavier-than-physical light quark masses~\cite{Barca:2024hrl}, Ref.~\cite{Barca:2025det} suggested that the ESC of the correlation function $\langle {\cal N}_X(t_f){\cal O}_X(t){\cal N}(0) \rangle=V\langle {\cal N}(t_f){\cal N}(0) \rangle\langle {\cal O}_X(t_f){\cal O}_X(t)\rangle +{\cal O}(1)$, which contains a disconnected quark diagram, can be enhanced by the spatial volume $V$ compared to that in  the standard three-point function $\langle {\cal N}(t_f){\cal O}_X(t){\cal N}(0) \rangle$. This volume-enhanced effect can be the dominant source of ESC, surpassing conventional ones. Consequently, even a small $c_X$ in the improved interpolating field ${\cal N}+c_X {\cal N}_X$ can significantly alter the observed ESC in the extraction of $g_X$. The contribution from the three-point function $\langle {\cal N}_X(t_f){\cal O}_X(t){\cal N}_X(0) \rangle$ is confirmed to be negligible, even though it entails 138 quark diagrams~\cite{supplemental}. This result was enabled by an automated and optimized contraction framework for arbitrary correlation functions using the blending method.

In the upper panel of Fig.~\ref{fig:gevp_main}, we show the ratio ${\cal R}^{\rm mid}_{T}\equiv {\cal R}_{T}(t_f,t_f/2)$ (dots) and the Feynman-Hellmann inspired combination ${\cal R}_{X=T}^{{\rm FH}}(t_f)$~\cite{Bouchard:2016heu,Chang:2018uxx, He:2021yvm} (diamonds),
\begin{align}\label{eq:Ratio2}
    {\cal R}_X^{{\rm FH}}(t_f)&\equiv \sum_{t=t_c}^{t_f+a-t_c}{\cal R}_X(t_f+a,t)-\sum_{t=t_c}^{t_f-t_c}{\cal R}_X(t_f,t) \nonumber \\
    & = \langle {\cal O}_X \rangle_N +{\cal O}(e^{-\delta m t_f}),
\end{align}
on the physical pion mass ensemble F64P13 at $a=0.078$ fm,
for the distilled interpolation fields ${\cal N}$ (hollow data points) and also ${\cal N}+c_T {\cal N}_T$ (filled ones) with $c_T=-0.0015$. It is noteworthy that this work marks the first time such high-precision ${\cal R}_T$ (and also ${\cal R}_S$) has been obtained directly at the physical pion mass, thanks to the high-precision all-to-all propagators using the blending method. This achievement is crucial for suppressing the systematic uncertainty associated with chiral extrapolation.

We do a joint three-state fit to the two- and three-point functions for both interpolating fields, and show the fits with the colored bands which describe the data well, starting from a quite short time separation of $t=3a\sim 0.24$ fm. The extracted value of $g_T$ is indicated by the gray band, whose uncertainty is comparable to that of the midpoint estimate ${\cal R}_X(t_f,t_f/2)$ at $t_f\sim 0.9$ fm. Utilizing the improved interpolating field, which incorporates a small component of the ${\cal N}_T$ operator, both  ${\cal R}^{\rm mid}_{T}$ and ${\cal R}^{\rm FH}_{T}$ demonstrate suppression of ESC and exhibit improved convergence toward the ground-state value $g_T$ (gray band) in the fit range shown as the colored bands.

The $g_S$ case is shown in the lower panel of Fig.~\ref{fig:gevp_main} and shows a similar but minor improvement with $c_S=0.00015$ since ESC has been suppressed well in the distilled interpolation field ${\cal N}$, even though the impact can be more obvious on the other ensembles with heavier pion masses. 

We refer the reader to the supplemental materials~\cite{supplemental} for an extended discussion on {the current-involved interpolator used in this work (distinct from the standard ${\cal N}\pi$ one), including its application to other typical ensembles and illustrations} on the inefficiency of GEVP with ${\cal N}\pi$ interpolator in the current fit range.

\textit{Results}: {With significantly fewer inversions compared to prior studies~\cite{Alexandrou:2022dtc,Bali:2023sdi,Jang:2023zts} and negligible autocorrelation~\cite{supplemental}, the values of $g_{S,T}$ obtained on all ensembles are compiled in Table~\ref{tab:info}.} To extract the results at the physical point in the continuum and infinite-volume limits, we employ the following joint fit ansatz:
\begin{align}
\label{eq:global_fit}
g_X (a,m_\pi,L) = & g_{X}^{\rm QCD} \left(1 + \sum_{i=2,3}c_l^{(i)}(m_\pi^i-m_{\pi,\rm phy}^i) \right) \nonumber \\
& (1+c_Ve^{-m_\pi L}) +c_{a} a^2,
\end{align}
where $g_X^{\rm QCD}$ is the target physical value. The parameters $c_l$, $c_V$, and $c_a$ describe the pion mass dependence, finite-volume dependence, and discretization effects, respectively. We omit any dependence on the strange quark mass, as our current data lack the sensitivity to constrain such a term. 

\begin{figure}
    \centering
    \includegraphics[width=1.0\linewidth]{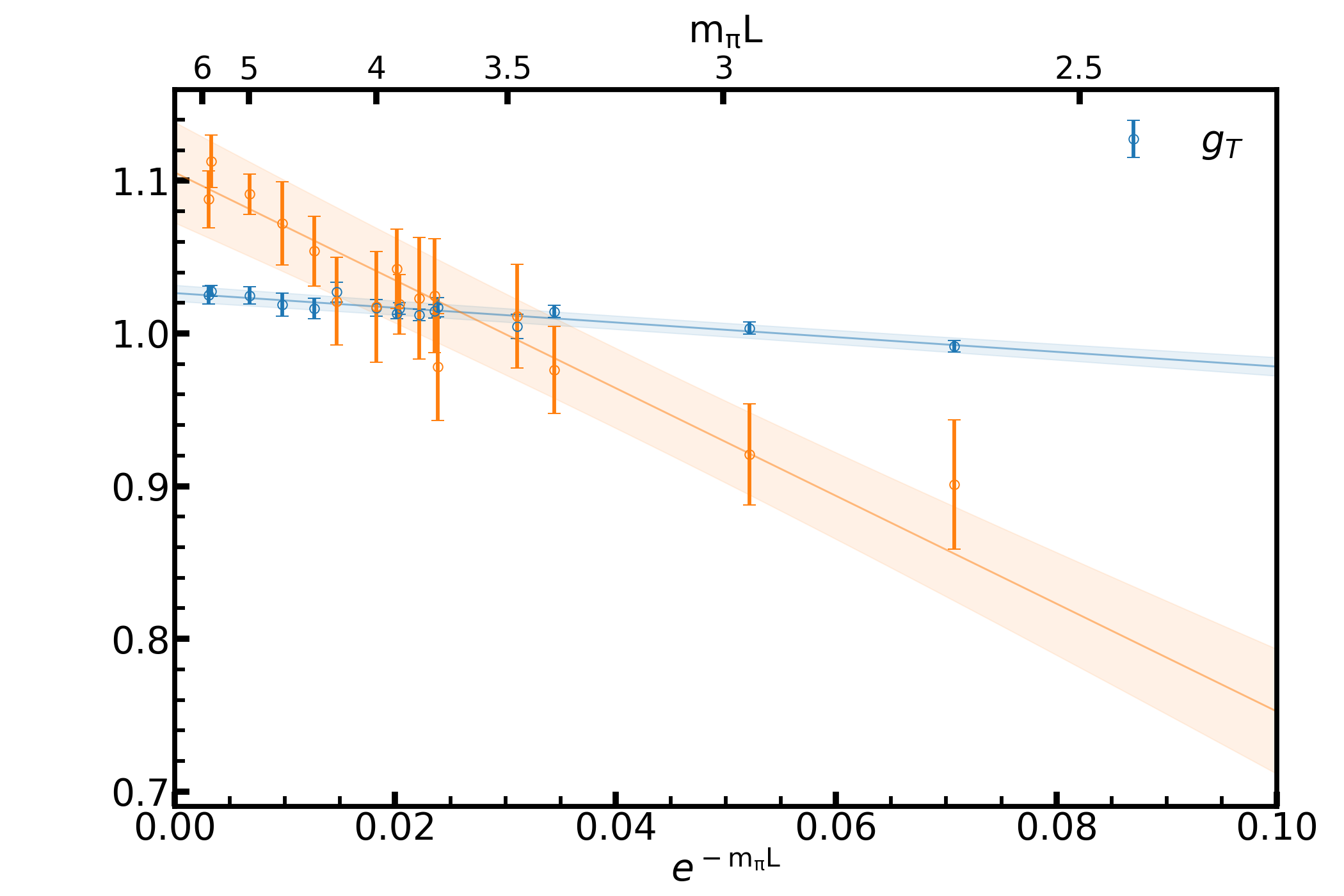}
    \caption{
        The finite spatial lattice length $L$ dependence of the \( g_S \) and \(g_T \) with data points corrected to the continuum limit and physical pion mass.
    }
    \label{fig:volume}
\end{figure}

Our analysis of $g_S$ reveals a strong preference for a phenomenological exponential finite-volume correction, $e^{-m_{\pi}L}$, over {the form $m_\pi^2 e^{-m_{\pi}L}/\sqrt{m_{\pi}L}$ motivated by HB$\chi$PT analysis in~\cite{Beane:2004rf}}. This is demonstrated by the corrected data (using parameters $c_l$ and $c_a$ from the global fit in Eq.~\ref{eq:global_fit}), which exhibit a clear linear dependence on $e^{-m_{\pi}L}$ (orange points, Fig.~\ref{fig:volume}). In contrast, while the FVE for $g_T$ (blue points) is also consistent with a linear $e^{-m_{\pi}L}$ dependence, it remains statistically compatible with the HB$\chi$PT ansatz due to its smaller overall magnitude.

\begin{figure}
    \centering
    \includegraphics[width=1.0\linewidth]{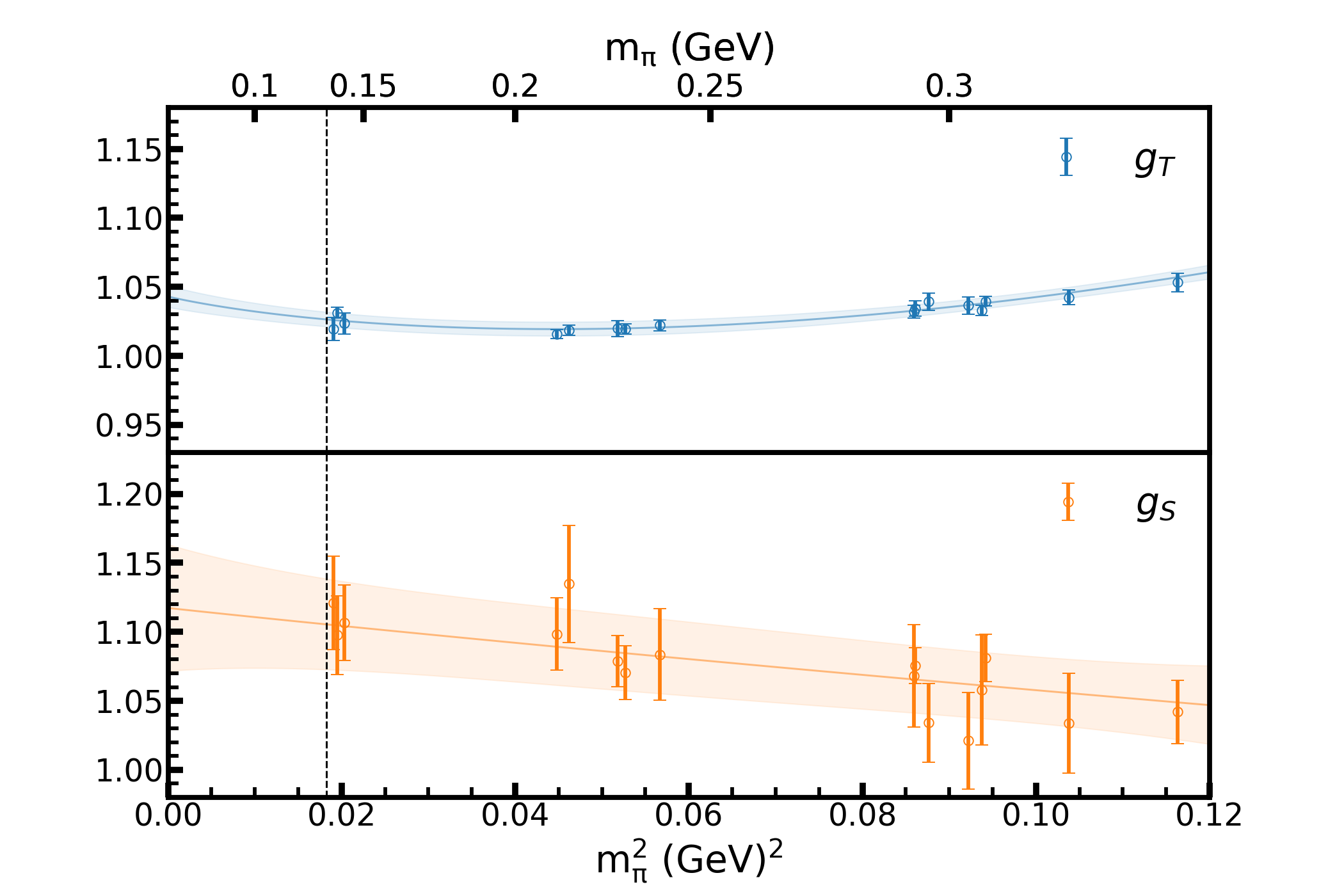}
    \caption{
        The pion mass dependence of the \( g_S \) and \(g_T \) with data points corrected to continuum and infinite volume limits. 
    }
    \label{fig:mpi}
\end{figure}

The pion mass dependence in the continuum and infinite-volume limits is shown in Fig.~\ref{fig:mpi}, using similar corrections with $c_V$ and $c_a$ obtained from the joint fit with the ansatz of Eq.~\ref{eq:global_fit}. The results for both $g_T$ and $g_S$ demonstrate a mild dependence on the pion mass. We incorporate the $m_{\pi}^3$ term to provide an improved description of data at relatively heavy pion masses. The presence of precise results at the physical pion mass for multiple lattice spacings and volumes ensures that the systematic uncertainty originating from the chiral extrapolation remains minor.

The systematic errors associated with infinite volume, continuum and chiral extrapolations are estimated by combining different fit models using the Akaike Information Criterion (AIC)~\cite{BMW:2014pzb}. The systematic uncertainty from ESC is estimated by taking the difference between the central values obtained from the three-state and two-state fits, and that from the renormalization constant is also integrated through the strategy used in our previous works for the quark masses~\cite{CLQCD:2023sdb,CLQCD:2024yyn}. The figure for the lattice spacing dependence of the results and also details on the estimate of the systematic uncertainties are provided in the Supplemental Material~\cite{supplemental}. 

Eventually we predict $g_{T,S}^{\rm QCD}$ with statistical and systematic errors to be:
\begin{align}
    & g_T^{\rm QCD} = 1.0264[77]_{\rm tot}(53)_{\rm stat}   (13)_{a}  (46)_{\rm FV}  (01)_\chi (28)_{\rm ex} (04)_{\rm re}, \nonumber \\
    & g_S^{\rm QCD} = 1.106[43]_{\rm tot}(31)_{\rm stat}   (03)_{a}  (28)_{\rm FV}  (01)_\chi (08)_{\rm ex} (08)_{\rm re},
\end{align}
where the label ``stat'' denotes the statistical error, while $a$, FV, $\chi$ , $\rm ex$ and ``$\rm re$" represent the systematic errors of the continuum, infinite volume, chiral extrapolation, excited state contamination and renormalization constant, respectively. ``tot'' is the total error combining the statistical and systematic ones. 

\begin{figure}
    \centering
    \includegraphics[width=1.0\linewidth]{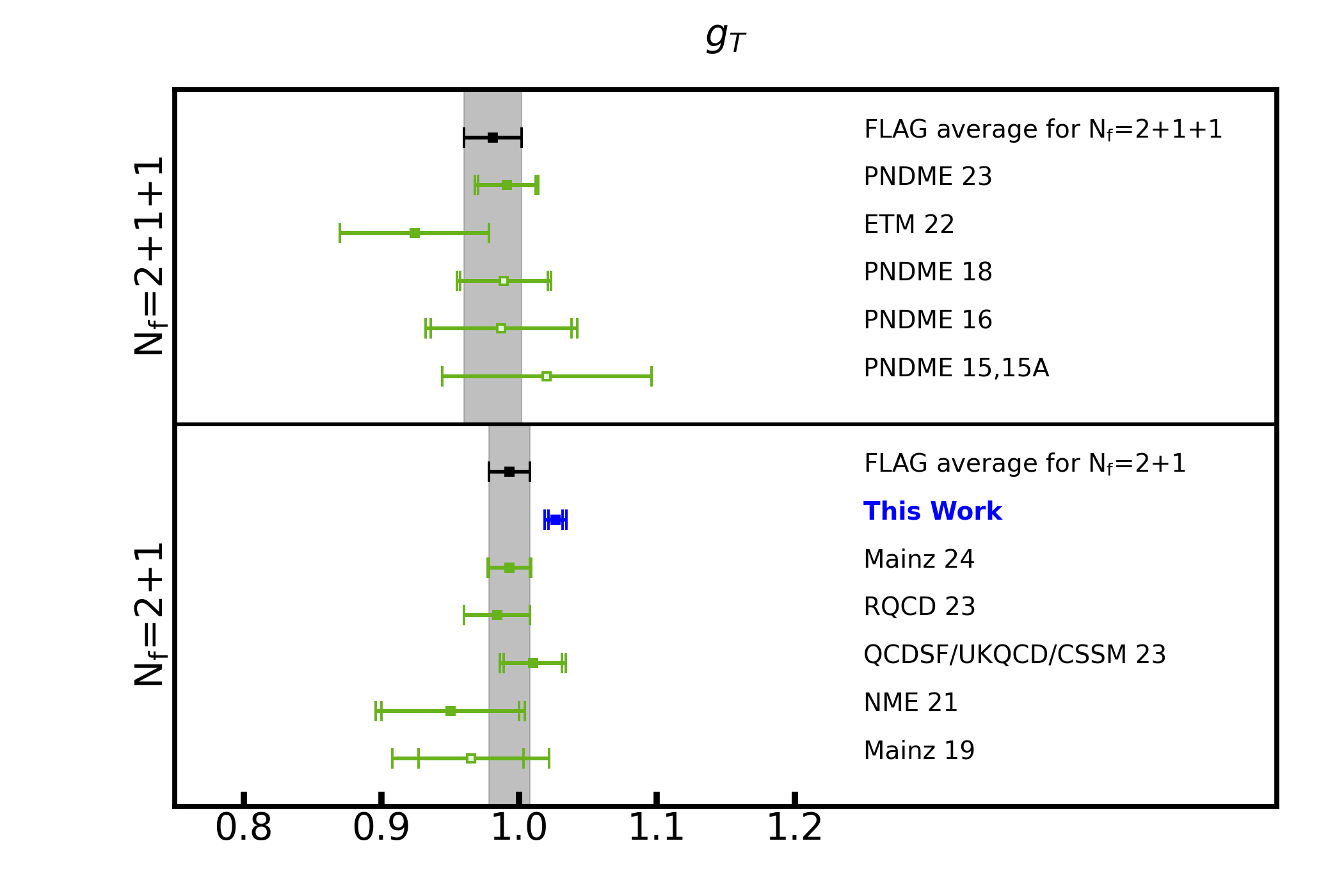}
    \includegraphics[width=1.0\linewidth]{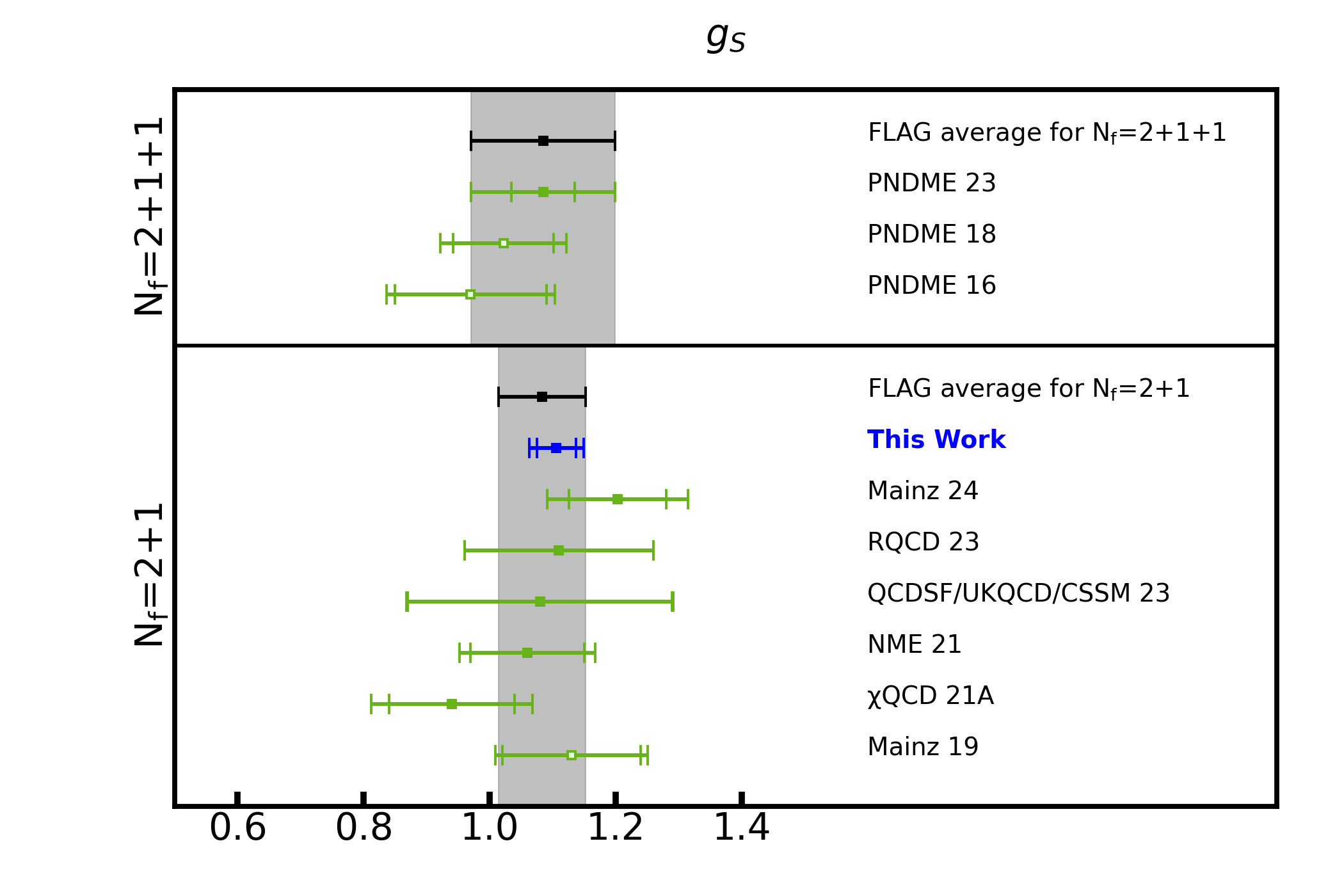}
    \caption{
Comparison of $g_T$ (upper panel) and $g_S$ (lower panel) from this work, PNEME 15~\cite{Bhattacharya:2015esa,Bhattacharya:2015wna}/16~\cite{Bhattacharya:2016zcn}/18~\cite{Gupta:2018qil}/23~\cite{Jang:2023zts}, ETM22~\cite{Alexandrou:2022dtc}, Mainz 19~\cite{Harris:2019bih}/24~\cite{Mainz24}, RQCD 23~\cite{Bali:2023sdi}, QCDSF/UUKQCD/CSSM 23~\cite{QCDSFUKQCDCSSM:2023qlx}, NME 21~\cite{Park:2021ypf}, $\chi$QCD 21~\cite{Liu:2021irg} and also the FLAG averages~\cite{FlavourLatticeAveragingGroupFLAG:2024oxs}. There are also lattice calculations~\cite{Yamanaka:2018uud, Hasan:2019noy, Abramczyk:2019fnf, Tsuji:2022ric, Alexandrou:2019brg} which are not included by FLAG due to certain uncontrollable systematics. All the values are renormalized at $\overline{\mathrm{MS}}(2~\mathrm{GeV})$.
    }
    \label{fig:compare}
\end{figure}

In Fig.~\ref{fig:compare}, we compare our results with those from the studies~\cite{Bhattacharya:2015esa,Bhattacharya:2015wna,Bhattacharya:2016zcn,Gupta:2018qil,Jang:2023zts,Alexandrou:2022dtc,Harris:2019bih,Mainz24,Bali:2023sdi,QCDSFUKQCDCSSM:2023qlx,Park:2021ypf,Liu:2021irg} that contribute to the current FLAG averages~\cite{FlavourLatticeAveragingGroupFLAG:2024oxs}. We achieve a statistical precision for $g_{T/S}$ that is improved by a factor of three or more over all previous works, with the most substantial gains on physical-point ensembles. This high precision enables a more robust and conservative evaluation of systematic errors. Our final value has a total uncertainty one-half smaller than the previous results.

Using $m_d-m_u=2.35(12)$ MeV--derived from the $N_f=2+1$ FLAG average~\cite{FlavourLatticeAveragingGroupFLAG:2024oxs} of $m_u/m_d$ and $m_l=(m_u+m_d)/2$--and the QED correction $-1.00(7)(14)$ MeV from Ref.~\cite{BMW:2014pzb}, we predict the neutron-proton mass difference\footnote{{In the calculation of $g_S$, we don't consider $u,d$ quark mass difference. This does not affect $m_n-m_p$ at $O(m_d-m_u)$ accuracy, but only contributes to $O[(m_d-m_u)^3]$ correction that is typically ignored in the study of nucleon mass splitting effect.}} to be~\cite{Gonzalez-Alonso:2013ura}:
\begin{align}
m_n-m_p=1.60[0.23]_{\rm tot}(0.11)_{g_S}(0.13)_{\rm ISB}(0.16)_{\rm QED} \ {\rm MeV}.
\end{align}
This result is in agreement with the experimental value of 1.293 MeV within $1.3\sigma$. Employing the newer phenomenological QED correction of $-0.58(16)$ MeV~\cite{Gasser:2020mzy} yields a prediction roughly 3$\sigma$ higher than experiment. {In contrast, other QED calculations~\cite{Walker-Loud:2012ift, Thomas:2014dxa} yield results consistent with experiment, albeit with larger uncertainties. This ambiguity} underscores the importance of an updated direct lattice QCD+QED calculation of the QED correction. 

\textit{Summary}: 
We report a high-precision calculation of the nucleon isovector scalar ($g_S$) and tensor ($g_T$) charges, employing the blending method with current-involved interpolation fields. Our results achieve a significant improvement in precision over the current FLAG averages. Furthermore, they provide robust numerical evidence that the combination of standard and current-involved interpolation fields universally suppresses excited-state contamination (ESC). This suppression is observed to be effective across different current operators, pion masses, volumes, and lattice spacings.

A major strength of the blending method is its capacity to evaluate the efficacy of various interpolation field combinations at arbitrary source-sink separations, without requiring additional propagator generation. The existing propagators produced by this method can, in principle, be used to construct a more complete set of interpolation fields for linear combination of different interpolation operators including the current-involved one, and also for GEVP studies, offering a pathway to further suppress ESC.

Furthermore, our high-precision data reveal a crucial finding: the finite-volume effects (FVEs) for $g_S$ are clearly described by the phenomenological ansatz $e^{-m_\pi L}$. This form is decisively favored over the functional form $m_\pi^2 e^{-m_{\pi}L}/\sqrt{m_\pi L}$ suggested by HB$\chi$PT. {Usage of this chiral suppressed ansatz} would introduce a systematic bias into the extrapolated physical results. {We note that if the FVE of the nucleon mass scales as $m_\pi^2 e^{-m_{\pi}L}$ as suggested by HB$\chi$PT, then taking the derivative with respect to $m_q$ (or equivalently $m_\pi^2$, according to the Feynman-Hellmann theorem $g_{S,q}=\frac{\partial m_N}{\partial m_q}$) will naturally generate a term proportional to $e^{-m_\pi L}/\sqrt{m_\pi L}$.}

Conservative estimation of the FVE systematic uncertainty using the AIC with several ansatzes including both the {$m_\pi^2 e^{-m_{\pi}L}/\sqrt{m_\pi L}$} and also the naive $e^{-m_\pi L}$ introduces a systematic uncertainty comparable to the statistical uncertainty. This uncertainty may be universal for a wide range of hadronic matrix elements, including the non-local matrix elements required for parton distribution function (PDF) calculations within the Large Momentum Effective Theory (LaMET) framework~\cite{Ji:2013dva,Ji:2020ect} {, and is worth follow-up studies.} Consequently, previous studies of nucleon matrix elements that used the {chiral suppressed ansatz} may need to reassess this systematic uncertainty.

\section*{Acknowledgments}

We thank the CLQCD collaborations for providing us their gauge configurations with dynamical fermions~\cite{CLQCD:2023sdb}, which are generated on HPC Cluster of ITP-CAS, IHEP-CAS and CSNS-CAS, the Southern Nuclear Science Computing Center(SNSC) and the Siyuan-1 cluster supported by the Center for High Performance Computing at Shanghai Jiao Tong University. 
We thank Rui Zhang and Srijit Paul for valuable comments and suggestions.
The calculations were performed using the PyQUDA package~\cite{Jiang:2024lto} with QUDA~\cite{Clark:2009wm,Babich:2011np,Clark:2016rdz} through HIP programming model~\cite{Bi:2020wpt}. The numerical calculation were carried out on the ORISE Supercomputer, HPC Cluster of ITP-CAS and Advanced Computing East China Sub-center. This work is supported in part by National Key R\&D Program of China No.2024YFE0109800, NSFC grants No. 12293060, 12293062, 12525504, 12435002 and 12447101, the science and education integration young faculty project of University of Chinese Academy of Sciences, the Strategic Priority Research Program of Chinese Academy of Sciences, Grant No.\ XDB34030303 and YSBR-101.

\bibliographystyle{apsrev4-1}
\bibliography{ref}

\begin{thebibliography}{70}%
\makeatletter
\providecommand \@ifxundefined [1]{%
 \@ifx{#1\undefined}
}%
\providecommand \@ifnum [1]{%
 \ifnum #1\expandafter \@firstoftwo
 \else \expandafter \@secondoftwo
 \fi
}%
\providecommand \@ifx [1]{%
 \ifx #1\expandafter \@firstoftwo
 \else \expandafter \@secondoftwo
 \fi
}%
\providecommand \natexlab [1]{#1}%
\providecommand \enquote  [1]{``#1''}%
\providecommand \bibnamefont  [1]{#1}%
\providecommand \bibfnamefont [1]{#1}%
\providecommand \citenamefont [1]{#1}%
\providecommand \href@noop [0]{\@secondoftwo}%
\providecommand \href [0]{\begingroup \@sanitize@url \@href}%
\providecommand \@href[1]{\@@startlink{#1}\@@href}%
\providecommand \@@href[1]{\endgroup#1\@@endlink}%
\providecommand \@sanitize@url [0]{\catcode `\\12\catcode `\$12\catcode
  `\&12\catcode `\#12\catcode `\^12\catcode `\_12\catcode `\%12\relax}%
\providecommand \@@startlink[1]{}%
\providecommand \@@endlink[0]{}%
\providecommand \url  [0]{\begingroup\@sanitize@url \@url }%
\providecommand \@url [1]{\endgroup\@href {#1}{\urlprefix }}%
\providecommand \urlprefix  [0]{URL }%
\providecommand \Eprint [0]{\href }%
\providecommand \doibase [0]{http://dx.doi.org/}%
\providecommand \selectlanguage [0]{\@gobble}%
\providecommand \bibinfo  [0]{\@secondoftwo}%
\providecommand \bibfield  [0]{\@secondoftwo}%
\providecommand \translation [1]{[#1]}%
\providecommand \BibitemOpen [0]{}%
\providecommand \bibitemStop [0]{}%
\providecommand \bibitemNoStop [0]{.\EOS\space}%
\providecommand \EOS [0]{\spacefactor3000\relax}%
\providecommand \BibitemShut  [1]{\csname bibitem#1\endcsname}%
\let\auto@bib@innerbib\@empty
\bibitem [{\citenamefont {Navas}\ \emph {et~al.}(2024)\citenamefont {Navas}
  \emph {et~al.}}]{ParticleDataGroup:2024cfk}%
  \BibitemOpen
  \bibfield  {author} {\bibinfo {author} {\bibfnamefont {S.}~\bibnamefont
  {Navas}} \emph {et~al.} (\bibinfo {collaboration} {Particle Data Group}),\
  }\href {\doibase 10.1103/PhysRevD.110.030001} {\bibfield  {journal} {\bibinfo
   {journal} {Phys. Rev. D}\ }\textbf {\bibinfo {volume} {110}},\ \bibinfo
  {pages} {030001} (\bibinfo {year} {2024})}\BibitemShut {NoStop}%
\bibitem [{\citenamefont {Bhattacharya}\ \emph {et~al.}(2012)\citenamefont
  {Bhattacharya}, \citenamefont {Cirigliano}, \citenamefont {Cohen},
  \citenamefont {Filipuzzi}, \citenamefont {Gonzalez-Alonso}, \citenamefont
  {Graesser}, \citenamefont {Gupta},\ and\ \citenamefont
  {Lin}}]{Bhattacharya:2011qm}%
  \BibitemOpen
  \bibfield  {author} {\bibinfo {author} {\bibfnamefont {T.}~\bibnamefont
  {Bhattacharya}}, \bibinfo {author} {\bibfnamefont {V.}~\bibnamefont
  {Cirigliano}}, \bibinfo {author} {\bibfnamefont {S.~D.}\ \bibnamefont
  {Cohen}}, \bibinfo {author} {\bibfnamefont {A.}~\bibnamefont {Filipuzzi}},
  \bibinfo {author} {\bibfnamefont {M.}~\bibnamefont {Gonzalez-Alonso}},
  \bibinfo {author} {\bibfnamefont {M.~L.}\ \bibnamefont {Graesser}}, \bibinfo
  {author} {\bibfnamefont {R.}~\bibnamefont {Gupta}}, \ and\ \bibinfo {author}
  {\bibfnamefont {H.-W.}\ \bibnamefont {Lin}},\ }\href {\doibase
  10.1103/PhysRevD.85.054512} {\bibfield  {journal} {\bibinfo  {journal} {Phys.
  Rev. D}\ }\textbf {\bibinfo {volume} {85}},\ \bibinfo {pages} {054512}
  (\bibinfo {year} {2012})},\ \Eprint {http://arxiv.org/abs/1110.6448}
  {arXiv:1110.6448 [hep-ph]} \BibitemShut {NoStop}%
\bibitem [{\citenamefont {Bhattacharya}\ \emph {et~al.}(2016)\citenamefont
  {Bhattacharya}, \citenamefont {Cirigliano}, \citenamefont {Cohen},
  \citenamefont {Gupta}, \citenamefont {Lin},\ and\ \citenamefont
  {Yoon}}]{Bhattacharya:2016zcn}%
  \BibitemOpen
  \bibfield  {author} {\bibinfo {author} {\bibfnamefont {T.}~\bibnamefont
  {Bhattacharya}}, \bibinfo {author} {\bibfnamefont {V.}~\bibnamefont
  {Cirigliano}}, \bibinfo {author} {\bibfnamefont {S.}~\bibnamefont {Cohen}},
  \bibinfo {author} {\bibfnamefont {R.}~\bibnamefont {Gupta}}, \bibinfo
  {author} {\bibfnamefont {H.-W.}\ \bibnamefont {Lin}}, \ and\ \bibinfo
  {author} {\bibfnamefont {B.}~\bibnamefont {Yoon}},\ }\href {\doibase
  10.1103/PhysRevD.94.054508} {\bibfield  {journal} {\bibinfo  {journal} {Phys.
  Rev. D}\ }\textbf {\bibinfo {volume} {94}},\ \bibinfo {pages} {054508}
  (\bibinfo {year} {2016})},\ \Eprint {http://arxiv.org/abs/1606.07049}
  {arXiv:1606.07049 [hep-lat]} \BibitemShut {NoStop}%
\bibitem [{\citenamefont {Borsanyi}\ \emph {et~al.}(2015)\citenamefont
  {Borsanyi} \emph {et~al.}}]{BMW:2014pzb}%
  \BibitemOpen
  \bibfield  {author} {\bibinfo {author} {\bibfnamefont {S.}~\bibnamefont
  {Borsanyi}} \emph {et~al.} (\bibinfo {collaboration} {BMW}),\ }\href
  {\doibase 10.1126/science.1257050} {\bibfield  {journal} {\bibinfo  {journal}
  {Science}\ }\textbf {\bibinfo {volume} {347}},\ \bibinfo {pages} {1452}
  (\bibinfo {year} {2015})},\ \Eprint {http://arxiv.org/abs/1406.4088}
  {arXiv:1406.4088 [hep-lat]} \BibitemShut {NoStop}%
\bibitem [{\citenamefont {Lin}\ \emph {et~al.}(2018)\citenamefont {Lin},
  \citenamefont {Melnitchouk}, \citenamefont {Prokudin}, \citenamefont {Sato},\
  and\ \citenamefont {Shows}}]{Lin:2017stx}%
  \BibitemOpen
  \bibfield  {author} {\bibinfo {author} {\bibfnamefont {H.-W.}\ \bibnamefont
  {Lin}}, \bibinfo {author} {\bibfnamefont {W.}~\bibnamefont {Melnitchouk}},
  \bibinfo {author} {\bibfnamefont {A.}~\bibnamefont {Prokudin}}, \bibinfo
  {author} {\bibfnamefont {N.}~\bibnamefont {Sato}}, \ and\ \bibinfo {author}
  {\bibfnamefont {H.}~\bibnamefont {Shows}},\ }\href {\doibase
  10.1103/PhysRevLett.120.152502} {\bibfield  {journal} {\bibinfo  {journal}
  {Phys. Rev. Lett.}\ }\textbf {\bibinfo {volume} {120}},\ \bibinfo {pages}
  {152502} (\bibinfo {year} {2018})},\ \Eprint
  {http://arxiv.org/abs/1710.09858} {arXiv:1710.09858 [hep-ph]} \BibitemShut
  {NoStop}%
\bibitem [{\citenamefont {Yao}\ \emph {et~al.}(2023)\citenamefont {Yao} \emph
  {et~al.}}]{LatticeParton:2022xsd}%
  \BibitemOpen
  \bibfield  {author} {\bibinfo {author} {\bibfnamefont {F.}~\bibnamefont
  {Yao}} \emph {et~al.} (\bibinfo {collaboration} {Lattice Parton}),\ }\href
  {\doibase 10.1103/PhysRevLett.131.261901} {\bibfield  {journal} {\bibinfo
  {journal} {Phys. Rev. Lett.}\ }\textbf {\bibinfo {volume} {131}},\ \bibinfo
  {pages} {261901} (\bibinfo {year} {2023})},\ \Eprint
  {http://arxiv.org/abs/2208.08008} {arXiv:2208.08008 [hep-lat]} \BibitemShut
  {NoStop}%
\bibitem [{\citenamefont {Harris}\ \emph {et~al.}(2019)\citenamefont {Harris},
  \citenamefont {von Hippel}, \citenamefont {Junnarkar}, \citenamefont {Meyer},
  \citenamefont {Ottnad}, \citenamefont {Wilhelm}, \citenamefont {Wittig},\
  and\ \citenamefont {Wrang}}]{Harris:2019bih}%
  \BibitemOpen
  \bibfield  {author} {\bibinfo {author} {\bibfnamefont {T.}~\bibnamefont
  {Harris}}, \bibinfo {author} {\bibfnamefont {G.}~\bibnamefont {von Hippel}},
  \bibinfo {author} {\bibfnamefont {P.}~\bibnamefont {Junnarkar}}, \bibinfo
  {author} {\bibfnamefont {H.~B.}\ \bibnamefont {Meyer}}, \bibinfo {author}
  {\bibfnamefont {K.}~\bibnamefont {Ottnad}}, \bibinfo {author} {\bibfnamefont
  {J.}~\bibnamefont {Wilhelm}}, \bibinfo {author} {\bibfnamefont
  {H.}~\bibnamefont {Wittig}}, \ and\ \bibinfo {author} {\bibfnamefont
  {L.}~\bibnamefont {Wrang}},\ }\href {\doibase 10.1103/PhysRevD.100.034513}
  {\bibfield  {journal} {\bibinfo  {journal} {Phys. Rev. D}\ }\textbf {\bibinfo
  {volume} {100}},\ \bibinfo {pages} {034513} (\bibinfo {year} {2019})},\
  \Eprint {http://arxiv.org/abs/1905.01291} {arXiv:1905.01291 [hep-lat]}
  \BibitemShut {NoStop}%
\bibitem [{\citenamefont {Bali}\ \emph {et~al.}(2023)\citenamefont {Bali},
  \citenamefont {Collins}, \citenamefont {Heybrock}, \citenamefont {L\"offler},
  \citenamefont {R\"odl}, \citenamefont {S\"oldner},\ and\ \citenamefont
  {Weish\"aupl}}]{Bali:2023sdi}%
  \BibitemOpen
  \bibfield  {author} {\bibinfo {author} {\bibfnamefont {G.~S.}\ \bibnamefont
  {Bali}}, \bibinfo {author} {\bibfnamefont {S.}~\bibnamefont {Collins}},
  \bibinfo {author} {\bibfnamefont {S.}~\bibnamefont {Heybrock}}, \bibinfo
  {author} {\bibfnamefont {M.}~\bibnamefont {L\"offler}}, \bibinfo {author}
  {\bibfnamefont {R.}~\bibnamefont {R\"odl}}, \bibinfo {author} {\bibfnamefont
  {W.}~\bibnamefont {S\"oldner}}, \ and\ \bibinfo {author} {\bibfnamefont
  {S.}~\bibnamefont {Weish\"aupl}} (\bibinfo {collaboration} {RQCD}),\ }\href
  {\doibase 10.1103/PhysRevD.108.034512} {\bibfield  {journal} {\bibinfo
  {journal} {Phys. Rev. D}\ }\textbf {\bibinfo {volume} {108}},\ \bibinfo
  {pages} {034512} (\bibinfo {year} {2023})},\ \Eprint
  {http://arxiv.org/abs/2305.04717} {arXiv:2305.04717 [hep-lat]} \BibitemShut
  {NoStop}%
\bibitem [{\citenamefont {{[LHPC 19] N. Hasan}}\ \emph
  {et~al.}(2019)\citenamefont {{[LHPC 19] N. Hasan}}, \citenamefont {Green},
  \citenamefont {Meinel}, \citenamefont {Engelhardt}, \citenamefont {Krieg},
  \citenamefont {Negele}, \citenamefont {Pochinsky},\ and\ \citenamefont
  {Syritsyn}}]{Hasan:2019noy}%
  \BibitemOpen
  \bibfield  {author} {\bibinfo {author} {\bibnamefont {{[LHPC 19] N. Hasan}}},
  \bibinfo {author} {\bibfnamefont {J.}~\bibnamefont {Green}}, \bibinfo
  {author} {\bibfnamefont {S.}~\bibnamefont {Meinel}}, \bibinfo {author}
  {\bibfnamefont {M.}~\bibnamefont {Engelhardt}}, \bibinfo {author}
  {\bibfnamefont {S.}~\bibnamefont {Krieg}}, \bibinfo {author} {\bibfnamefont
  {J.}~\bibnamefont {Negele}}, \bibinfo {author} {\bibfnamefont
  {A.}~\bibnamefont {Pochinsky}}, \ and\ \bibinfo {author} {\bibfnamefont
  {S.}~\bibnamefont {Syritsyn}},\ }\href {\doibase 10.1103/PhysRevD.99.114505}
  {\bibfield  {journal} {\bibinfo  {journal} {Phys. Rev. D}\ }\textbf {\bibinfo
  {volume} {99}},\ \bibinfo {pages} {114505} (\bibinfo {year} {2019})},\
  \Eprint {http://arxiv.org/abs/1903.06487} {arXiv:1903.06487 [hep-lat]}
  \BibitemShut {NoStop}%
\bibitem [{\citenamefont {{[ETM 19] C. Alexandrou}}\ \emph
  {et~al.}(2020)\citenamefont {{[ETM 19] C. Alexandrou}}, \citenamefont
  {Bacchio}, \citenamefont {Constantinou}, \citenamefont {Finkenrath},
  \citenamefont {Hadjiyiannakou}, \citenamefont {Jansen}, \citenamefont
  {Koutsou},\ and\ \citenamefont {Vaquero Aviles-Casco}}]{Alexandrou:2019brg}%
  \BibitemOpen
  \bibfield  {author} {\bibinfo {author} {\bibnamefont {{[ETM 19] C.
  Alexandrou}}}, \bibinfo {author} {\bibfnamefont {S.}~\bibnamefont {Bacchio}},
  \bibinfo {author} {\bibfnamefont {M.}~\bibnamefont {Constantinou}}, \bibinfo
  {author} {\bibfnamefont {J.}~\bibnamefont {Finkenrath}}, \bibinfo {author}
  {\bibfnamefont {K.}~\bibnamefont {Hadjiyiannakou}}, \bibinfo {author}
  {\bibfnamefont {K.}~\bibnamefont {Jansen}}, \bibinfo {author} {\bibfnamefont
  {G.}~\bibnamefont {Koutsou}}, \ and\ \bibinfo {author} {\bibfnamefont
  {A.}~\bibnamefont {Vaquero Aviles-Casco}},\ }\href {\doibase
  10.1103/PhysRevD.102.054517} {\bibfield  {journal} {\bibinfo  {journal}
  {Phys. Rev. D}\ }\textbf {\bibinfo {volume} {102}},\ \bibinfo {pages}
  {054517} (\bibinfo {year} {2020})},\ \Eprint
  {http://arxiv.org/abs/1909.00485} {arXiv:1909.00485 [hep-lat]} \BibitemShut
  {NoStop}%
\bibitem [{\citenamefont {Ottnad}(2021)}]{Ottnad:2020qbw}%
  \BibitemOpen
  \bibfield  {author} {\bibinfo {author} {\bibfnamefont {K.}~\bibnamefont
  {Ottnad}},\ }\href {\doibase 10.1140/epja/s10050-021-00355-5} {\bibfield
  {journal} {\bibinfo  {journal} {Eur. Phys. J. A}\ }\textbf {\bibinfo {volume}
  {57}},\ \bibinfo {pages} {50} (\bibinfo {year} {2021})},\ \Eprint
  {http://arxiv.org/abs/2011.12471} {arXiv:2011.12471 [hep-lat]} \BibitemShut
  {NoStop}%
\bibitem [{\citenamefont {Alexandrou}\ \emph {et~al.}(2025)\citenamefont
  {Alexandrou}, \citenamefont {Koutsou}, \citenamefont {Li}, \citenamefont
  {Petschlies},\ and\ \citenamefont {Pittler}}]{Alexandrou:2024tps}%
  \BibitemOpen
  \bibfield  {author} {\bibinfo {author} {\bibfnamefont {C.}~\bibnamefont
  {Alexandrou}}, \bibinfo {author} {\bibfnamefont {G.}~\bibnamefont {Koutsou}},
  \bibinfo {author} {\bibfnamefont {Y.}~\bibnamefont {Li}}, \bibinfo {author}
  {\bibfnamefont {M.}~\bibnamefont {Petschlies}}, \ and\ \bibinfo {author}
  {\bibfnamefont {F.}~\bibnamefont {Pittler}},\ }\href {\doibase
  10.22323/1.466.0317} {\bibfield  {journal} {\bibinfo  {journal} {PoS}\
  }\textbf {\bibinfo {volume} {LATTICE2024}},\ \bibinfo {pages} {317} (\bibinfo
  {year} {2025})},\ \Eprint {http://arxiv.org/abs/2412.07263} {arXiv:2412.07263
  [hep-lat]} \BibitemShut {NoStop}%
\bibitem [{\citenamefont {Wang}\ \emph {et~al.}(2024)\citenamefont {Wang},
  \citenamefont {Zhang}, \citenamefont {Cao}, \citenamefont {Fan},
  \citenamefont {Feng}, \citenamefont {Gao}, \citenamefont {Jin},\ and\
  \citenamefont {Liu}}]{Wang:2023omf}%
  \BibitemOpen
  \bibfield  {author} {\bibinfo {author} {\bibfnamefont {X.-H.}\ \bibnamefont
  {Wang}}, \bibinfo {author} {\bibfnamefont {Z.-L.}\ \bibnamefont {Zhang}},
  \bibinfo {author} {\bibfnamefont {X.-H.}\ \bibnamefont {Cao}}, \bibinfo
  {author} {\bibfnamefont {C.-L.}\ \bibnamefont {Fan}}, \bibinfo {author}
  {\bibfnamefont {X.}~\bibnamefont {Feng}}, \bibinfo {author} {\bibfnamefont
  {Y.-S.}\ \bibnamefont {Gao}}, \bibinfo {author} {\bibfnamefont {L.-C.}\
  \bibnamefont {Jin}}, \ and\ \bibinfo {author} {\bibfnamefont
  {C.}~\bibnamefont {Liu}},\ }\href {\doibase 10.1103/PhysRevLett.133.141901}
  {\bibfield  {journal} {\bibinfo  {journal} {Phys. Rev. Lett.}\ }\textbf
  {\bibinfo {volume} {133}},\ \bibinfo {pages} {141901} (\bibinfo {year}
  {2024})},\ \Eprint {http://arxiv.org/abs/2310.01168} {arXiv:2310.01168
  [hep-lat]} \BibitemShut {NoStop}%
\bibitem [{\citenamefont {Barca}\ \emph {et~al.}(2025)\citenamefont {Barca},
  \citenamefont {Bali},\ and\ \citenamefont {Collins}}]{Barca:2024hrl}%
  \BibitemOpen
  \bibfield  {author} {\bibinfo {author} {\bibfnamefont {L.}~\bibnamefont
  {Barca}}, \bibinfo {author} {\bibfnamefont {G.}~\bibnamefont {Bali}}, \ and\
  \bibinfo {author} {\bibfnamefont {S.}~\bibnamefont {Collins}},\ }\href
  {\doibase 10.1103/PhysRevD.111.L031505} {\bibfield  {journal} {\bibinfo
  {journal} {Phys. Rev. D}\ }\textbf {\bibinfo {volume} {111}},\ \bibinfo
  {pages} {L031505} (\bibinfo {year} {2025})},\ \Eprint
  {http://arxiv.org/abs/2412.13138} {arXiv:2412.13138 [hep-lat]} \BibitemShut
  {NoStop}%
\bibitem [{\citenamefont {Fucito}\ \emph {et~al.}(1982)\citenamefont {Fucito},
  \citenamefont {Parisi},\ and\ \citenamefont {Petrarca}}]{Fucito:1982ff}%
  \BibitemOpen
  \bibfield  {author} {\bibinfo {author} {\bibfnamefont {F.}~\bibnamefont
  {Fucito}}, \bibinfo {author} {\bibfnamefont {G.}~\bibnamefont {Parisi}}, \
  and\ \bibinfo {author} {\bibfnamefont {S.}~\bibnamefont {Petrarca}},\ }\href
  {\doibase 10.1016/0370-2693(82)90816-4} {\bibfield  {journal} {\bibinfo
  {journal} {Phys. Lett. B}\ }\textbf {\bibinfo {volume} {115}},\ \bibinfo
  {pages} {148} (\bibinfo {year} {1982})}\BibitemShut {NoStop}%
\bibitem [{\citenamefont {Capitani}\ \emph {et~al.}(2012)\citenamefont
  {Capitani}, \citenamefont {Della~Morte}, \citenamefont {von Hippel},
  \citenamefont {Jager}, \citenamefont {Juttner}, \citenamefont {Knippschild},
  \citenamefont {Meyer},\ and\ \citenamefont {Wittig}}]{Capitani:2012gj}%
  \BibitemOpen
  \bibfield  {author} {\bibinfo {author} {\bibfnamefont {S.}~\bibnamefont
  {Capitani}}, \bibinfo {author} {\bibfnamefont {M.}~\bibnamefont
  {Della~Morte}}, \bibinfo {author} {\bibfnamefont {G.}~\bibnamefont {von
  Hippel}}, \bibinfo {author} {\bibfnamefont {B.}~\bibnamefont {Jager}},
  \bibinfo {author} {\bibfnamefont {A.}~\bibnamefont {Juttner}}, \bibinfo
  {author} {\bibfnamefont {B.}~\bibnamefont {Knippschild}}, \bibinfo {author}
  {\bibfnamefont {H.~B.}\ \bibnamefont {Meyer}}, \ and\ \bibinfo {author}
  {\bibfnamefont {H.}~\bibnamefont {Wittig}},\ }\href {\doibase
  10.1103/PhysRevD.86.074502} {\bibfield  {journal} {\bibinfo  {journal} {Phys.
  Rev. D}\ }\textbf {\bibinfo {volume} {86}},\ \bibinfo {pages} {074502}
  (\bibinfo {year} {2012})},\ \Eprint {http://arxiv.org/abs/1205.0180}
  {arXiv:1205.0180 [hep-lat]} \BibitemShut {NoStop}%
\bibitem [{\citenamefont {de~Divitiis}\ \emph {et~al.}(2012)\citenamefont
  {de~Divitiis}, \citenamefont {Petronzio},\ and\ \citenamefont
  {Tantalo}}]{deDivitiis:2012vs}%
  \BibitemOpen
  \bibfield  {author} {\bibinfo {author} {\bibfnamefont {G.~M.}\ \bibnamefont
  {de~Divitiis}}, \bibinfo {author} {\bibfnamefont {R.}~\bibnamefont
  {Petronzio}}, \ and\ \bibinfo {author} {\bibfnamefont {N.}~\bibnamefont
  {Tantalo}},\ }\href {\doibase 10.1016/j.physletb.2012.10.035} {\bibfield
  {journal} {\bibinfo  {journal} {Phys. Lett. B}\ }\textbf {\bibinfo {volume}
  {718}},\ \bibinfo {pages} {589} (\bibinfo {year} {2012})},\ \Eprint
  {http://arxiv.org/abs/1208.5914} {arXiv:1208.5914 [hep-lat]} \BibitemShut
  {NoStop}%
\bibitem [{\citenamefont {Bouchard}\ \emph {et~al.}(2017)\citenamefont
  {Bouchard}, \citenamefont {Chang}, \citenamefont {Kurth}, \citenamefont
  {Orginos},\ and\ \citenamefont {Walker-Loud}}]{Bouchard:2016heu}%
  \BibitemOpen
  \bibfield  {author} {\bibinfo {author} {\bibfnamefont {C.}~\bibnamefont
  {Bouchard}}, \bibinfo {author} {\bibfnamefont {C.~C.}\ \bibnamefont {Chang}},
  \bibinfo {author} {\bibfnamefont {T.}~\bibnamefont {Kurth}}, \bibinfo
  {author} {\bibfnamefont {K.}~\bibnamefont {Orginos}}, \ and\ \bibinfo
  {author} {\bibfnamefont {A.}~\bibnamefont {Walker-Loud}},\ }\href {\doibase
  10.1103/PhysRevD.96.014504} {\bibfield  {journal} {\bibinfo  {journal} {Phys.
  Rev. D}\ }\textbf {\bibinfo {volume} {96}},\ \bibinfo {pages} {014504}
  (\bibinfo {year} {2017})},\ \Eprint {http://arxiv.org/abs/1612.06963}
  {arXiv:1612.06963 [hep-lat]} \BibitemShut {NoStop}%
\bibitem [{\citenamefont {Chang}\ \emph {et~al.}(2018)\citenamefont {Chang}
  \emph {et~al.}}]{Chang:2018uxx}%
  \BibitemOpen
  \bibfield  {author} {\bibinfo {author} {\bibfnamefont {C.~C.}\ \bibnamefont
  {Chang}} \emph {et~al.},\ }\href {\doibase 10.1038/s41586-018-0161-8}
  {\bibfield  {journal} {\bibinfo  {journal} {Nature}\ }\textbf {\bibinfo
  {volume} {558}},\ \bibinfo {pages} {91} (\bibinfo {year} {2018})},\ \Eprint
  {http://arxiv.org/abs/1805.12130} {arXiv:1805.12130 [hep-lat]} \BibitemShut
  {NoStop}%
\bibitem [{\citenamefont {He}\ \emph {et~al.}(2022)\citenamefont {He} \emph
  {et~al.}}]{He:2021yvm}%
  \BibitemOpen
  \bibfield  {author} {\bibinfo {author} {\bibfnamefont {J.}~\bibnamefont {He}}
  \emph {et~al.},\ }\href {\doibase 10.1103/PhysRevC.105.065203} {\bibfield
  {journal} {\bibinfo  {journal} {Phys. Rev. C}\ }\textbf {\bibinfo {volume}
  {105}},\ \bibinfo {pages} {065203} (\bibinfo {year} {2022})},\ \Eprint
  {http://arxiv.org/abs/2104.05226} {arXiv:2104.05226 [hep-lat]} \BibitemShut
  {NoStop}%
\bibitem [{\citenamefont {Aoki}\ \emph {et~al.}(2024)\citenamefont {Aoki} \emph
  {et~al.}}]{FlavourLatticeAveragingGroupFLAG:2024oxs}%
  \BibitemOpen
  \bibfield  {author} {\bibinfo {author} {\bibfnamefont {Y.}~\bibnamefont
  {Aoki}} \emph {et~al.} (\bibinfo {collaboration} {Flavour Lattice Averaging
  Group (FLAG)}),\ }\href@noop {} {\  (\bibinfo {year} {2024})},\ \Eprint
  {http://arxiv.org/abs/2411.04268} {arXiv:2411.04268 [hep-lat]} \BibitemShut
  {NoStop}%
\bibitem [{\citenamefont {Bali}\ \emph {et~al.}(2015)\citenamefont {Bali},
  \citenamefont {Collins}, \citenamefont {Gl{\"a}ssle}, \citenamefont
  {G{\"o}ckeler}, \citenamefont {Najjar}, \citenamefont {R{\"o}dl},
  \citenamefont {Sch{\"a}fer}, \citenamefont {Schiel}, \citenamefont
  {S{\"o}ldner},\ and\ \citenamefont {Sternbeck}}]{Bali:2014nma}%
  \BibitemOpen
  \bibfield  {author} {\bibinfo {author} {\bibfnamefont {G.~S.}\ \bibnamefont
  {Bali}}, \bibinfo {author} {\bibfnamefont {S.}~\bibnamefont {Collins}},
  \bibinfo {author} {\bibfnamefont {B.}~\bibnamefont {Gl{\"a}ssle}}, \bibinfo
  {author} {\bibfnamefont {M.}~\bibnamefont {G{\"o}ckeler}}, \bibinfo {author}
  {\bibfnamefont {J.}~\bibnamefont {Najjar}}, \bibinfo {author} {\bibfnamefont
  {R.~H.}\ \bibnamefont {R{\"o}dl}}, \bibinfo {author} {\bibfnamefont
  {A.}~\bibnamefont {Sch{\"a}fer}}, \bibinfo {author} {\bibfnamefont {R.~W.}\
  \bibnamefont {Schiel}}, \bibinfo {author} {\bibfnamefont {W.}~\bibnamefont
  {S{\"o}ldner}}, \ and\ \bibinfo {author} {\bibfnamefont {A.}~\bibnamefont
  {Sternbeck}},\ }\href {\doibase 10.1103/PhysRevD.91.054501} {\bibfield
  {journal} {\bibinfo  {journal} {Phys. Rev. D}\ }\textbf {\bibinfo {volume}
  {91}},\ \bibinfo {pages} {054501} (\bibinfo {year} {2015})},\ \Eprint
  {http://arxiv.org/abs/1412.7336} {arXiv:1412.7336 [hep-lat]} \BibitemShut
  {NoStop}%
\bibitem [{\citenamefont {Capitani}\ \emph {et~al.}(2019)\citenamefont
  {Capitani}, \citenamefont {Della~Morte}, \citenamefont {Djukanovic},
  \citenamefont {von Hippel}, \citenamefont {Hua}, \citenamefont {J{\"a}ger},
  \citenamefont {Junnarkar}, \citenamefont {Meyer}, \citenamefont {Rae},\ and\
  \citenamefont {Wittig}}]{Capitani:2017qpc}%
  \BibitemOpen
  \bibfield  {author} {\bibinfo {author} {\bibfnamefont {S.}~\bibnamefont
  {Capitani}}, \bibinfo {author} {\bibfnamefont {M.}~\bibnamefont
  {Della~Morte}}, \bibinfo {author} {\bibfnamefont {D.}~\bibnamefont
  {Djukanovic}}, \bibinfo {author} {\bibfnamefont {G.~M.}\ \bibnamefont {von
  Hippel}}, \bibinfo {author} {\bibfnamefont {J.}~\bibnamefont {Hua}}, \bibinfo
  {author} {\bibfnamefont {B.}~\bibnamefont {J{\"a}ger}}, \bibinfo {author}
  {\bibfnamefont {P.~M.}\ \bibnamefont {Junnarkar}}, \bibinfo {author}
  {\bibfnamefont {H.~B.}\ \bibnamefont {Meyer}}, \bibinfo {author}
  {\bibfnamefont {T.~D.}\ \bibnamefont {Rae}}, \ and\ \bibinfo {author}
  {\bibfnamefont {H.}~\bibnamefont {Wittig}},\ }\href {\doibase
  10.1142/S0217751X1950009X} {\bibfield  {journal} {\bibinfo  {journal} {Int.
  J. Mod. Phys. A}\ }\textbf {\bibinfo {volume} {34}},\ \bibinfo {pages}
  {1950009} (\bibinfo {year} {2019})},\ \Eprint
  {http://arxiv.org/abs/1705.06186} {arXiv:1705.06186 [hep-lat]} \BibitemShut
  {NoStop}%
\bibitem [{\citenamefont {Djukanovic}\ \emph {et~al.}(2024)\citenamefont
  {Djukanovic}, \citenamefont {von Hippel}, \citenamefont {Meyer},
  \citenamefont {Ottnad},\ and\ \citenamefont {Wittig}}]{Mainz24}%
  \BibitemOpen
  \bibfield  {author} {\bibinfo {author} {\bibfnamefont {D.}~\bibnamefont
  {Djukanovic}}, \bibinfo {author} {\bibfnamefont {G.}~\bibnamefont {von
  Hippel}}, \bibinfo {author} {\bibfnamefont {H.~B.}\ \bibnamefont {Meyer}},
  \bibinfo {author} {\bibfnamefont {K.}~\bibnamefont {Ottnad}}, \ and\ \bibinfo
  {author} {\bibfnamefont {H.}~\bibnamefont {Wittig}},\ }\href {\doibase
  10.1103/PhysRevD.109.074507} {\bibfield  {journal} {\bibinfo  {journal}
  {Phys. Rev. D}\ }\textbf {\bibinfo {volume} {109}},\ \bibinfo {pages}
  {074507} (\bibinfo {year} {2024})},\ \Eprint
  {http://arxiv.org/abs/2402.03024} {arXiv:2402.03024 [hep-lat]} \BibitemShut
  {NoStop}%
\bibitem [{\citenamefont {Jang}\ \emph {et~al.}(2024)\citenamefont {Jang},
  \citenamefont {Gupta}, \citenamefont {Bhattacharya}, \citenamefont {Yoon},\
  and\ \citenamefont {Lin}}]{Jang:2023zts}%
  \BibitemOpen
  \bibfield  {author} {\bibinfo {author} {\bibfnamefont {Y.-C.}\ \bibnamefont
  {Jang}}, \bibinfo {author} {\bibfnamefont {R.}~\bibnamefont {Gupta}},
  \bibinfo {author} {\bibfnamefont {T.}~\bibnamefont {Bhattacharya}}, \bibinfo
  {author} {\bibfnamefont {B.}~\bibnamefont {Yoon}}, \ and\ \bibinfo {author}
  {\bibfnamefont {H.-W.}\ \bibnamefont {Lin}} (\bibinfo {collaboration}
  {Precision Neutron Decay Matrix Elements (PNDME)}),\ }\href {\doibase
  10.1103/PhysRevD.109.014503} {\bibfield  {journal} {\bibinfo  {journal}
  {Phys. Rev. D}\ }\textbf {\bibinfo {volume} {109}},\ \bibinfo {pages}
  {014503} (\bibinfo {year} {2024})},\ \Eprint
  {http://arxiv.org/abs/2305.11330} {arXiv:2305.11330 [hep-lat]} \BibitemShut
  {NoStop}%
\bibitem [{\citenamefont {Beane}\ and\ \citenamefont
  {Savage}(2004)}]{Beane:2004rf}%
  \BibitemOpen
  \bibfield  {author} {\bibinfo {author} {\bibfnamefont {S.~R.}\ \bibnamefont
  {Beane}}\ and\ \bibinfo {author} {\bibfnamefont {M.~J.}\ \bibnamefont
  {Savage}},\ }\href {\doibase 10.1103/PhysRevD.70.074029} {\bibfield
  {journal} {\bibinfo  {journal} {Phys. Rev. D}\ }\textbf {\bibinfo {volume}
  {70}},\ \bibinfo {pages} {074029} (\bibinfo {year} {2004})},\ \Eprint
  {http://arxiv.org/abs/hep-ph/0404131} {arXiv:hep-ph/0404131} \BibitemShut
  {NoStop}%
\bibitem [{\citenamefont {Hu}\ \emph {et~al.}(2025)\citenamefont {Hu},
  \citenamefont {Wang}, \citenamefont {Jiang}, \citenamefont {Liu},
  \citenamefont {Su}, \citenamefont {Sun},\ and\ \citenamefont
  {Yang}}]{Hu:2025vhd}%
  \BibitemOpen
  \bibfield  {author} {\bibinfo {author} {\bibfnamefont {Z.-C.}\ \bibnamefont
  {Hu}}, \bibinfo {author} {\bibfnamefont {J.-H.}\ \bibnamefont {Wang}},
  \bibinfo {author} {\bibfnamefont {X.}~\bibnamefont {Jiang}}, \bibinfo
  {author} {\bibfnamefont {L.}~\bibnamefont {Liu}}, \bibinfo {author}
  {\bibfnamefont {S.-H.}\ \bibnamefont {Su}}, \bibinfo {author} {\bibfnamefont
  {P.}~\bibnamefont {Sun}}, \ and\ \bibinfo {author} {\bibfnamefont {Y.-B.}\
  \bibnamefont {Yang}},\ }\href@noop {} {\  (\bibinfo {year} {2025})},\ \Eprint
  {http://arxiv.org/abs/2505.01719} {arXiv:2505.01719 [hep-lat]} \BibitemShut
  {NoStop}%
\bibitem [{\citenamefont {Barca}(2025)}]{Barca:2025det}%
  \BibitemOpen
  \bibfield  {author} {\bibinfo {author} {\bibfnamefont {L.}~\bibnamefont
  {Barca}},\ }\href@noop {} {\  (\bibinfo {year} {2025})},\ \Eprint
  {http://arxiv.org/abs/2508.09006} {arXiv:2508.09006 [hep-lat]} \BibitemShut
  {NoStop}%
\bibitem [{gA_()}]{gA_work}%
  \BibitemOpen
  \href@noop {} {\bibinfo  {journal} {Z.-C. Hu et al., in preparation}\
  }\BibitemShut {NoStop}%
\bibitem [{\citenamefont {Cirigliano}\ \emph {et~al.}(2022)\citenamefont
  {Cirigliano}, \citenamefont {de~Vries}, \citenamefont {Hayen}, \citenamefont
  {Mereghetti},\ and\ \citenamefont {Walker-Loud}}]{Cirigliano:2022hob}%
  \BibitemOpen
\bibfield  {journal} {  }\bibfield  {author} {\bibinfo {author} {\bibfnamefont
  {V.}~\bibnamefont {Cirigliano}}, \bibinfo {author} {\bibfnamefont
  {J.}~\bibnamefont {de~Vries}}, \bibinfo {author} {\bibfnamefont
  {L.}~\bibnamefont {Hayen}}, \bibinfo {author} {\bibfnamefont
  {E.}~\bibnamefont {Mereghetti}}, \ and\ \bibinfo {author} {\bibfnamefont
  {A.}~\bibnamefont {Walker-Loud}},\ }\href {\doibase
  10.1103/PhysRevLett.129.121801} {\bibfield  {journal} {\bibinfo  {journal}
  {Phys. Rev. Lett.}\ }\textbf {\bibinfo {volume} {129}},\ \bibinfo {pages}
  {121801} (\bibinfo {year} {2022})},\ \Eprint
  {http://arxiv.org/abs/2202.10439} {arXiv:2202.10439 [nucl-th]} \BibitemShut
  {NoStop}%
\bibitem [{\citenamefont {Tomalak}(2025)}]{Tomalak:2025jtn}%
  \BibitemOpen
  \bibfield  {author} {\bibinfo {author} {\bibfnamefont {O.}~\bibnamefont
  {Tomalak}},\ }\href@noop {} {\  (\bibinfo {year} {2025})},\ \Eprint
  {http://arxiv.org/abs/2512.07956} {arXiv:2512.07956 [hep-ph]} \BibitemShut
  {NoStop}%
\bibitem [{\citenamefont {Hall}\ \emph {et~al.}(2025)\citenamefont {Hall} \emph
  {et~al.}}]{Hall:2025ytt}%
  \BibitemOpen
  \bibfield  {author} {\bibinfo {author} {\bibfnamefont {Z.~B.}\ \bibnamefont
  {Hall}} \emph {et~al.},\ }\href@noop {} {\  (\bibinfo {year} {2025})},\
  \Eprint {http://arxiv.org/abs/2503.09891} {arXiv:2503.09891 [hep-lat]}
  \BibitemShut {NoStop}%
\bibitem [{\citenamefont {Hu}\ \emph {et~al.}(2024)\citenamefont {Hu} \emph
  {et~al.}}]{CLQCD:2023sdb}%
  \BibitemOpen
  \bibfield  {author} {\bibinfo {author} {\bibfnamefont {Z.-C.}\ \bibnamefont
  {Hu}} \emph {et~al.} (\bibinfo {collaboration} {CLQCD}),\ }\href {\doibase
  10.1103/PhysRevD.109.054507} {\bibfield  {journal} {\bibinfo  {journal}
  {Phys. Rev. D}\ }\textbf {\bibinfo {volume} {109}},\ \bibinfo {pages}
  {054507} (\bibinfo {year} {2024})},\ \Eprint
  {http://arxiv.org/abs/2310.00814} {arXiv:2310.00814 [hep-lat]} \BibitemShut
  {NoStop}%
\bibitem [{\citenamefont {Du}\ \emph {et~al.}(2025)\citenamefont {Du} \emph
  {et~al.}}]{CLQCD:2024yyn}%
  \BibitemOpen
  \bibfield  {author} {\bibinfo {author} {\bibfnamefont {H.-Y.}\ \bibnamefont
  {Du}} \emph {et~al.} (\bibinfo {collaboration} {CLQCD}),\ }\href {\doibase
  10.1103/PhysRevD.111.054504} {\bibfield  {journal} {\bibinfo  {journal}
  {Phys. Rev. D}\ }\textbf {\bibinfo {volume} {111}},\ \bibinfo {pages}
  {054504} (\bibinfo {year} {2025})},\ \Eprint
  {http://arxiv.org/abs/2408.03548} {arXiv:2408.03548 [hep-lat]} \BibitemShut
  {NoStop}%
\bibitem [{sup()}]{supplemental}%
  \BibitemOpen
  \href@noop {} {\bibinfo  {journal} {Supplemental materials}\ }\BibitemShut
  {NoStop}%
\bibitem [{\citenamefont {Peardon}\ \emph {et~al.}(2009)\citenamefont
  {Peardon}, \citenamefont {Bulava}, \citenamefont {Foley}, \citenamefont
  {Morningstar}, \citenamefont {Dudek}, \citenamefont {Edwards}, \citenamefont
  {Joo}, \citenamefont {Lin}, \citenamefont {Richards},\ and\ \citenamefont
  {Juge}}]{HadronSpectrum:2009krc}%
  \BibitemOpen
\bibfield  {journal} {  }\bibfield  {author} {\bibinfo {author} {\bibfnamefont
  {M.}~\bibnamefont {Peardon}}, \bibinfo {author} {\bibfnamefont
  {J.}~\bibnamefont {Bulava}}, \bibinfo {author} {\bibfnamefont
  {J.}~\bibnamefont {Foley}}, \bibinfo {author} {\bibfnamefont
  {C.}~\bibnamefont {Morningstar}}, \bibinfo {author} {\bibfnamefont
  {J.}~\bibnamefont {Dudek}}, \bibinfo {author} {\bibfnamefont {R.~G.}\
  \bibnamefont {Edwards}}, \bibinfo {author} {\bibfnamefont {B.}~\bibnamefont
  {Joo}}, \bibinfo {author} {\bibfnamefont {H.-W.}\ \bibnamefont {Lin}},
  \bibinfo {author} {\bibfnamefont {D.~G.}\ \bibnamefont {Richards}}, \ and\
  \bibinfo {author} {\bibfnamefont {K.~J.}\ \bibnamefont {Juge}} (\bibinfo
  {collaboration} {Hadron Spectrum}),\ }\href {\doibase
  10.1103/PhysRevD.80.054506} {\bibfield  {journal} {\bibinfo  {journal} {Phys.
  Rev. D}\ }\textbf {\bibinfo {volume} {80}},\ \bibinfo {pages} {054506}
  (\bibinfo {year} {2009})},\ \Eprint {http://arxiv.org/abs/0905.2160}
  {arXiv:0905.2160 [hep-lat]} \BibitemShut {NoStop}%
\bibitem [{\citenamefont {Morningstar}\ \emph {et~al.}(2011)\citenamefont
  {Morningstar}, \citenamefont {Bulava}, \citenamefont {Foley}, \citenamefont
  {Juge}, \citenamefont {Lenkner}, \citenamefont {Peardon},\ and\ \citenamefont
  {Wong}}]{Morningstar:2011ka}%
  \BibitemOpen
  \bibfield  {author} {\bibinfo {author} {\bibfnamefont {C.}~\bibnamefont
  {Morningstar}}, \bibinfo {author} {\bibfnamefont {J.}~\bibnamefont {Bulava}},
  \bibinfo {author} {\bibfnamefont {J.}~\bibnamefont {Foley}}, \bibinfo
  {author} {\bibfnamefont {K.~J.}\ \bibnamefont {Juge}}, \bibinfo {author}
  {\bibfnamefont {D.}~\bibnamefont {Lenkner}}, \bibinfo {author} {\bibfnamefont
  {M.}~\bibnamefont {Peardon}}, \ and\ \bibinfo {author} {\bibfnamefont
  {C.~H.}\ \bibnamefont {Wong}},\ }\href {\doibase 10.1103/PhysRevD.83.114505}
  {\bibfield  {journal} {\bibinfo  {journal} {Phys. Rev. D}\ }\textbf {\bibinfo
  {volume} {83}},\ \bibinfo {pages} {114505} (\bibinfo {year} {2011})},\
  \Eprint {http://arxiv.org/abs/1104.3870} {arXiv:1104.3870 [hep-lat]}
  \BibitemShut {NoStop}%
\bibitem [{\citenamefont {{[ETM 22] C. Alexandrou}}\ \emph
  {et~al.}(2023)\citenamefont {{[ETM 22] C. Alexandrou}} \emph
  {et~al.}}]{Alexandrou:2022dtc}%
  \BibitemOpen
  \bibfield  {author} {\bibinfo {author} {\bibnamefont {{[ETM 22] C.
  Alexandrou}}} \emph {et~al.},\ }\href {\doibase 10.1103/PhysRevD.107.054504}
  {\bibfield  {journal} {\bibinfo  {journal} {Phys. Rev. D}\ }\textbf {\bibinfo
  {volume} {107}},\ \bibinfo {pages} {054504} (\bibinfo {year} {2023})},\
  \Eprint {http://arxiv.org/abs/2202.09871} {arXiv:2202.09871 [hep-lat]}
  \BibitemShut {NoStop}%
\bibitem [{\citenamefont {{[PNDME 15] T. Bhattacharya}}\ \emph
  {et~al.}(2015)\citenamefont {{[PNDME 15] T. Bhattacharya}}, \citenamefont
  {Cirigliano}, \citenamefont {Gupta}, \citenamefont {Lin},\ and\ \citenamefont
  {Yoon}}]{Bhattacharya:2015esa}%
  \BibitemOpen
  \bibfield  {author} {\bibinfo {author} {\bibnamefont {{[PNDME 15] T.
  Bhattacharya}}}, \bibinfo {author} {\bibfnamefont {V.}~\bibnamefont
  {Cirigliano}}, \bibinfo {author} {\bibfnamefont {R.}~\bibnamefont {Gupta}},
  \bibinfo {author} {\bibfnamefont {H.-W.}\ \bibnamefont {Lin}}, \ and\
  \bibinfo {author} {\bibfnamefont {B.}~\bibnamefont {Yoon}},\ }\href {\doibase
  10.1103/PhysRevLett.115.212002} {\bibfield  {journal} {\bibinfo  {journal}
  {Phys. Rev. Lett.}\ }\textbf {\bibinfo {volume} {115}},\ \bibinfo {pages}
  {212002} (\bibinfo {year} {2015})},\ \Eprint
  {http://arxiv.org/abs/1506.04196} {arXiv:1506.04196 [hep-lat]} \BibitemShut
  {NoStop}%
\bibitem [{\citenamefont {{[PNDME 15A] T. Bhattacharya}}\ \emph
  {et~al.}(2015)\citenamefont {{[PNDME 15A] T. Bhattacharya}}, \citenamefont
  {Cirigliano}, \citenamefont {Cohen}, \citenamefont {Gupta}, \citenamefont
  {Joseph}, \citenamefont {Lin},\ and\ \citenamefont
  {Yoon}}]{Bhattacharya:2015wna}%
  \BibitemOpen
  \bibfield  {author} {\bibinfo {author} {\bibnamefont {{[PNDME 15A] T.
  Bhattacharya}}}, \bibinfo {author} {\bibfnamefont {V.}~\bibnamefont
  {Cirigliano}}, \bibinfo {author} {\bibfnamefont {S.}~\bibnamefont {Cohen}},
  \bibinfo {author} {\bibfnamefont {R.}~\bibnamefont {Gupta}}, \bibinfo
  {author} {\bibfnamefont {A.}~\bibnamefont {Joseph}}, \bibinfo {author}
  {\bibfnamefont {H.-W.}\ \bibnamefont {Lin}}, \ and\ \bibinfo {author}
  {\bibfnamefont {B.}~\bibnamefont {Yoon}},\ }\href {\doibase
  10.1103/PhysRevD.92.094511} {\bibfield  {journal} {\bibinfo  {journal} {Phys.
  Rev.}\ }\textbf {\bibinfo {volume} {D92}},\ \bibinfo {pages} {094511}
  (\bibinfo {year} {2015})},\ \Eprint {http://arxiv.org/abs/1506.06411}
  {arXiv:1506.06411 [hep-lat]} \BibitemShut {NoStop}%
\bibitem [{\citenamefont {{[PNDME 18] R. Gupta}}\ \emph
  {et~al.}(2018)\citenamefont {{[PNDME 18] R. Gupta}}, \citenamefont {Jang},
  \citenamefont {Yoon}, \citenamefont {Lin}, \citenamefont {Cirigliano},\ and\
  \citenamefont {Bhattacharya}}]{Gupta:2018qil}%
  \BibitemOpen
  \bibfield  {author} {\bibinfo {author} {\bibnamefont {{[PNDME 18] R.
  Gupta}}}, \bibinfo {author} {\bibfnamefont {Y.-C.}\ \bibnamefont {Jang}},
  \bibinfo {author} {\bibfnamefont {B.}~\bibnamefont {Yoon}}, \bibinfo {author}
  {\bibfnamefont {H.-W.}\ \bibnamefont {Lin}}, \bibinfo {author} {\bibfnamefont
  {V.}~\bibnamefont {Cirigliano}}, \ and\ \bibinfo {author} {\bibfnamefont
  {T.}~\bibnamefont {Bhattacharya}},\ }\href {\doibase
  10.1103/PhysRevD.98.034503} {\bibfield  {journal} {\bibinfo  {journal} {Phys.
  Rev.}\ }\textbf {\bibinfo {volume} {D98}},\ \bibinfo {pages} {034503}
  (\bibinfo {year} {2018})},\ \Eprint {http://arxiv.org/abs/1806.09006}
  {arXiv:1806.09006 [hep-lat]} \BibitemShut {NoStop}%
\bibitem [{\citenamefont {{[QCDSF/UKQCD/CSSM 23] R.~E.~Smail}}\ \emph
  {et~al.}(2023)\citenamefont {{[QCDSF/UKQCD/CSSM 23] R.~E.~Smail}} \emph
  {et~al.}}]{QCDSFUKQCDCSSM:2023qlx}%
  \BibitemOpen
  \bibfield  {author} {\bibinfo {author} {\bibnamefont {{[QCDSF/UKQCD/CSSM 23]
  R.~E.~Smail}}} \emph {et~al.},\ }\href {\doibase 10.1103/PhysRevD.108.094511}
  {\bibfield  {journal} {\bibinfo  {journal} {Phys. Rev. D}\ }\textbf {\bibinfo
  {volume} {108}},\ \bibinfo {pages} {094511} (\bibinfo {year} {2023})},\
  \Eprint {http://arxiv.org/abs/2304.02866} {arXiv:2304.02866 [hep-lat]}
  \BibitemShut {NoStop}%
\bibitem [{\citenamefont {{[NME 21] S. Park}}\ \emph
  {et~al.}(2022)\citenamefont {{[NME 21] S. Park}}, \citenamefont {Gupta},
  \citenamefont {Yoon}, \citenamefont {Mondal}, \citenamefont {Bhattacharya},
  \citenamefont {Jang}, \citenamefont {Jo\'o},\ and\ \citenamefont
  {Winter}}]{Park:2021ypf}%
  \BibitemOpen
  \bibfield  {author} {\bibinfo {author} {\bibnamefont {{[NME 21] S. Park}}},
  \bibinfo {author} {\bibfnamefont {R.}~\bibnamefont {Gupta}}, \bibinfo
  {author} {\bibfnamefont {B.}~\bibnamefont {Yoon}}, \bibinfo {author}
  {\bibfnamefont {S.}~\bibnamefont {Mondal}}, \bibinfo {author} {\bibfnamefont
  {T.}~\bibnamefont {Bhattacharya}}, \bibinfo {author} {\bibfnamefont {Y.-C.}\
  \bibnamefont {Jang}}, \bibinfo {author} {\bibfnamefont {B.}~\bibnamefont
  {Jo\'o}}, \ and\ \bibinfo {author} {\bibfnamefont {F.}~\bibnamefont
  {Winter}},\ }\href {\doibase 10.1103/PhysRevD.105.054505} {\bibfield
  {journal} {\bibinfo  {journal} {Phys. Rev. D}\ }\textbf {\bibinfo {volume}
  {105}},\ \bibinfo {pages} {054505} (\bibinfo {year} {2022})},\ \Eprint
  {http://arxiv.org/abs/2103.05599} {arXiv:2103.05599 [hep-lat]} \BibitemShut
  {NoStop}%
\bibitem [{\citenamefont {{[$\chi$QCD 21A] L. Liu}}\ \emph
  {et~al.}(2021)\citenamefont {{[$\chi$QCD 21A] L. Liu}}, \citenamefont {Chen},
  \citenamefont {Draper}, \citenamefont {Liang}, \citenamefont {Liu},
  \citenamefont {Wang},\ and\ \citenamefont {Yang}}]{Liu:2021irg}%
  \BibitemOpen
  \bibfield  {author} {\bibinfo {author} {\bibnamefont {{[$\chi$QCD 21A] L.
  Liu}}}, \bibinfo {author} {\bibfnamefont {T.}~\bibnamefont {Chen}}, \bibinfo
  {author} {\bibfnamefont {T.}~\bibnamefont {Draper}}, \bibinfo {author}
  {\bibfnamefont {J.}~\bibnamefont {Liang}}, \bibinfo {author} {\bibfnamefont
  {K.-F.}\ \bibnamefont {Liu}}, \bibinfo {author} {\bibfnamefont
  {G.}~\bibnamefont {Wang}}, \ and\ \bibinfo {author} {\bibfnamefont {Y.-B.}\
  \bibnamefont {Yang}},\ }\href {\doibase 10.1103/PhysRevD.104.094503}
  {\bibfield  {journal} {\bibinfo  {journal} {Phys. Rev. D}\ }\textbf {\bibinfo
  {volume} {104}},\ \bibinfo {pages} {094503} (\bibinfo {year} {2021})},\
  \Eprint {http://arxiv.org/abs/2103.12933} {arXiv:2103.12933 [hep-lat]}
  \BibitemShut {NoStop}%
\bibitem [{\citenamefont {{[JLQCD 18] N. Yamanaka}}\ \emph
  {et~al.}(2018)\citenamefont {{[JLQCD 18] N. Yamanaka}}, \citenamefont
  {Hashimoto}, \citenamefont {Kaneko},\ and\ \citenamefont
  {Ohki}}]{Yamanaka:2018uud}%
  \BibitemOpen
  \bibfield  {author} {\bibinfo {author} {\bibnamefont {{[JLQCD 18] N.
  Yamanaka}}}, \bibinfo {author} {\bibfnamefont {S.}~\bibnamefont {Hashimoto}},
  \bibinfo {author} {\bibfnamefont {T.}~\bibnamefont {Kaneko}}, \ and\ \bibinfo
  {author} {\bibfnamefont {H.}~\bibnamefont {Ohki}},\ }\href {\doibase
  10.1103/PhysRevD.98.054516} {\bibfield  {journal} {\bibinfo  {journal} {Phys.
  Rev.}\ }\textbf {\bibinfo {volume} {D98}},\ \bibinfo {pages} {054516}
  (\bibinfo {year} {2018})},\ \Eprint {http://arxiv.org/abs/1805.10507}
  {arXiv:1805.10507 [hep-lat]} \BibitemShut {NoStop}%
\bibitem [{\citenamefont {{[RBC/UKQCD 19] M. Abramczyk}}\ \emph
  {et~al.}(2020)\citenamefont {{[RBC/UKQCD 19] M. Abramczyk}}, \citenamefont
  {Blum}, \citenamefont {Izubuchi}, \citenamefont {Jung}, \citenamefont {Lin},
  \citenamefont {Lytle}, \citenamefont {Ohta},\ and\ \citenamefont
  {Shintani}}]{Abramczyk:2019fnf}%
  \BibitemOpen
  \bibfield  {author} {\bibinfo {author} {\bibnamefont {{[RBC/UKQCD 19] M.
  Abramczyk}}}, \bibinfo {author} {\bibfnamefont {T.}~\bibnamefont {Blum}},
  \bibinfo {author} {\bibfnamefont {T.}~\bibnamefont {Izubuchi}}, \bibinfo
  {author} {\bibfnamefont {C.}~\bibnamefont {Jung}}, \bibinfo {author}
  {\bibfnamefont {M.}~\bibnamefont {Lin}}, \bibinfo {author} {\bibfnamefont
  {A.}~\bibnamefont {Lytle}}, \bibinfo {author} {\bibfnamefont
  {S.}~\bibnamefont {Ohta}}, \ and\ \bibinfo {author} {\bibfnamefont
  {E.}~\bibnamefont {Shintani}},\ }\href {\doibase 10.1103/PhysRevD.101.034510}
  {\bibfield  {journal} {\bibinfo  {journal} {Phys. Rev. D}\ }\textbf {\bibinfo
  {volume} {101}},\ \bibinfo {pages} {034510} (\bibinfo {year} {2020})},\
  \Eprint {http://arxiv.org/abs/1911.03524} {arXiv:1911.03524 [hep-lat]}
  \BibitemShut {NoStop}%
\bibitem [{\citenamefont {{[PACS 22B] R.~Tsuji}}\ \emph
  {et~al.}(2022)\citenamefont {{[PACS 22B] R.~Tsuji}}, \citenamefont
  {Tsukamoto}, \citenamefont {Aoki}, \citenamefont {Ishikawa}, \citenamefont
  {Kuramashi}, \citenamefont {Sasaki}, \citenamefont {Shintani},\ and\
  \citenamefont {Yamazaki}}]{Tsuji:2022ric}%
  \BibitemOpen
  \bibfield  {author} {\bibinfo {author} {\bibnamefont {{[PACS 22B]
  R.~Tsuji}}}, \bibinfo {author} {\bibfnamefont {N.}~\bibnamefont {Tsukamoto}},
  \bibinfo {author} {\bibfnamefont {Y.}~\bibnamefont {Aoki}}, \bibinfo {author}
  {\bibfnamefont {K.-I.}\ \bibnamefont {Ishikawa}}, \bibinfo {author}
  {\bibfnamefont {Y.}~\bibnamefont {Kuramashi}}, \bibinfo {author}
  {\bibfnamefont {S.}~\bibnamefont {Sasaki}}, \bibinfo {author} {\bibfnamefont
  {E.}~\bibnamefont {Shintani}}, \ and\ \bibinfo {author} {\bibfnamefont
  {T.}~\bibnamefont {Yamazaki}},\ }\href {\doibase 10.1103/PhysRevD.106.094505}
  {\bibfield  {journal} {\bibinfo  {journal} {Phys. Rev. D}\ }\textbf {\bibinfo
  {volume} {106}},\ \bibinfo {pages} {094505} (\bibinfo {year} {2022})},\
  \Eprint {http://arxiv.org/abs/2207.11914} {arXiv:2207.11914 [hep-lat]}
  \BibitemShut {NoStop}%
\bibitem [{\citenamefont {Gonz{\'a}lez-Alonso}\ and\ \citenamefont
  {Martin~Camalich}(2014)}]{Gonzalez-Alonso:2013ura}%
  \BibitemOpen
  \bibfield  {author} {\bibinfo {author} {\bibfnamefont {M.}~\bibnamefont
  {Gonz{\'a}lez-Alonso}}\ and\ \bibinfo {author} {\bibfnamefont
  {J.}~\bibnamefont {Martin~Camalich}},\ }\href {\doibase
  10.1103/PhysRevLett.112.042501} {\bibfield  {journal} {\bibinfo  {journal}
  {Phys. Rev. Lett.}\ }\textbf {\bibinfo {volume} {112}},\ \bibinfo {pages}
  {042501} (\bibinfo {year} {2014})},\ \Eprint {http://arxiv.org/abs/1309.4434}
  {arXiv:1309.4434 [hep-ph]} \BibitemShut {NoStop}%
\bibitem [{\citenamefont {Gasser}\ \emph {et~al.}(2021)\citenamefont {Gasser},
  \citenamefont {Leutwyler},\ and\ \citenamefont {Rusetsky}}]{Gasser:2020mzy}%
  \BibitemOpen
  \bibfield  {author} {\bibinfo {author} {\bibfnamefont {J.}~\bibnamefont
  {Gasser}}, \bibinfo {author} {\bibfnamefont {H.}~\bibnamefont {Leutwyler}}, \
  and\ \bibinfo {author} {\bibfnamefont {A.}~\bibnamefont {Rusetsky}},\ }\href
  {\doibase 10.1016/j.physletb.2021.136087} {\bibfield  {journal} {\bibinfo
  {journal} {Phys. Lett. B}\ }\textbf {\bibinfo {volume} {814}},\ \bibinfo
  {pages} {136087} (\bibinfo {year} {2021})},\ \Eprint
  {http://arxiv.org/abs/2003.13612} {arXiv:2003.13612 [hep-ph]} \BibitemShut
  {NoStop}%
\bibitem [{\citenamefont {Walker-Loud}\ \emph {et~al.}(2012)\citenamefont
  {Walker-Loud}, \citenamefont {Carlson},\ and\ \citenamefont
  {Miller}}]{Walker-Loud:2012ift}%
  \BibitemOpen
  \bibfield  {author} {\bibinfo {author} {\bibfnamefont {A.}~\bibnamefont
  {Walker-Loud}}, \bibinfo {author} {\bibfnamefont {C.~E.}\ \bibnamefont
  {Carlson}}, \ and\ \bibinfo {author} {\bibfnamefont {G.~A.}\ \bibnamefont
  {Miller}},\ }\href {\doibase 10.1103/PhysRevLett.108.232301} {\bibfield
  {journal} {\bibinfo  {journal} {Phys. Rev. Lett.}\ }\textbf {\bibinfo
  {volume} {108}},\ \bibinfo {pages} {232301} (\bibinfo {year} {2012})},\
  \Eprint {http://arxiv.org/abs/1203.0254} {arXiv:1203.0254 [nucl-th]}
  \BibitemShut {NoStop}%
\bibitem [{\citenamefont {Thomas}\ \emph {et~al.}(2015)\citenamefont {Thomas},
  \citenamefont {Wang},\ and\ \citenamefont {Young}}]{Thomas:2014dxa}%
  \BibitemOpen
  \bibfield  {author} {\bibinfo {author} {\bibfnamefont {A.~W.}\ \bibnamefont
  {Thomas}}, \bibinfo {author} {\bibfnamefont {X.~G.}\ \bibnamefont {Wang}}, \
  and\ \bibinfo {author} {\bibfnamefont {R.~D.}\ \bibnamefont {Young}},\ }\href
  {\doibase 10.1103/PhysRevC.91.015209} {\bibfield  {journal} {\bibinfo
  {journal} {Phys. Rev. C}\ }\textbf {\bibinfo {volume} {91}},\ \bibinfo
  {pages} {015209} (\bibinfo {year} {2015})},\ \Eprint
  {http://arxiv.org/abs/1406.4579} {arXiv:1406.4579 [nucl-th]} \BibitemShut
  {NoStop}%
\bibitem [{\citenamefont {Ji}(2013)}]{Ji:2013dva}%
  \BibitemOpen
  \bibfield  {author} {\bibinfo {author} {\bibfnamefont {X.}~\bibnamefont
  {Ji}},\ }\href {\doibase 10.1103/PhysRevLett.110.262002} {\bibfield
  {journal} {\bibinfo  {journal} {Phys. Rev. Lett.}\ }\textbf {\bibinfo
  {volume} {110}},\ \bibinfo {pages} {262002} (\bibinfo {year} {2013})},\
  \Eprint {http://arxiv.org/abs/1305.1539} {arXiv:1305.1539 [hep-ph]}
  \BibitemShut {NoStop}%
\bibitem [{\citenamefont {Ji}\ \emph {et~al.}(2021)\citenamefont {Ji},
  \citenamefont {Liu}, \citenamefont {Liu}, \citenamefont {Zhang},\ and\
  \citenamefont {Zhao}}]{Ji:2020ect}%
  \BibitemOpen
  \bibfield  {author} {\bibinfo {author} {\bibfnamefont {X.}~\bibnamefont
  {Ji}}, \bibinfo {author} {\bibfnamefont {Y.-S.}\ \bibnamefont {Liu}},
  \bibinfo {author} {\bibfnamefont {Y.}~\bibnamefont {Liu}}, \bibinfo {author}
  {\bibfnamefont {J.-H.}\ \bibnamefont {Zhang}}, \ and\ \bibinfo {author}
  {\bibfnamefont {Y.}~\bibnamefont {Zhao}},\ }\href {\doibase
  10.1103/RevModPhys.93.035005} {\bibfield  {journal} {\bibinfo  {journal}
  {Rev. Mod. Phys.}\ }\textbf {\bibinfo {volume} {93}},\ \bibinfo {pages}
  {035005} (\bibinfo {year} {2021})},\ \Eprint
  {http://arxiv.org/abs/2004.03543} {arXiv:2004.03543 [hep-ph]} \BibitemShut
  {NoStop}%
\bibitem [{\citenamefont {Jiang}\ \emph {et~al.}(2024)\citenamefont {Jiang},
  \citenamefont {Shi}, \citenamefont {Chen}, \citenamefont {Gong},\ and\
  \citenamefont {Yang}}]{Jiang:2024lto}%
  \BibitemOpen
  \bibfield  {author} {\bibinfo {author} {\bibfnamefont {X.}~\bibnamefont
  {Jiang}}, \bibinfo {author} {\bibfnamefont {C.}~\bibnamefont {Shi}}, \bibinfo
  {author} {\bibfnamefont {Y.}~\bibnamefont {Chen}}, \bibinfo {author}
  {\bibfnamefont {M.}~\bibnamefont {Gong}}, \ and\ \bibinfo {author}
  {\bibfnamefont {Y.-B.}\ \bibnamefont {Yang}},\ }\href@noop {} {\  (\bibinfo
  {year} {2024})},\ \Eprint {http://arxiv.org/abs/2411.08461} {arXiv:2411.08461
  [hep-lat]} \BibitemShut {NoStop}%
\bibitem [{\citenamefont {Clark}\ \emph {et~al.}(2010)\citenamefont {Clark},
  \citenamefont {Babich}, \citenamefont {Barros}, \citenamefont {Brower},\ and\
  \citenamefont {Rebbi}}]{Clark:2009wm}%
  \BibitemOpen
  \bibfield  {author} {\bibinfo {author} {\bibfnamefont {M.~A.}\ \bibnamefont
  {Clark}}, \bibinfo {author} {\bibfnamefont {R.}~\bibnamefont {Babich}},
  \bibinfo {author} {\bibfnamefont {K.}~\bibnamefont {Barros}}, \bibinfo
  {author} {\bibfnamefont {R.~C.}\ \bibnamefont {Brower}}, \ and\ \bibinfo
  {author} {\bibfnamefont {C.}~\bibnamefont {Rebbi}},\ }\href {\doibase
  10.1016/j.cpc.2010.05.002} {\bibfield  {journal} {\bibinfo  {journal}
  {Comput. Phys. Commun.}\ }\textbf {\bibinfo {volume} {181}},\ \bibinfo
  {pages} {1517} (\bibinfo {year} {2010})},\ \Eprint
  {http://arxiv.org/abs/0911.3191} {arXiv:0911.3191 [hep-lat]} \BibitemShut
  {NoStop}%
\bibitem [{\citenamefont {Babich}\ \emph {et~al.}(2011)\citenamefont {Babich},
  \citenamefont {Clark}, \citenamefont {Joo}, \citenamefont {Shi},
  \citenamefont {Brower},\ and\ \citenamefont {Gottlieb}}]{Babich:2011np}%
  \BibitemOpen
  \bibfield  {author} {\bibinfo {author} {\bibfnamefont {R.}~\bibnamefont
  {Babich}}, \bibinfo {author} {\bibfnamefont {M.~A.}\ \bibnamefont {Clark}},
  \bibinfo {author} {\bibfnamefont {B.}~\bibnamefont {Joo}}, \bibinfo {author}
  {\bibfnamefont {G.}~\bibnamefont {Shi}}, \bibinfo {author} {\bibfnamefont
  {R.~C.}\ \bibnamefont {Brower}}, \ and\ \bibinfo {author} {\bibfnamefont
  {S.}~\bibnamefont {Gottlieb}},\ }in\ \href {\doibase 10.1145/2063384.2063478}
  {\emph {\bibinfo {booktitle} {{SC11 International Conference for High
  Performance Computing, Networking, Storage and Analysis Seattle, Washington,
  November 12-18, 2011}}}}\ (\bibinfo {year} {2011})\ \Eprint
  {http://arxiv.org/abs/1109.2935} {arXiv:1109.2935 [hep-lat]} \BibitemShut
  {NoStop}%
\bibitem [{\citenamefont {Clark}\ \emph {et~al.}(2016)\citenamefont {Clark},
  \citenamefont {Jo}, \citenamefont {Strelchenko}, \citenamefont {Cheng},
  \citenamefont {Gambhir},\ and\ \citenamefont {Brower}}]{Clark:2016rdz}%
  \BibitemOpen
  \bibfield  {author} {\bibinfo {author} {\bibfnamefont {M.~A.}\ \bibnamefont
  {Clark}}, \bibinfo {author} {\bibfnamefont {B.}~\bibnamefont {Jo}}, \bibinfo
  {author} {\bibfnamefont {A.}~\bibnamefont {Strelchenko}}, \bibinfo {author}
  {\bibfnamefont {M.}~\bibnamefont {Cheng}}, \bibinfo {author} {\bibfnamefont
  {A.}~\bibnamefont {Gambhir}}, \ and\ \bibinfo {author} {\bibfnamefont
  {R.}~\bibnamefont {Brower}},\ }\href@noop {} {\  (\bibinfo {year} {2016})},\
  \Eprint {http://arxiv.org/abs/1612.07873} {arXiv:1612.07873 [hep-lat]}
  \BibitemShut {NoStop}%
\bibitem [{\citenamefont {Bi}\ \emph {et~al.}(2020)\citenamefont {Bi},
  \citenamefont {Xiao}, \citenamefont {Gong}, \citenamefont {Guo},
  \citenamefont {Sun}, \citenamefont {Xu},\ and\ \citenamefont
  {Yang}}]{Bi:2020wpt}%
  \BibitemOpen
  \bibfield  {author} {\bibinfo {author} {\bibfnamefont {Y.-J.}\ \bibnamefont
  {Bi}}, \bibinfo {author} {\bibfnamefont {Y.}~\bibnamefont {Xiao}}, \bibinfo
  {author} {\bibfnamefont {M.}~\bibnamefont {Gong}}, \bibinfo {author}
  {\bibfnamefont {W.-Y.}\ \bibnamefont {Guo}}, \bibinfo {author} {\bibfnamefont
  {P.}~\bibnamefont {Sun}}, \bibinfo {author} {\bibfnamefont {S.}~\bibnamefont
  {Xu}}, \ and\ \bibinfo {author} {\bibfnamefont {Y.-B.}\ \bibnamefont
  {Yang}},\ }\bibfield  {booktitle} {\emph {\bibinfo {booktitle} {{Proceedings,
  37th International Symposium on Lattice Field Theory (Lattice 2019): Wuhan,
  China, June 16-22 2019}}},\ }\href {\doibase 10.22323/1.363.0286} {\bibfield
  {journal} {\bibinfo  {journal} {PoS}\ }\textbf {\bibinfo {volume}
  {LATTICE2019}},\ \bibinfo {pages} {286} (\bibinfo {year} {2020})},\ \Eprint
  {http://arxiv.org/abs/2001.05706} {arXiv:2001.05706 [hep-lat]} \BibitemShut
  {NoStop}%
\bibitem [{\citenamefont {L{\"u}scher}\ and\ \citenamefont
  {Wolff}(1990)}]{LuscherWolff1990}%
  \BibitemOpen
  \bibfield  {author} {\bibinfo {author} {\bibfnamefont {M.}~\bibnamefont
  {L{\"u}scher}}\ and\ \bibinfo {author} {\bibfnamefont {U.}~\bibnamefont
  {Wolff}},\ }\href {\doibase 10.1016/0550-3213(90)90540-T} {\bibfield
  {journal} {\bibinfo  {journal} {Nucl. Phys. B}\ }\textbf {\bibinfo {volume}
  {339}},\ \bibinfo {pages} {222} (\bibinfo {year} {1990})}\BibitemShut
  {NoStop}%
\bibitem [{\citenamefont {Blossier}\ \emph {et~al.}(2009)\citenamefont
  {Blossier}, \citenamefont {Della~Morte}, \citenamefont {von Hippel},
  \citenamefont {Mendes},\ and\ \citenamefont {Sommer}}]{Blossier:2009kd}%
  \BibitemOpen
  \bibfield  {author} {\bibinfo {author} {\bibfnamefont {B.}~\bibnamefont
  {Blossier}}, \bibinfo {author} {\bibfnamefont {M.}~\bibnamefont
  {Della~Morte}}, \bibinfo {author} {\bibfnamefont {G.}~\bibnamefont {von
  Hippel}}, \bibinfo {author} {\bibfnamefont {T.}~\bibnamefont {Mendes}}, \
  and\ \bibinfo {author} {\bibfnamefont {R.}~\bibnamefont {Sommer}},\ }\href
  {\doibase 10.1088/1126-6708/2009/04/094} {\bibfield  {journal} {\bibinfo
  {journal} {JHEP}\ }\textbf {\bibinfo {volume} {04}},\ \bibinfo {pages} {094}
  (\bibinfo {year} {2009})},\ \Eprint {http://arxiv.org/abs/0902.1265}
  {arXiv:0902.1265 [hep-lat]} \BibitemShut {NoStop}%
\bibitem [{\citenamefont {Barca}\ \emph {et~al.}(2023)\citenamefont {Barca},
  \citenamefont {Bali},\ and\ \citenamefont {Collins}}]{Barca:2022uhi}%
  \BibitemOpen
  \bibfield  {author} {\bibinfo {author} {\bibfnamefont {L.}~\bibnamefont
  {Barca}}, \bibinfo {author} {\bibfnamefont {G.}~\bibnamefont {Bali}}, \ and\
  \bibinfo {author} {\bibfnamefont {S.}~\bibnamefont {Collins}},\ }\href
  {\doibase 10.1103/PhysRevD.107.L051505} {\bibfield  {journal} {\bibinfo
  {journal} {Phys. Rev. D}\ }\textbf {\bibinfo {volume} {107}},\ \bibinfo
  {pages} {L051505} (\bibinfo {year} {2023})},\ \Eprint
  {http://arxiv.org/abs/2211.12278} {arXiv:2211.12278 [hep-lat]} \BibitemShut
  {NoStop}%
\bibitem [{\citenamefont {Grebe}\ and\ \citenamefont
  {Wagman}(2024)}]{Grebe:2023tfx}%
  \BibitemOpen
  \bibfield  {author} {\bibinfo {author} {\bibfnamefont {A.~V.}\ \bibnamefont
  {Grebe}}\ and\ \bibinfo {author} {\bibfnamefont {M.}~\bibnamefont {Wagman}},\
  }\href {\doibase 10.22323/1.453.0049} {\bibfield  {journal} {\bibinfo
  {journal} {PoS}\ }\textbf {\bibinfo {volume} {LATTICE2023}},\ \bibinfo
  {pages} {049} (\bibinfo {year} {2024})},\ \Eprint
  {http://arxiv.org/abs/2312.00321} {arXiv:2312.00321 [hep-lat]} \BibitemShut
  {NoStop}%
\bibitem [{\citenamefont {Alexandrou}\ \emph {et~al.}(2024)\citenamefont
  {Alexandrou}, \citenamefont {Koutsou}, \citenamefont {Li}, \citenamefont
  {Petschlies},\ and\ \citenamefont {Pittler}}]{Alexandrou:2024tin}%
  \BibitemOpen
  \bibfield  {author} {\bibinfo {author} {\bibfnamefont {C.}~\bibnamefont
  {Alexandrou}}, \bibinfo {author} {\bibfnamefont {G.}~\bibnamefont {Koutsou}},
  \bibinfo {author} {\bibfnamefont {Y.}~\bibnamefont {Li}}, \bibinfo {author}
  {\bibfnamefont {M.}~\bibnamefont {Petschlies}}, \ and\ \bibinfo {author}
  {\bibfnamefont {F.}~\bibnamefont {Pittler}},\ }\href {\doibase
  10.1103/PhysRevD.110.094514} {\bibfield  {journal} {\bibinfo  {journal}
  {Phys. Rev. D}\ }\textbf {\bibinfo {volume} {110}},\ \bibinfo {pages}
  {094514} (\bibinfo {year} {2024})},\ \Eprint
  {http://arxiv.org/abs/2408.03893} {arXiv:2408.03893 [hep-lat]} \BibitemShut
  {NoStop}%
\bibitem [{\citenamefont {Hackl}\ and\ \citenamefont
  {Lehner}(2024)}]{Hackl:2024whw}%
  \BibitemOpen
  \bibfield  {author} {\bibinfo {author} {\bibfnamefont {A.}~\bibnamefont
  {Hackl}}\ and\ \bibinfo {author} {\bibfnamefont {C.}~\bibnamefont {Lehner}},\
  }\href@noop {} {\  (\bibinfo {year} {2024})},\ \Eprint
  {http://arxiv.org/abs/2412.17442} {arXiv:2412.17442 [hep-lat]} \BibitemShut
  {NoStop}%
\bibitem [{\citenamefont {Gao}\ \emph {et~al.}(2025)\citenamefont {Gao},
  \citenamefont {Zhang}, \citenamefont {Feng}, \citenamefont {Jin},
  \citenamefont {Liu},\ and\ \citenamefont {Mei{\ss}ner}}]{Gao:2025loz}%
  \BibitemOpen
  \bibfield  {author} {\bibinfo {author} {\bibfnamefont {Y.-S.}\ \bibnamefont
  {Gao}}, \bibinfo {author} {\bibfnamefont {Z.-L.}\ \bibnamefont {Zhang}},
  \bibinfo {author} {\bibfnamefont {X.}~\bibnamefont {Feng}}, \bibinfo {author}
  {\bibfnamefont {L.-C.}\ \bibnamefont {Jin}}, \bibinfo {author} {\bibfnamefont
  {C.}~\bibnamefont {Liu}}, \ and\ \bibinfo {author} {\bibfnamefont {U.-G.}\
  \bibnamefont {Mei{\ss}ner}},\ }\href {\doibase
  10.1103/PhysRevLett.134.171904} {\bibfield  {journal} {\bibinfo  {journal}
  {Phys. Rev. Lett.}\ }\textbf {\bibinfo {volume} {134}},\ \bibinfo {pages}
  {171904} (\bibinfo {year} {2025})},\ \Eprint
  {http://arxiv.org/abs/2502.12074} {arXiv:2502.12074 [hep-lat]} \BibitemShut
  {NoStop}%
\bibitem [{\citenamefont {Prelovsek}\ \emph {et~al.}(2017)\citenamefont
  {Prelovsek}, \citenamefont {Skerbis},\ and\ \citenamefont
  {Lang}}]{Prelovsek:2016iyo}%
  \BibitemOpen
  \bibfield  {author} {\bibinfo {author} {\bibfnamefont {S.}~\bibnamefont
  {Prelovsek}}, \bibinfo {author} {\bibfnamefont {U.}~\bibnamefont {Skerbis}},
  \ and\ \bibinfo {author} {\bibfnamefont {C.~B.}\ \bibnamefont {Lang}},\
  }\href {\doibase 10.1007/JHEP01(2017)129} {\bibfield  {journal} {\bibinfo
  {journal} {JHEP}\ }\textbf {\bibinfo {volume} {01}},\ \bibinfo {pages} {129}
  (\bibinfo {year} {2017})},\ \Eprint {http://arxiv.org/abs/1607.06738}
  {arXiv:1607.06738 [hep-lat]} \BibitemShut {NoStop}%
\bibitem [{\citenamefont {Detmold}\ \emph {et~al.}(2024)\citenamefont
  {Detmold}, \citenamefont {Jay}, \citenamefont {Kanwar}, \citenamefont
  {Shanahan},\ and\ \citenamefont {Wagman}}]{Detmold:2024ifm}%
  \BibitemOpen
  \bibfield  {author} {\bibinfo {author} {\bibfnamefont {W.}~\bibnamefont
  {Detmold}}, \bibinfo {author} {\bibfnamefont {W.~I.}\ \bibnamefont {Jay}},
  \bibinfo {author} {\bibfnamefont {G.}~\bibnamefont {Kanwar}}, \bibinfo
  {author} {\bibfnamefont {P.~E.}\ \bibnamefont {Shanahan}}, \ and\ \bibinfo
  {author} {\bibfnamefont {M.~L.}\ \bibnamefont {Wagman}},\ }\href {\doibase
  10.1103/PhysRevD.109.094516} {\bibfield  {journal} {\bibinfo  {journal}
  {Phys. Rev. D}\ }\textbf {\bibinfo {volume} {109}},\ \bibinfo {pages}
  {094516} (\bibinfo {year} {2024})},\ \Eprint
  {http://arxiv.org/abs/2403.00672} {arXiv:2403.00672 [hep-lat]} \BibitemShut
  {NoStop}%
\bibitem [{\citenamefont {Hasenbusch}(2001)}]{Hasenbusch:2001ne}%
  \BibitemOpen
  \bibfield  {author} {\bibinfo {author} {\bibfnamefont {M.}~\bibnamefont
  {Hasenbusch}},\ }\href {\doibase 10.1016/S0370-2693(01)01102-9} {\bibfield
  {journal} {\bibinfo  {journal} {Phys. Lett. B}\ }\textbf {\bibinfo {volume}
  {519}},\ \bibinfo {pages} {177} (\bibinfo {year} {2001})},\ \Eprint
  {http://arxiv.org/abs/hep-lat/0107019} {arXiv:hep-lat/0107019} \BibitemShut
  {NoStop}%
\bibitem [{\citenamefont {Borsanyi}\ \emph {et~al.}(2021)\citenamefont
  {Borsanyi} \emph {et~al.}}]{Borsanyi:2020mff}%
  \BibitemOpen
  \bibfield  {author} {\bibinfo {author} {\bibfnamefont {S.}~\bibnamefont
  {Borsanyi}} \emph {et~al.},\ }\href {\doibase 10.1038/s41586-021-03418-1}
  {\bibfield  {journal} {\bibinfo  {journal} {Nature}\ }\textbf {\bibinfo
  {volume} {593}},\ \bibinfo {pages} {51} (\bibinfo {year} {2021})},\ \Eprint
  {http://arxiv.org/abs/2002.12347} {arXiv:2002.12347 [hep-lat]} \BibitemShut
  {NoStop}%
\bibitem [{\citenamefont {Capitani}\ \emph {et~al.}(2001)\citenamefont
  {Capitani}, \citenamefont {Gockeler}, \citenamefont {Horsley}, \citenamefont
  {Perlt}, \citenamefont {Rakow}, \citenamefont {Schierholz},\ and\
  \citenamefont {Schiller}}]{Capitani:2000xi}%
  \BibitemOpen
  \bibfield  {author} {\bibinfo {author} {\bibfnamefont {S.}~\bibnamefont
  {Capitani}}, \bibinfo {author} {\bibfnamefont {M.}~\bibnamefont {Gockeler}},
  \bibinfo {author} {\bibfnamefont {R.}~\bibnamefont {Horsley}}, \bibinfo
  {author} {\bibfnamefont {H.}~\bibnamefont {Perlt}}, \bibinfo {author}
  {\bibfnamefont {P.~E.~L.}\ \bibnamefont {Rakow}}, \bibinfo {author}
  {\bibfnamefont {G.}~\bibnamefont {Schierholz}}, \ and\ \bibinfo {author}
  {\bibfnamefont {A.}~\bibnamefont {Schiller}},\ }\href {\doibase
  10.1016/S0550-3213(00)00590-3} {\bibfield  {journal} {\bibinfo  {journal}
  {Nucl. Phys. B}\ }\textbf {\bibinfo {volume} {593}},\ \bibinfo {pages} {183}
  (\bibinfo {year} {2001})},\ \Eprint {http://arxiv.org/abs/hep-lat/0007004}
  {arXiv:hep-lat/0007004} \BibitemShut {NoStop}%
\end{thebibliography}%

\clearpage
\begin{widetext}

\section*{Supplemental materials}

\subsection{Brief review of the blending method and verification from $g_V$}

The blending method~\cite{Hu:2025vhd} splits the vector space associated with the lattice sites and color degrees of freedom, with dimension \( [\mathcal{L}] = N_c N_L^3 \) into two subspaces:  

1. Subspace \( \mathcal{L}_1 \) (that used in the distillation method~\cite{HadronSpectrum:2009krc}): The span of the \( N_{\rm e} \) low-lying eigenvectors \( \{ v_{\lambda_i} \}_{i=1}^{N_{\rm e}} \) of the discrete Laplace operator;

2. Complement subspace \( \mathcal{L}_2 \): The orthogonal complement of \( \mathcal{L}_1 \) in \( \mathcal{L} \), i.e., \( \mathcal{L}_2 = \mathcal{L}/ \mathcal{L}_1 \), with dimension \([\mathcal{L}_2] = N_c N_L^3 - N_{\rm e} \).  

A ``blending space" is constructed by combining the eigenvectors \( \{ \phi_i       \equiv v_{\lambda_i} \}_{i=1}^{N_{\rm e}} \) spanning \( \mathcal{L}_1 \), and  
a set of \( N_{\rm st} \) orthogonal random vectors \(\{ \phi_{N_{\rm e}+j} \equiv\eta_j \}_{j=1}^{N_{\rm st}} \) sampled uniformly from \( \mathcal{L}_2 \). Then the identity operator \( \hat{I} \) on \( \mathcal{L} \) can be unbiasedly approximated as:  
\begin{align}
\hat{I} = \sum_{i=1}^{[\mathcal{L}]} | V_i \rangle \langle V_i | = \lim_{N_{\rm st} \to [\mathcal{L}_2]} \sum_{k=1}^{N_{\rm e} + N_{\rm st}} \Omega^{(1)}_k | \phi_k \rangle \langle \phi_k |,\ \Omega^{(1)}_k = 
\begin{cases} 
1 & \text{for } k \leq N_{\rm e}, \\
[\mathcal{L}_2]/N_{\rm st} & \text{for } k > N_{\rm e},
\end{cases}
\end{align} 
where \( \{ |V_i\rangle \} \) is a complete orthonormal basis of \( \mathcal{L} \). Then the all-to-all propagator $S$ can be projected onto the blending space,
\begin{align}
    S_{ij}(t_1,t_2)=& \int \mathrm{d}^3 x \mathrm{d}^3 y  \langle 
 \phi_i(\vec{x},t_1)|S(\vec{x},t_1;\vec{y},t_2)|\phi_j(\vec{y},t_2)\rangle,\label{eq:prop}
\end{align}
and then compressed from $(4N_cN_TN_L^3)^2$ to $(4N_T(N_{\rm e}+N_{\rm st}))^2$ (projected propagator) + $N_TN_cN_L^3(N_{\rm e}+N_{\rm st})$ (vectors in the blending space).

More details about the mathematical proof and numerical justification of the blending method can be found in Ref.~\cite{Hu:2025vhd}.

To further validate the results from the blending method, we compute the nucleon iso-vector vector charge, defined by the matrix element $g_V \equiv \langle \bar{u}\gamma_4 u - \bar{d}\gamma_4 d  \rangle_N$
using the blending method. With a hadron-independent normalization, $g_V$ is expected to be 1, up to discretization effects and within statistical uncertainties.

In practice, $g_V$ is extracted from the standard ratio of correlation functions at finite source-sink separation $t_f$, current insertion time $t$, interpolation field ${\cal N}$, and current operator ${\cal O}_X=\bar{u}\gamma_4 u - \bar{d}\gamma_4 d$,
\begin{align}\label{eq:errorso}
{\cal R}_{X}(t_f,t;{\cal N})& = \frac{\int \text{d}^3x\text{d}^3y\text{d}^3z\langle {\cal N}(\vec{x},t_f) {\cal O}_{X}(\vec{y},t) {\cal N}^{\dagger}(\vec{z},0)\rangle}{\int \text{d}^3x\text{d}^3z \langle {\cal N}(\vec{x}, t_f) {\cal N}^{\dagger}(\vec{z}, 0) \rangle} \nonumber\\
=&\langle {\cal O}_X \rangle_N + \mathcal{O}(e^{-\delta m , t}, e^{-\delta m , (t_f-t)}, e^{-\delta m , t_f}).
\end{align}

\begin{figure}[!h]
    \centering
    \includegraphics[width=0.45\linewidth]{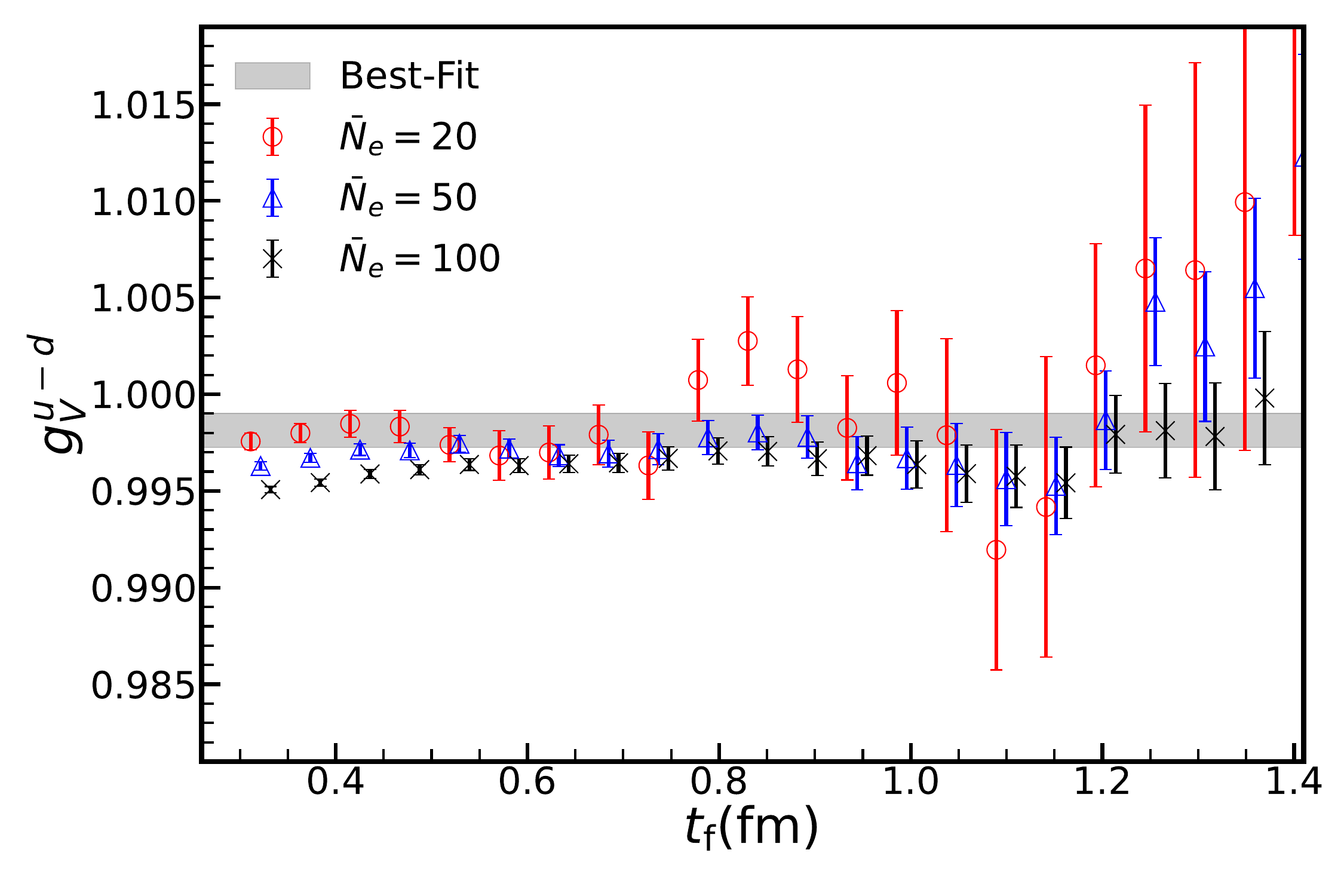}
    \includegraphics[width=0.45\linewidth]{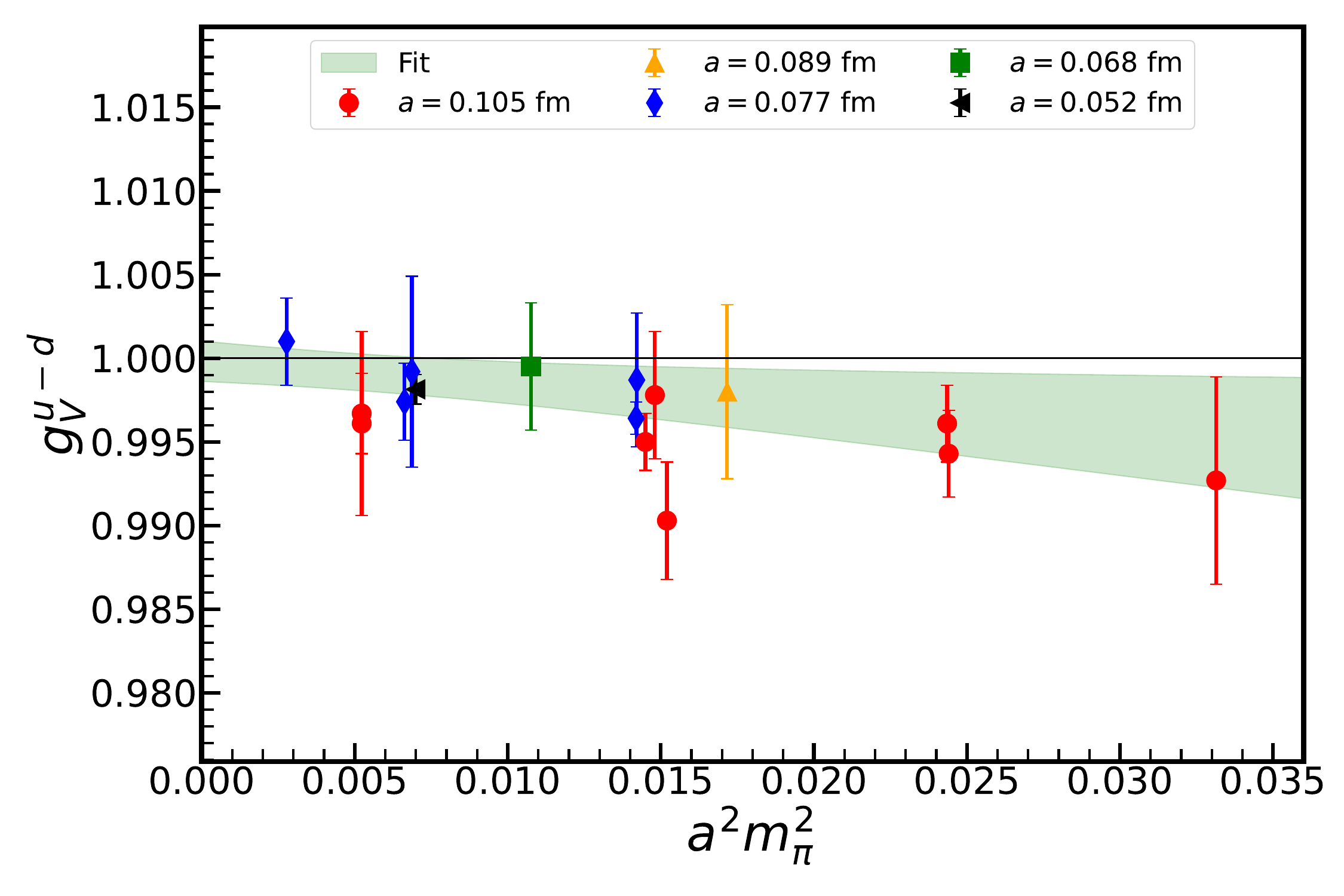}
    \caption{
        {\bf Left}: Source-sink separation $t_f$ dependence of $Z_VR_{V}(t_f,t_f/2,{\cal N})\ _{t_f\to\  \infty}\rightarrow g_V$ with different $\bar{N}_{\rm e}$, on the H48P32 ensemble, with $Z_V$ from the renormalization condition $Z_V\langle \pi^+|\bar{u}\gamma_t u|\pi^+\rangle=1$ of the pion matrix element of the vector current.
        {\bf Right}: $a^2m_\pi^2$ dependence of the  hadron-independent normalized $g_V$, and the result in the continuum limit is consistent with 1 within 0.12\% statistical uncertainty.
        }
    \label{fig:g_V}
\end{figure}

Our calculation is performed on the H48P32 ensemble ($a=0.052$ fm, $m{\pi}=320$ MeV) using 46 configurations, with $N_{\rm e}=100$ and $N_{\rm st}=200$ for the projected propagator generation via the blending method. For the external nucleon states, we employ $\bar{N}_{\rm e}=20$, 50, and 100, while using the full $N_{\rm e}=100$ for the vector current to ensure an unbiased reconstruction of the local current.

The left panel of Fig.~\ref{fig:g_V} shows the normalized ratio $R_{V,{\cal N}}^R(t_f)\equiv Z_V R_{V}(t_f,t_f/2,{\cal N})$, which tends to $g_V$ as $t_f \rightarrow \infty$. Here, $Z_V \equiv 1/\langle \pi^+ | \bar{u}\gamma_t u | \pi^+ \rangle$ is the vector renormalization constant determined via the traditional sequential method~\cite{CLQCD:2024yyn}. As expected, the statistical uncertainty of $R_{V,\cal N}^R(t_f)$ decreases with a larger $\bar{N}_e$, though the excited-state contamination becomes more pronounced at small $t_f$. A two-state fit yields $g_V = 0.9981(9)$, which deviates from 1 by approximately $2\sigma$ with a 0.1\% statistical uncertainty. This result confirms that the value of $g_V$ obtained with the blending method is consistent with the theoretical expectation at the 0.2\% level.

Since $g_V$ at finite lattice spacing is subject to discretization errors of order $a^2$ and $m_{\pi}^2a^2$, we perform a global extrapolation using the ansatz
\begin{align}
g^R_V(m_{\pi},a)=g_V^{\chi}(1+c_1m_{\pi}^2a^2+c_2 a^2),
\end{align}
based on the value of $g_V$ on other ensembles, as collected in Table~\ref{tab:two-state}. As shown in the right panel of Fig.~\ref{fig:g_V}, the resulting value in the continuum limit is $g^{\chi}_V=0.9998(12)$ with $\chi^2/{\rm d.o.f.} = 0.35$, and perfectly agrees with our expectation. 

{To quantify the computational efficiency of the blending method, we compare the cost—primarily measured by the number of quark propagator inversions required to reach a target statistical precision—against traditional methods (sequential source or stochastic) used in recent high-precision studies. As a benchmark, we target similar statistical precision for $g_{S,T}$ on physical-point ensembles with $a\approx 0.08\ \mathrm{fm}$ and $L\approx 5.1\ \mathrm{fm}$.

We list here how to get the total number of inversions here:
        \begin{itemize}
            \item  PNDME, ``Sequential Source Method'': $N_{\rm meas}^{\rm LP} \times (1+n_{\rm pol}\times n_{f} \times N_{\rm tsep}/N_{\rm sub})=165120\times(1+4 \times 2 \times 5/4)=1.8{\rm M}$
            \item  ETMC, ``Sequential Source Method'':  $\sum_{ t_{\rm sep}} N_{\rm t_{sep}}^{\rm meas}\times N_f \times N_{\rm pol} = ( (750 + 1500 + 3000 + 4500 + 120000 + 36000 + 48000) \times 2 \times 4 = 1.71{\rm M}   $
            \item  RQCD, ``Stochastic Sequential Source Method'':  $n_{\rm cfg} \times (N_{\rm tsep}+N_{\rm noise}/12)=1000\times(4+100/12)=0.012{\rm M}$
            \item  {\bf This work},  ``Blending Method'': 
            $n_{\rm cfg}\times N_T \times (N_e+N_{\rm st})/N_c = 46 \times 96 \times (140+60)/3  = 0.39{\rm M}$
        \end{itemize} 
}

    \begin{table}[htbp]
     \centering
     \caption{
        The total inversion cost for numcleon matrix element including $g_S^{u-d},g_T^{u-d}$ among different collaboration at physical point.
        For a fair comparison, we list the inverse cost upto same precision of $g_T$ of this work in the last column.
     }
    \resizebox{0.99\columnwidth}{!}{
\begin{tabular}{c|ccccc|c|cc|c|cc}
\hline
\hline
\multirow{2}{*}{Collaborations} & \multirow{2}{*}{Ensemble} & \multirow{2}{*}{L} & \multirow{2}{*}{T} & \multirow{2}{*}{a(fm)} & \multirow{2}{*}{$m_\pi$(MeV)} & \multirow{2}{*}{$n_{\rm cfg}$} & \multirow{2}{*}{$g_S$} & \multirow{2}{*}{$g_T$} & \multirow{2}{*}{Inversions} & \multicolumn{2}{c}{Inversion for our precision of} \\
\cline{11-12}
 & & & & & & & & & &  $g_S$ &   $g_T$ \\
\hline
ETMC(2020)   & cB211.072.64 & 64   & 128  & 0.08 & 139  & 750  & 1.35(17) & 0.939(027) & 1.71M & 32M & 92M \\
RQCD(2023)   & D452 & 64   & 128  & 0.076 & 156  & 1000 & -0.6(3.0) & 0.870(110) & 0.01M & 364M & 11M \\
PNDME(2023) & a09m130 & 64   & 96   & 0.09 & 138  & 1290 & 1.05(23) & 1.010(006) & 1.8M & 103M & 4.2M \\
\hline
\textbf{This Work} & F64P13 & 64 & 128 & 0.078 & 134 & 46 & 1.00(03) & 0.998(004) & 0.39M & 0.39M & 0.39M \\
\hline
\hline
\end{tabular}%
    }
     \label{tab:CostConnect1}%
   \end{table}%

{The comparison in Table~\ref{tab:CostConnect1} reveals our substantial efficiency gain. For five of the six comparable results from ETMC~\cite{Alexandrou:2022dtc}, RQCD~\cite{Bali:2023sdi}, and PNDME~\cite{Jang:2023zts}, we use a factor of 17 to 877 times fewer inversions to reach similar precision. The advantage is smaller but still significant (a factor of 6) for $g_T$ from PNDME~\cite{Jang:2023zts}.

These gains stem from the method's core design: it projects quark propagators into a low-mode subspace (like distillation) but uses stochastic techniques to estimate the full propagator within this space. This hybrid approach maximizes data reuse, allowing the calculation of N-point functions for arbitrary source-sink separations, inserted operator and hadron states, from a single set of generated ``distilled" propagators.

Regarding storage cost, the primary output is these distilled propagators, which are modest in size relative to full propagators. Crucially, this same set of propagators can be reused to compute matrix elements for various hadrons and currents, amortizing the storage cost per observable and making the framework highly efficient for multi-observable campaigns.

In summary, the blending method provides a computationally and storage-efficient pathway to high-precision hadron structure calculations by achieving superior statistical precision per inversion and enabling extensive data reuse.}

\subsection{Excited state contamination}

The extraction of the ground-state matrix element from lattice calculations is systematically affected by excited-state contamination (ESC). 
The leading ESCs of ${\cal R}_{X}(t_f,t;{\cal N})$ include terms of order $e^{-\delta m \, t}$ and $e^{-\delta m \, (t_f-t)}$ which originate from transitions between the ground state and excited states with mass gap $\delta m$, and also terms of order $e^{-\delta m \, t_f}$ which arise from matrix elements of ${\cal O}_X$ within the excited states themselves.

\subsubsection{Brief review of {\rm GEVP}}

Since different hadron interpolation operators have different overlap with ground and excited states, the generalized eigenvalue problem (GEVP) provides a framework to suppress the low energy excited states through the combination of different interpolation operators~\cite{LuscherWolff1990,Blossier:2009kd}.
Under GEVP, one introduces a tower of interpolation operators (GEVP basis),
\begin{align}
    \{{\cal O}_0, {\cal O}_1, {\cal O}_2,...,{\cal O}_{N-1} \} \ .
\end{align}
These interpolation operators furnish a matrix of two-point functions (2pts),
\begin{align}
C_{i,j}(t) = \langle {\cal O}_i(t) {\cal O}^{\dagger}_j(0) \rangle \ .
\end{align}
Since $C_{i,j}(t)$ is the transfer matrix $e^{-H t}$ of the QCD Hamiltonian in a truncated Hilbert space, its eigenstates are also approximations to the corresponding hadrons. Since the GEVP basis may not be orthonormal, one needs to solve a GEVP in order to find the eigenstates.
\begin{align}\label{eq:GEVP}
C_{i,j}(t') V_{j}^{\alpha} (t',t_0) = C_{i,j}(t_0) V_{j}^{\alpha} (t',t_0) \lambda^{\alpha}(t',t_0)  \ ,
\end{align}
where $V_{j}^{\alpha} (t',t_0)$ and $\lambda^{\alpha}(t',t_0)$ are the $\alpha$-th eigenstate and eigenvalue, respectively. By analyzing $\lambda^{\alpha}(t',t_0)$, one can find the energy of each eigenstate, where the ground state is labeled as $V_{j}^{0} (t',t_0)$. The optimized ground state hadron interpolation operator can be constructed as,
\begin{align}
{\cal O}^{\rm opt}(t',t_0) = \sum_{j=0}^{N-1}  V_{j}^{0*} (t',t_0) {\cal O}_{j} \ .
\end{align}
Then, this optimized operator can be utilized to calculate the matrix element. As shown in Ref.~\cite{Blossier:2009kd}, under proper choices of time slices, the ESC for the three-point functions behaves as ${\cal O} (e^{-\delta m' \ t_0})$,
where $\delta m' = m_{N}-m_{0}$ is an enlarged energy gap compared to ordinary methods with a single interpolation operator in Eq.~(\ref{eq:errorso}). Therefore, GEVP is a large time asymptotic analysis, which systematically approaches the desired ground state with an increased number of interpolation operators.

But the question is how to choose an efficient GEVP basis for a desired matrix element. From the point of view of large time asymptotic analysis, one should start from the lowest energy states and gradually approaches the higher energy states, in the subspace constrained by the quantum number of hadron interpolation operator. Therefore, the operator list should start from the operators trying to capture the lowest energy states and proceed higher ones. For example, for nucleon matrix elements, one should start from $N$, $N \pi$, $N \pi \pi$, $N \sigma$ and so on. Recent works have explored the ESC for nucleon matrix elements under GEVP~\cite{Barca:2022uhi,Grebe:2023tfx,Alexandrou:2024tin,Barca:2024hrl,Hackl:2024whw,Gao:2025loz}.

\subsubsection{current-involved excited state contamination}

The concept of current-meson dominance, introduced in Refs.~\cite{Barca:2024hrl,Barca:2025det}, posits that certain excited states are selectively enhanced by the inserted current, becoming the dominant source of contamination in the desired matrix element. Specifically, the energy gap for this dominant state is $\Delta E = m_X$, where $m_X$ is the mass of the ground state projected by the current operator $O_X$. Conventional nucleon interpolation operators are inefficient at isolating this particular contamination. The solution is to include the current-involved operator ${\cal N}_X(\vec{0}) \equiv {\cal N}(\vec{0})\sum_{\vec{z}}O_X(\vec{z})$, where ${\cal N} = \epsilon_{abc} u_{a} C\gamma_5 d_{b} u_c$ is the standard proton interpolating field.

{The essential difference is that \(\mathcal{N}_X\) yields a non-vanishing disconnected contraction, owing to the non-zero vacuum expectation value \(\langle O_X(t) O_X^\dagger(0) \rangle\).} Consider the correlation function of ${\cal N}(t_f)$ and ${\cal N}_X(0)$ with an inserted current $O_X(t)$, it can be decomposed into the quark line disconnected and connected diagrams\footnote{This equation is for the theoretical argument on the determination of current-enhanced excited states. In practical calculations, we consider all diagrams from the Wick theorem, including quark-line connected and disconnected contributions, as detailed in Sec.~\ref{sec:ContriNON}.},
\begin{align}
   \sum_{\vec{x},\vec{y}}\left\langle{\cal N}\left(t_f;\vec{x}\right) O_X\left(t;\vec{0}\right) {\cal N}_X^{\dagger}(0;\vec{y})\right\rangle&=V\sum_y\left\langle{\cal N}\left(t_f;\vec{0}\right) {\cal N}^{\dagger}(0;\vec{y})\right\rangle\sum_z\langle O_X\left(t;\vec{0}\right) O^{\dagger}_X(0;\vec{z}) \rangle \nonumber\\
   &\quad \quad +\mathrm{quark\ connected\ diagrams}\nonumber\\
   &={\cal O}(Ve^{-m_Nt_f}e^{-m_Xt})+\mathrm{quark\ connected\ diagrams}
\end{align}
where the first term in the right-hand side corresponds to the ESC we need and is enhanced by a spatial volume factor $V$. This significant enhancement underscores why the ${\cal N}_X$ operator must be included in the basis.

Here, we provide another argument by studying the ``linear error correction" to a nucleon matrix element. Denote the true nucleon state as $| N \rangle$ and the estimated nucleon state as $|N^{\rm est}\rangle$. Suppose they differ by a perturbation $| \delta N\rangle$ caused by ESC,
\begin{align}
 |N^{\rm est}\rangle = | N \rangle + | \delta N\rangle \ .
\end{align}
One calculates the matrix element using the estimated nucleon state,
\begin{align}
\langle N^{\rm est} | O_X |N^{\rm est}\rangle 
= \langle N | O_X |N \rangle + \langle \delta N | O_X |N \rangle + \langle N | O_X |\delta N \rangle + \langle \delta N | O_X | \delta N\rangle \ ,
\end{align}
which differs from the true nucleon matrix element $\langle N | O_X |N \rangle$ by linear and quadratic error corrections. We focus on the linear error correction $\langle N | O_X |\delta N \rangle$ or $\langle \delta N | O_X |N \rangle$ and ignore the quadratic correction. $|\delta N \rangle$ is a vector in Hilbert space that could point towards various possible directions. However, only the direction parallel to $O_X |N \rangle$ contributes to the linear error correction while the orthogonal directions vanish. Thus the second interpolation operator ${\cal N}_X$ should be a good choice to mimic this direction. 

In this work, we follow this idea to choose a two-operator basis as a preliminary investigation in this direction,
\begin{align}
    H_{X} = \{ {\cal N}, {\cal N}_X\} \ ,
\end{align}
and then the basis of the scalar charge case differs from the tensor one. For $O_S$, the proton is an unpolarized state; but for $O_T$ with $\epsilon^{ijk}\sigma_{ij}$, the proton is polarized along the $k$-direction so that ${\cal N}_T$ has the same polarization as ${\cal N}$. Since $O_S$ and $O_T$ are parity even operators, we consider the S-wave type scattering state and the momentum projections for all the operators are zero.

Then the optimal operator which can suppress the current-involved ESC can be expressed as a linear combination ${\cal N}^{\rm opt}={\cal N} + c^{\rm opt} {\cal N}_{X}$ with the coefficient $c$ to be determined. In our calculation, we use the interpolation operator ${\cal N}_X$ directly and not to project the reducible lattice representations onto the spin-$1/2$, isospin-$1/2$ subspace to obtain irreducible operators~\cite{Prelovsek:2016iyo,Detmold:2024ifm}. With our choice ${\cal N}_X$, excitations with higher spin or isospin are allowed and are handled effectively by varying the coefficient $c$ in the linear combination. In particular, both ${\cal N}_S$ and ${\cal N}_T$ also create isospin-$3/2$ states, which share the same energy with isospin-$1/2$ states because we take $m_u=m_d$. Moreover, ${\cal N}_T$ couples to both spin-$1/2$ and spin-$3/2$ $N\,b_1$ states; these correspond to different linear combinations of polarization components but are degenerate eigenstates of the Hamiltonian. Consequently, such higher-spin or higher-isospin excited states share the same Euclidean-time dependence in the linear error correction and can be absorbed by a rescaling of $c$.

\begin{figure}
    \centering
    \includegraphics[width=0.475\linewidth]{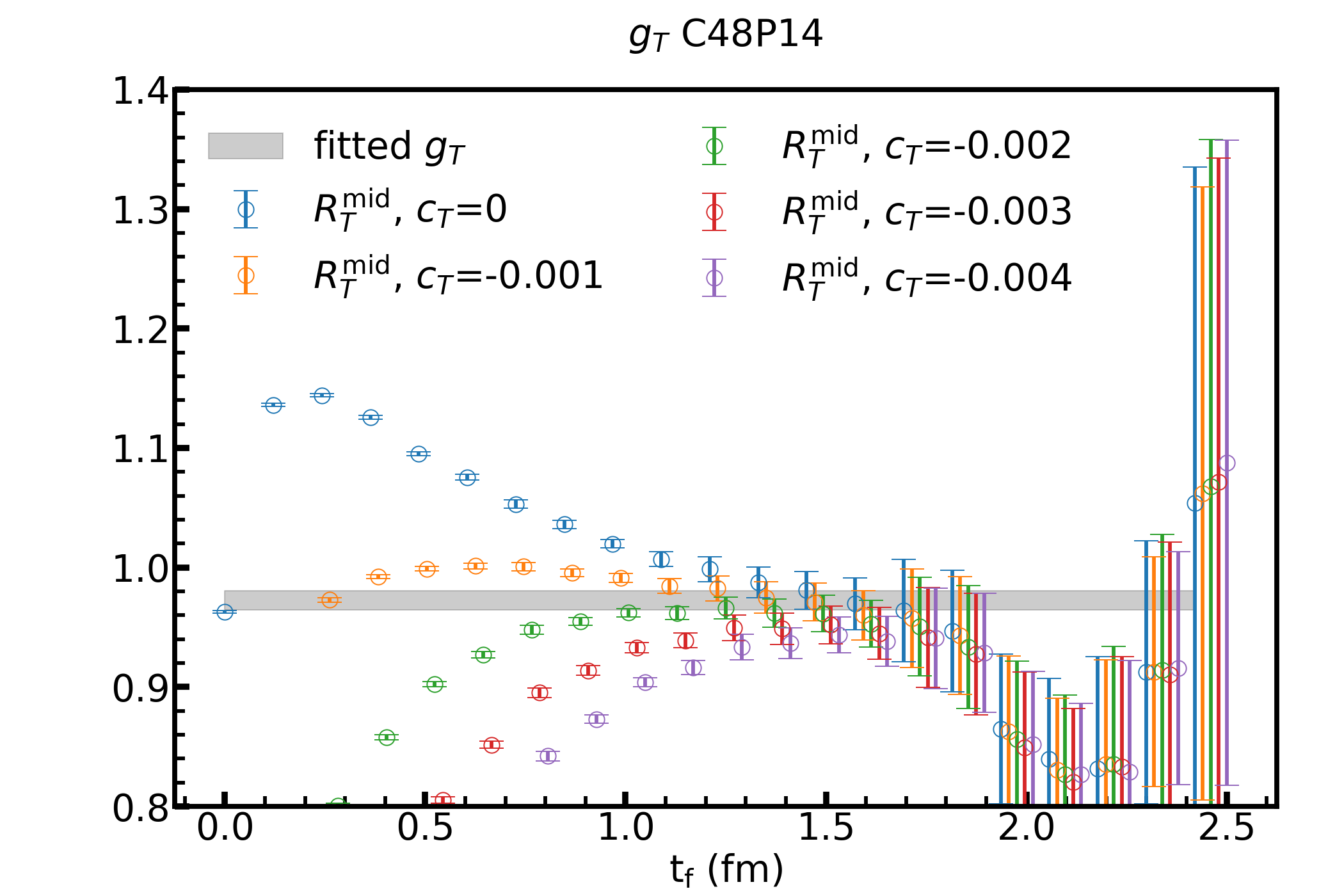}
    \includegraphics[width=0.475\linewidth]{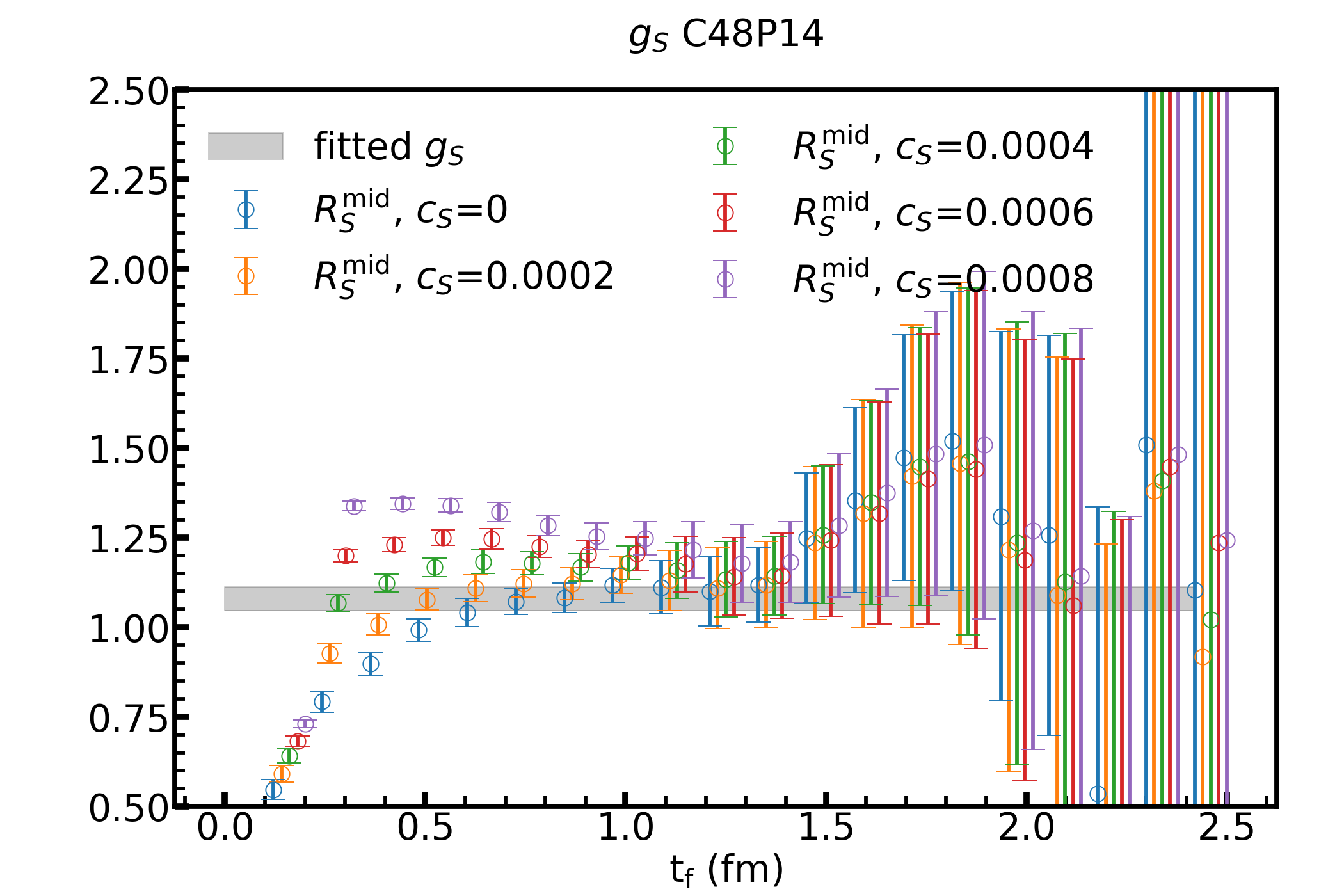}
    \caption{The dependence of the source-sink separation \( t_f \) for the ratios \( \mathcal{R}_X^{\rm mid}(t_f,{\cal N}(c))\equiv \mathcal{R}_X(t_f, t = t_f/2;{\cal N}(c)) \) under various coefficients $c$, for the iso-vector tensor charge (left panel) and scalar charge (right panel). Those ratios are generated on C48P14 with physical pion mass and $a=0.105$ fm and renormalized to $\overline{\mathrm{MS}}(2~\mathrm{GeV})$}. 
    \label{fig:gT_GEVP}
\end{figure}

We present the midpoint ratio ${\cal R}^{\rm mid}_{X}(t_f;c)\equiv {\cal R}_{X}(t_f,t_f/2;{\cal N}(c))$ on one of the physical pion mass ensembles, C48P14 ($a=0.105$ fm), in Fig.~\ref{fig:gT_GEVP}, exploring its dependence on the source-sink separation $t_f/a$ and the parameter $c$ in the interpolation operator combination ${\cal N}(c)\equiv {\cal N} + c {\cal N}_{X}$. The data show a strong $c$-dependence at small $t_f$ for both charges, despite converging to a common ground-state value (gray band) at large $t_f$. A notable finding is that the optimal $c$ values for good ESC suppression are very small—approximately -0.001 (tensor) and 0.0004 (scalar). Consequently, determining these small coefficients via a standard GEVP analysis in the $H_X$ basis is plagued by large relative uncertainties.

\begin{figure}
    \centering
    \includegraphics[width=0.475\linewidth]{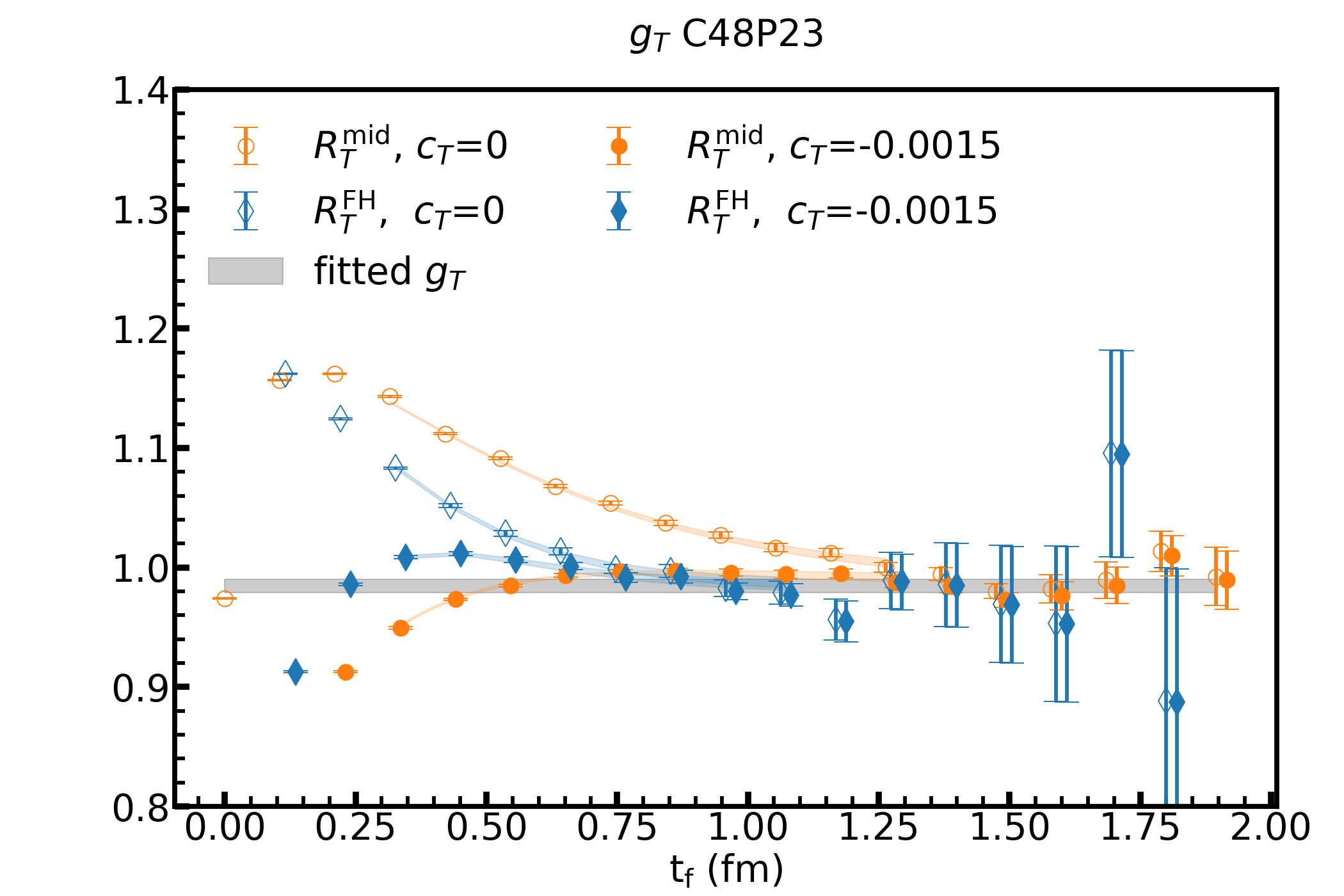}
    \includegraphics[width=0.475\linewidth]{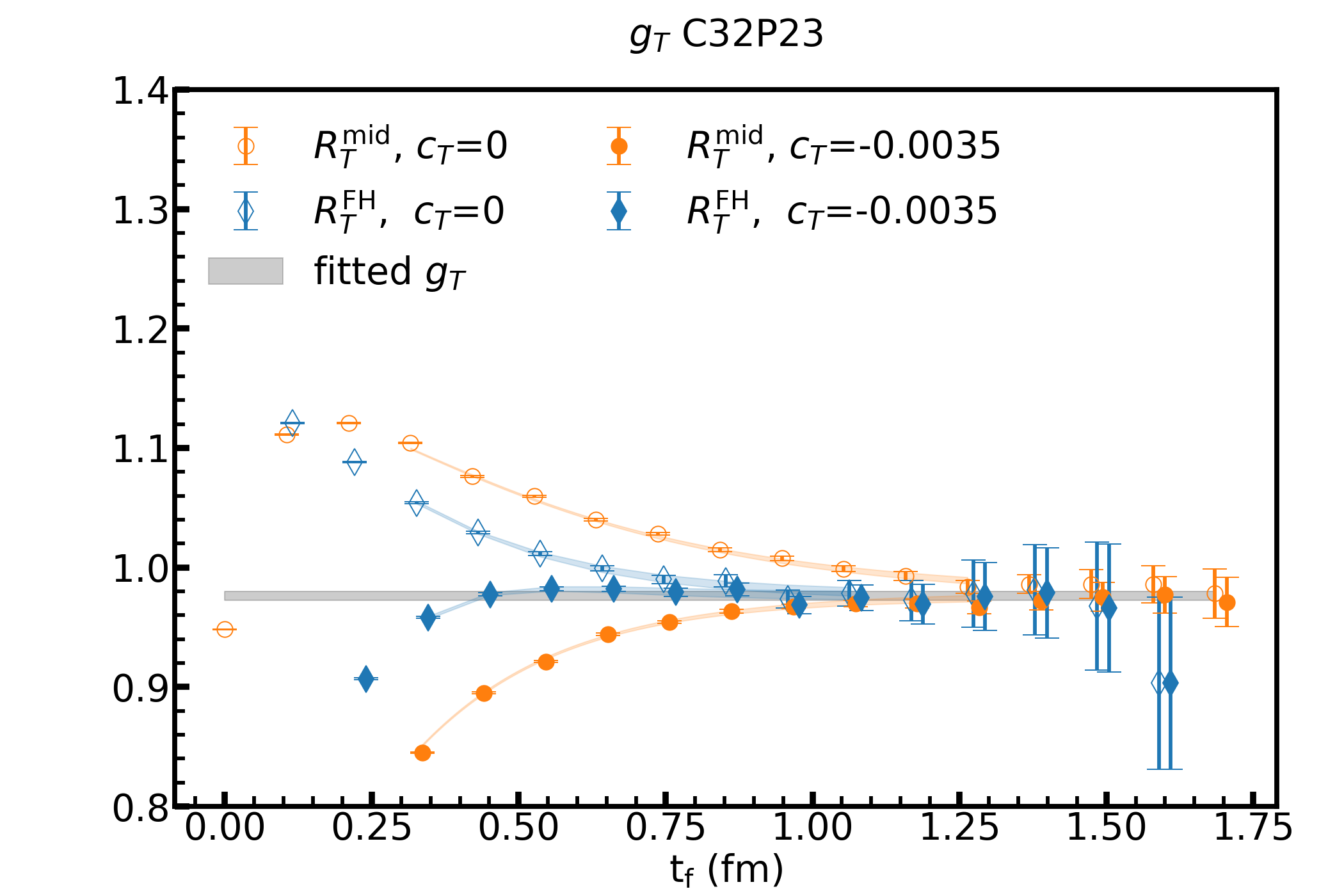}
    \includegraphics[width=0.475\linewidth]{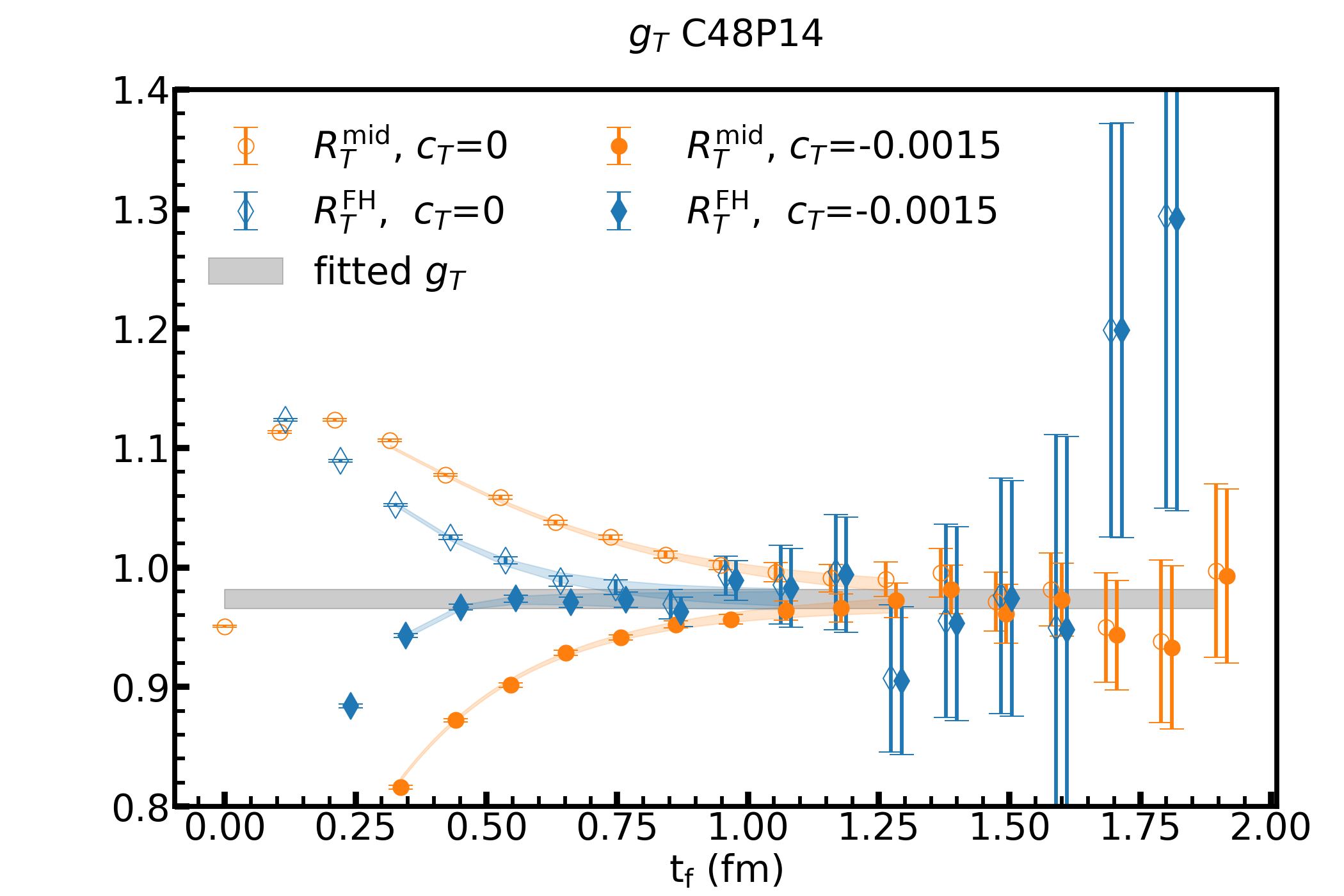}
    \includegraphics[width=0.475\linewidth]{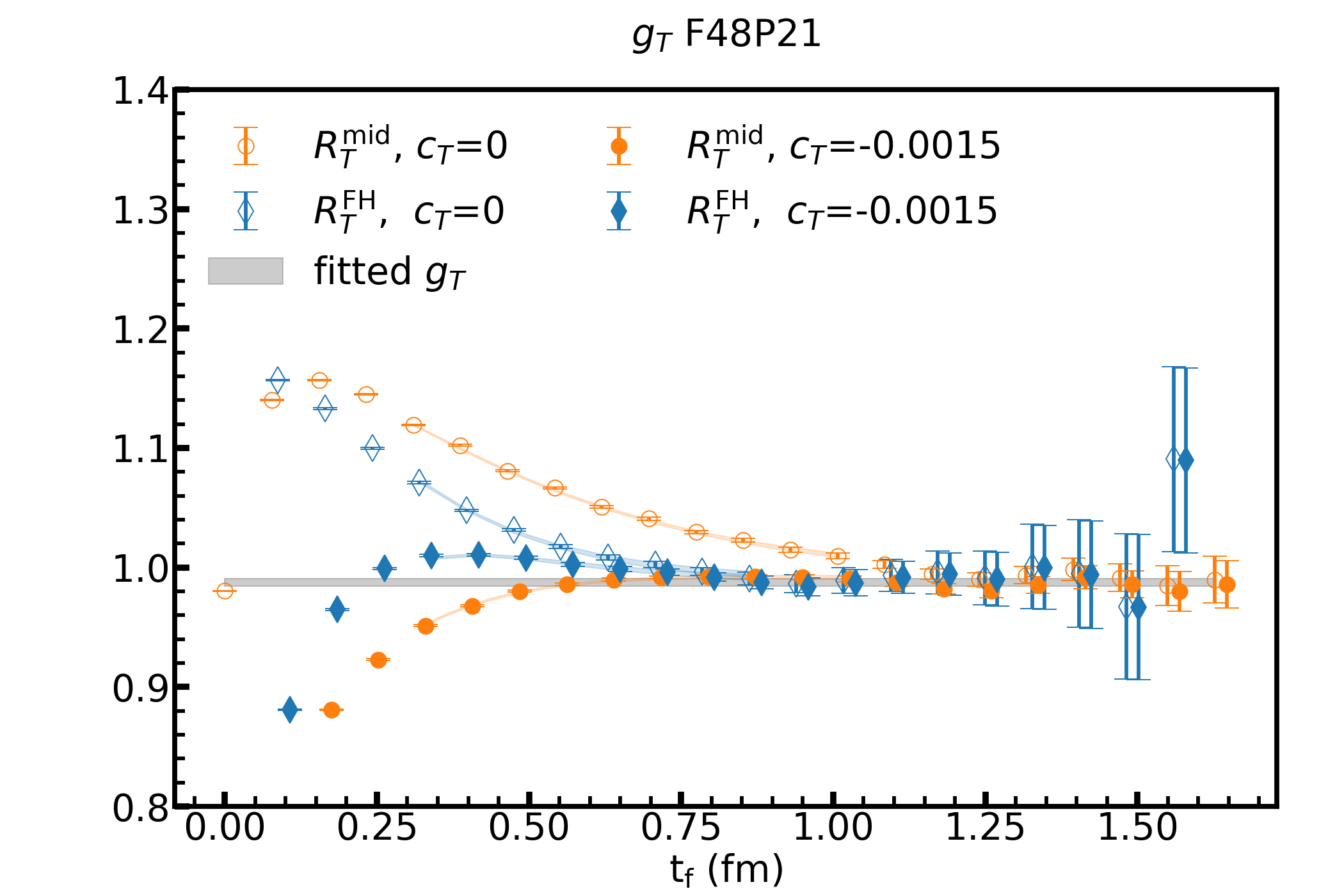}
    \includegraphics[width=0.475\linewidth]{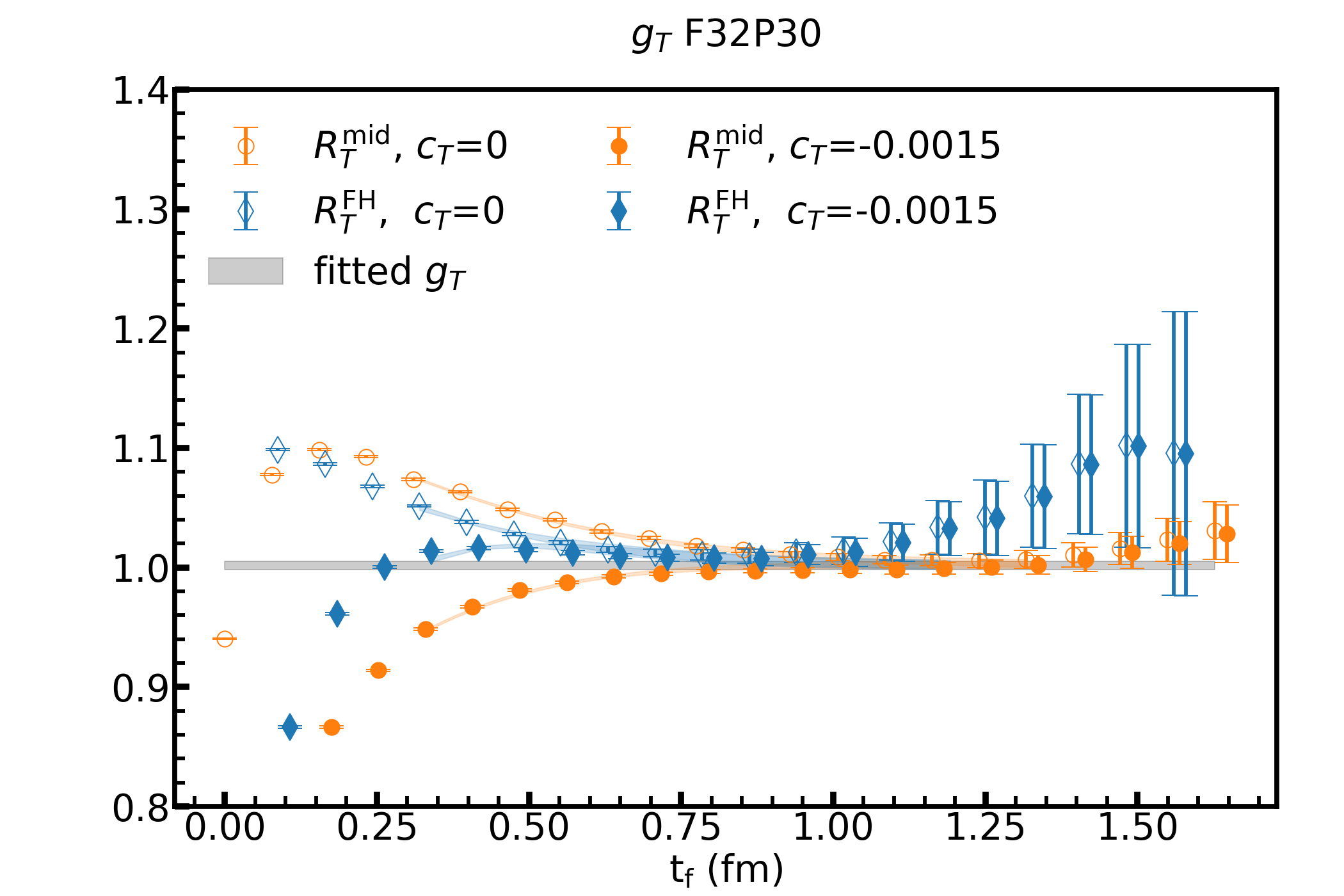}
    \includegraphics[width=0.475\linewidth]{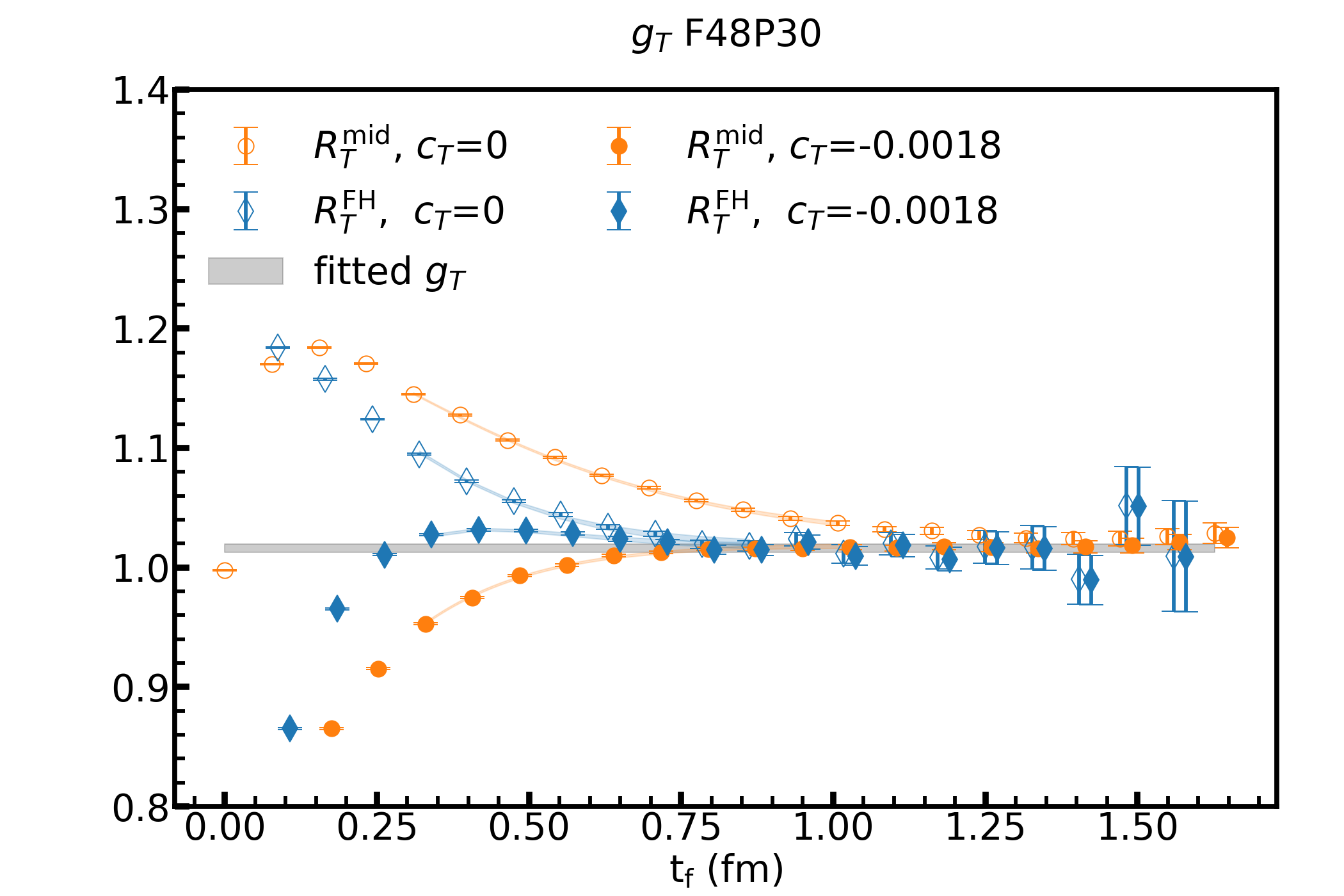}
    \caption{The dependence of the source-sink separation \( t_f \) for the ratios \( \mathcal{R}_X^{\rm mid}(t_f,c)\equiv \mathcal{R}_X(t_f, t = t_f/2;{\cal N}(c)) \) and Feynman-Hellmann fnunctions \( \mathcal{R}_X^{\text{FH}}(t_f) \) for the tensor charge $X=T$ on various ensembles: C48P23 (upper left), C32P23 (upper right), C48P14 (middle left), F48P21 (middle right),F48P30 (lower left), F48P30 (lower right), which cover different pion masses, volumes, and lattice spacings. The gray band represents the result from the joint three-state fit. The minimum source-sink separation for fitting is \( t_f^{\text{min}} \sim 0.3~\text{fm} \).}
    \label{fig:gTens}
\end{figure}

\begin{figure}
    \centering
    \includegraphics[width=0.475\linewidth]{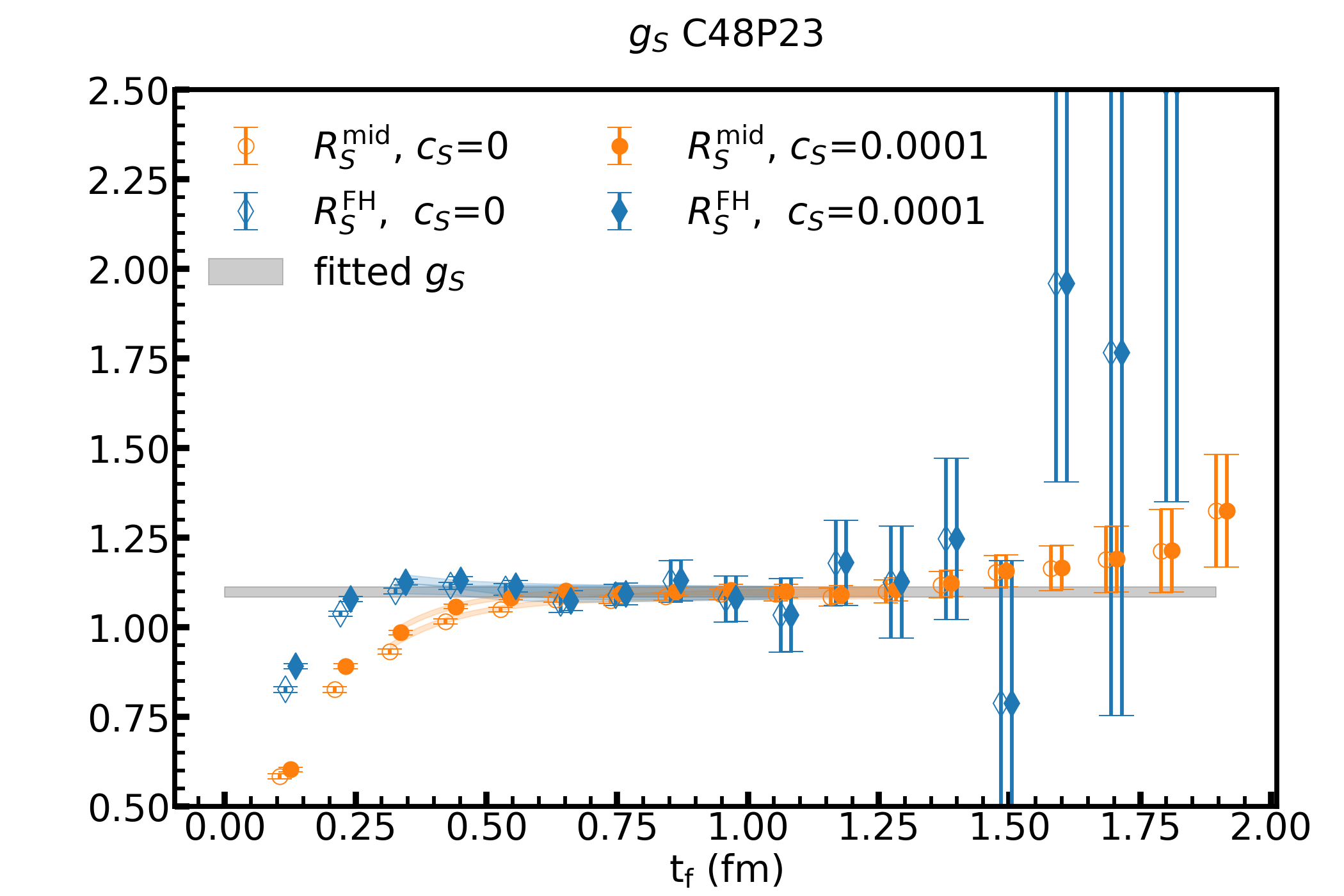}
    \includegraphics[width=0.475\linewidth]{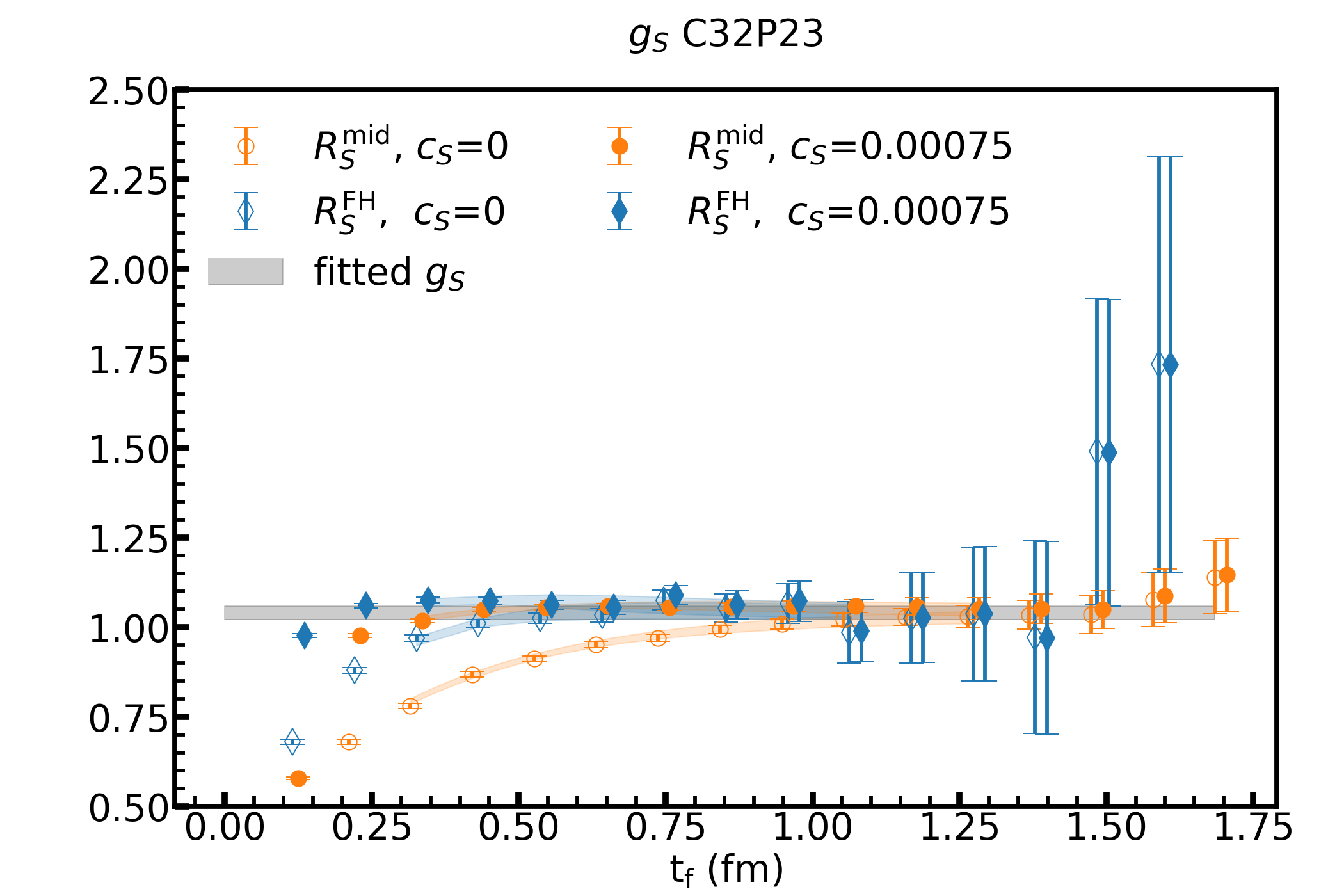}
    \includegraphics[width=0.475\linewidth]{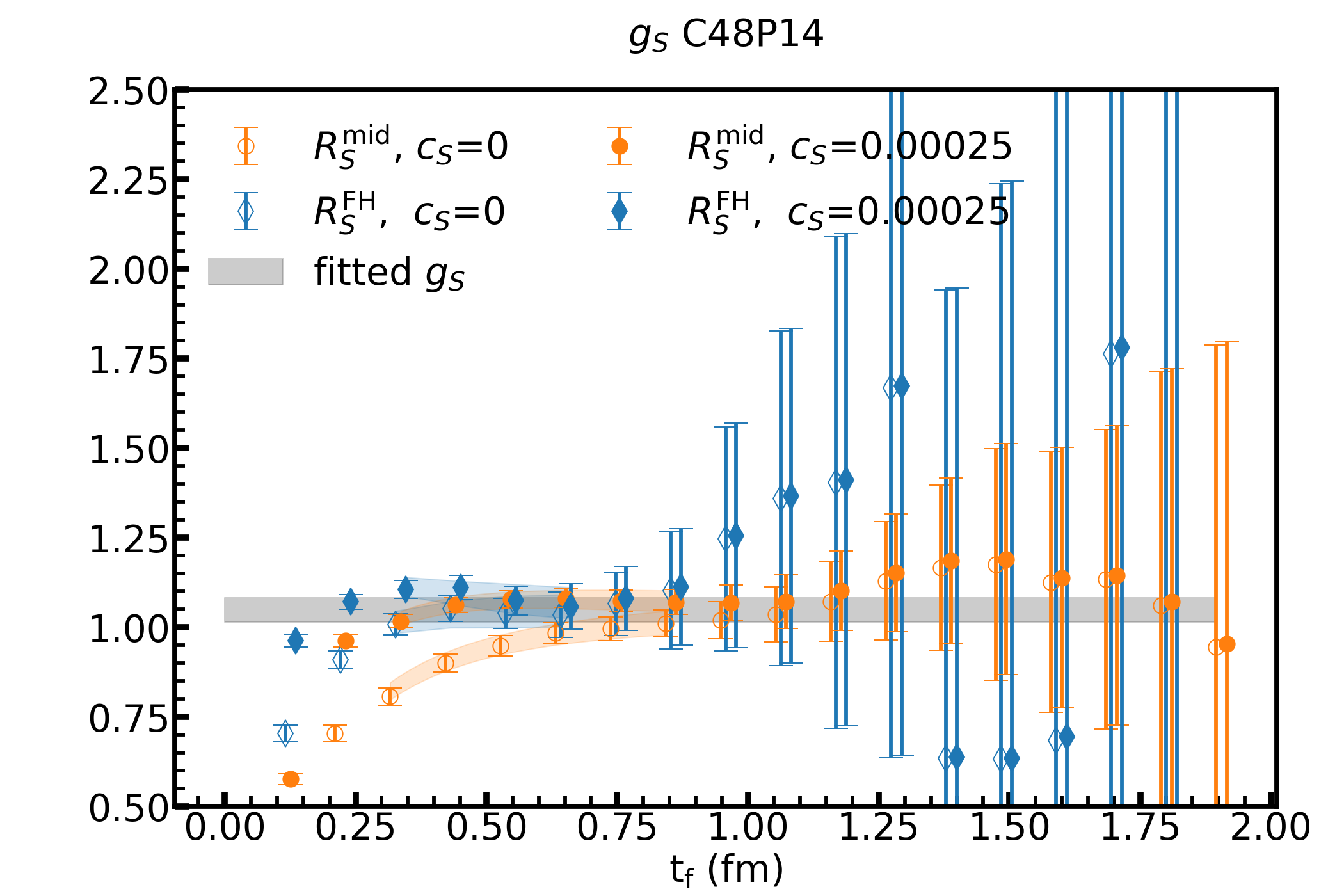}
    \includegraphics[width=0.475\linewidth]{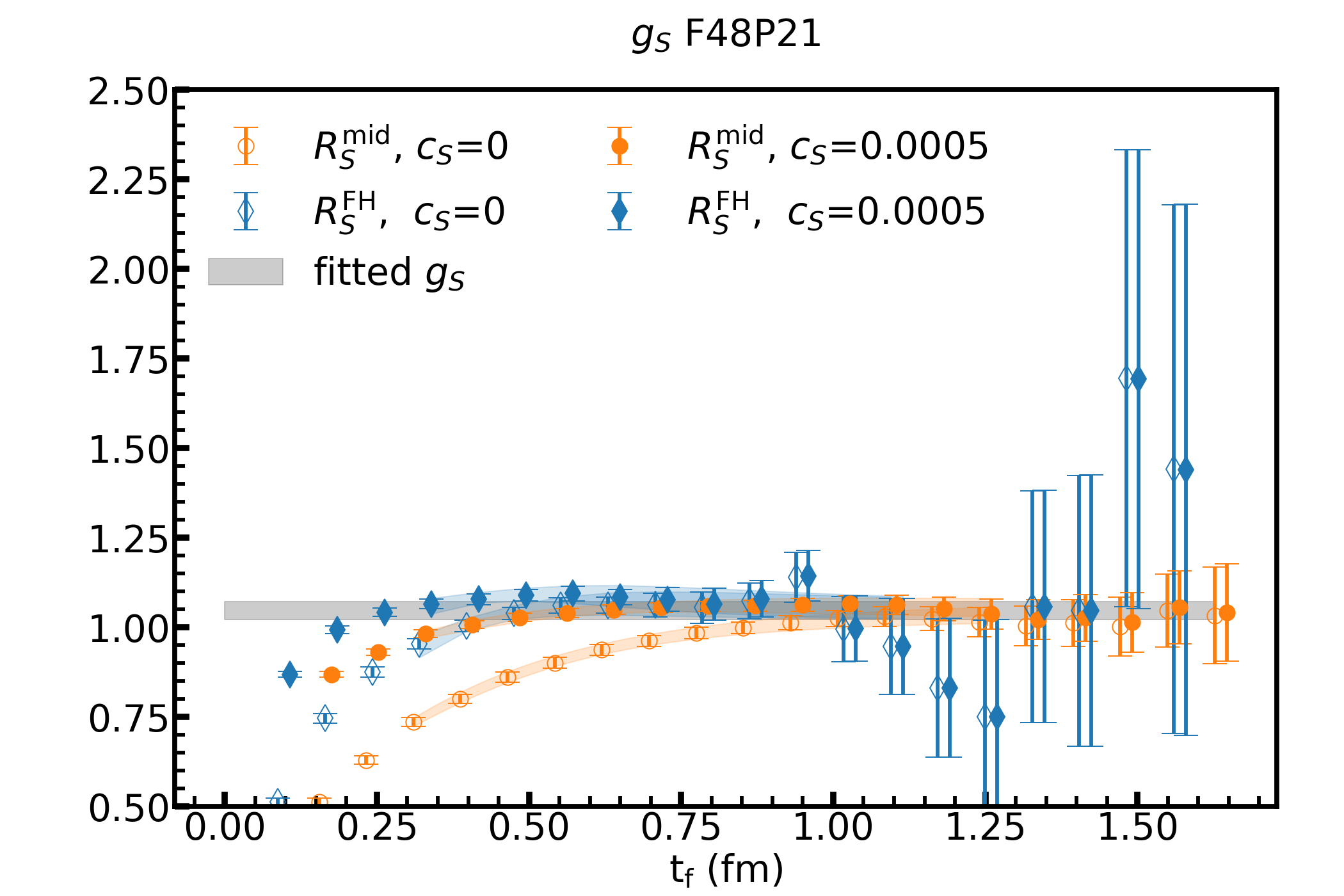}
    \includegraphics[width=0.475\linewidth]{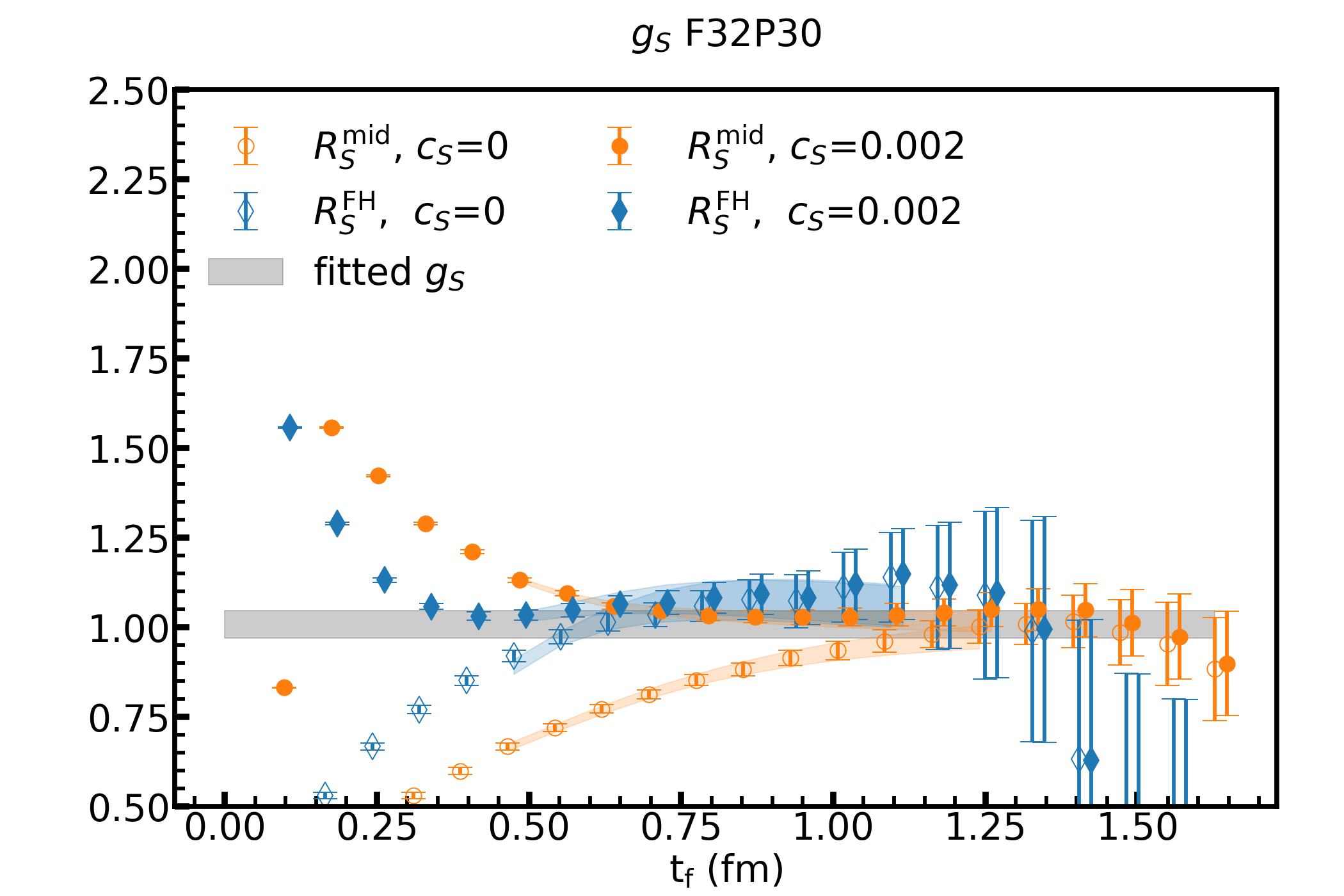}
    \includegraphics[width=0.475\linewidth]{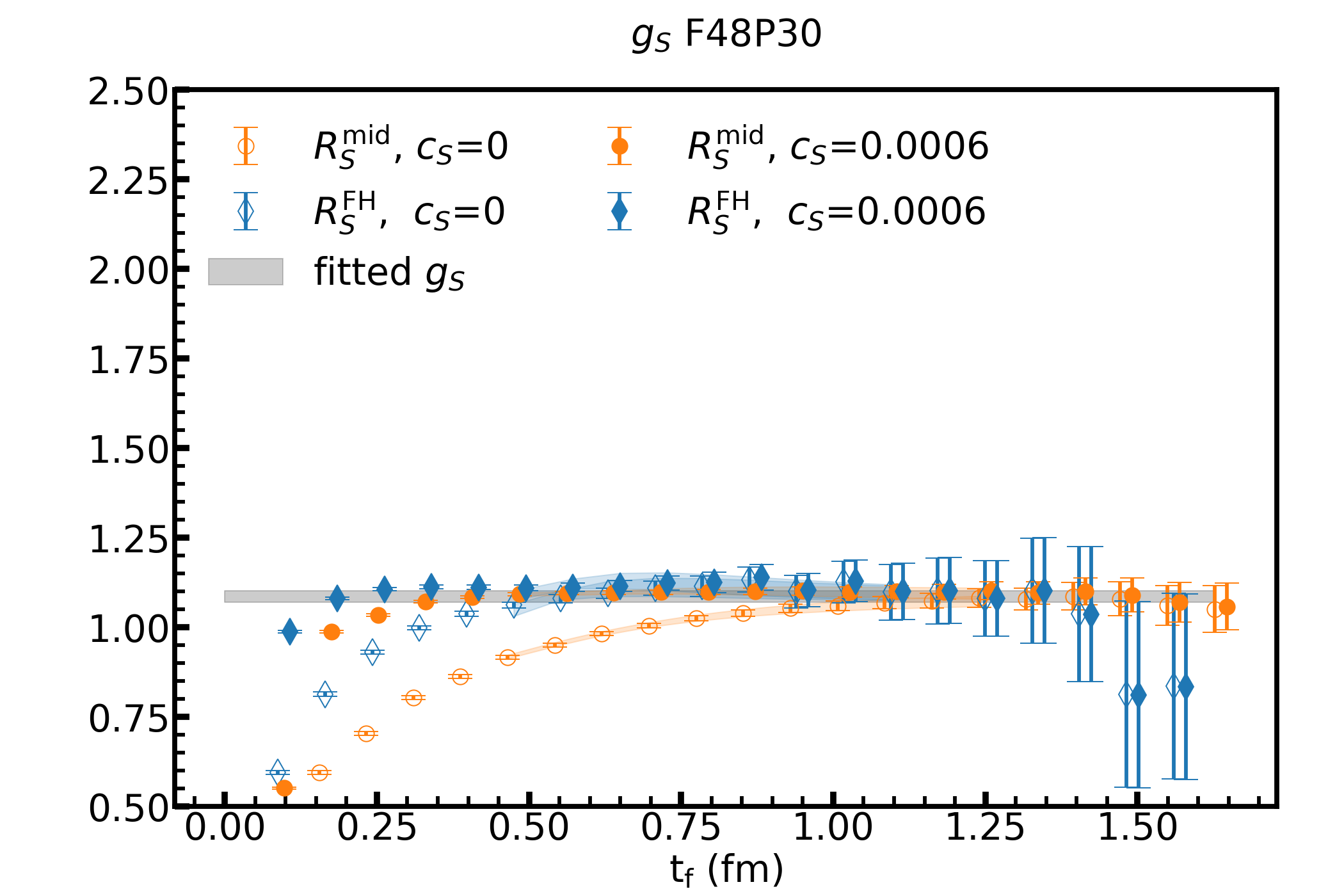}
    \caption{Similar to Fig.~\ref{fig:gTens} except for the scalar charge. }
    \label{fig:gSens}
\end{figure}

On each ensemble, we can find an optimized coefficient $c=c^{\rm opt}$ with mild time-dependencies in $R^{\rm mid}_{X}(t_f)$ and $R^{\rm FH}_{X}(t_f)$ in moderate and large $t_f$. Fig.~\ref{fig:gTens} displays the results for the tensor charge $g_T$ on six characteristic ensembles covering different pion masses (including the physical pion mass), volumes, and lattice spacings:

1) C48P23 with a=0.105 fm, $m_{\pi}\sim 230$ MeV, $L\sim$ 5.1 fm (upper left panel);

2) C32P23 with a=0.105 fm, $m_{\pi}\sim 230$ MeV, $L\sim$ 3.4 fm (upper right panel);

3) C48P14 with a=0.105 fm, $m_{\pi}\sim 135$ MeV, $L\sim$ 5.1 fm (middle left panel);

4) F48P21 with a=0.077 fm, $m_{\pi}\sim 210$ MeV, $L\sim$ 3.7 fm (middle right panel);

5) F32P30 with a=0.077 fm, $m_{\pi}\sim 300$ MeV, $L\sim$ 2.4 fm (lower left panel);

6) F48P30 with a=0.077 fm, $m_{\pi}\sim 300$ MeV, $L\sim$ 3.7 fm (lower right panel);

To illustrate the ESC, we also define another characteristic function
\begin{align}
    {\cal R}_X^{{\rm FH}}(t_f)&\equiv \sum_{t=t_c}^{t_f+a-t_c}{\cal R}_X(t_f+a,t)-\sum_{t=t_c}^{t_f-t_c}{\cal R}_X(t_f,t) = \langle {\cal O}_X \rangle_N +{\cal O}(e^{-\delta m \, t_f}),
\end{align}
which have different ESC compared to ${\cal R}^{\rm mid}_{T}= \langle {\cal O}_X \rangle_N +{\cal O}(e^{-\delta m \, t_f/2})$.
Among all of them, the ESCs of both ${\cal R}_X^{{\rm mid}}$ and ${\cal R}_X^{{\rm FH}}$ are dramatically suppressed after an optimized coefficient $c=c_X$ is chosen compared to the $c=0$ case. 

Similar improvements are observed for the scalar charge as shown in Fig.~\ref{fig:gSens}, except for C48P23 where the ESCs have been already efficiently suppressed with the distillated interpolation fields. Notably, the optimal parameter $c^{\mathrm{opt}}_X$ is consistently small across all cases examined, for both the tensor ($X=T$) and scalar ($X=S$) charges. However, a direct comparison of its values across ensembles is complicated by the fact that the distillation interpolation fields utilize different upper eigenvalue bands of the Laplace operator in each case. A systematic investigation into the lattice spacing, pion mass, and volume dependence of $c^{\mathrm{opt}}$ is therefore reserved for future study.

To obtain the ground state matrix element, we perform a joint fit combining $c=0$ and $c=c^{\rm opt}$ (the one with good suppression of ESC) with respect to $t$ and $t_f$ dependencies in ${\cal R}_{X}(t_f, \, t;{\cal N}(c))$ and $c_2(t_f;{\cal N}(c))$ with $c_2(t_f;{\cal N})\equiv \int \text{d}^3x\text{d}^3z \langle {\cal N}(\vec{x}, t_f) {\cal N}^{\dagger}(\vec{z}, 0) \rangle$, using a three-state parametrization (one ground state and two excited states),
\begin{align}
& c_2(t_f;c)=[ 1+d_1^2(c)e^{-\Delta_1 t_f}+d_2^2(c)e^{-\Delta_2 t_f}]Z(c) e^{-m_0 t_f}, \nonumber\\
&{\cal R}_{X}(t_f, \, t; c) = [b_{00}  + b_{10} d_1(c) (e^{-\Delta_1 t} + e^{-\Delta_1 (t_f-t)}) + b_{20} d_2(c)( e^{-\Delta_2 t} + e^{-\Delta_2 (t_f-t)}) \nonumber\\
&+ b_{11} d_1^2(c) e^{-\Delta_1 t_f} + b_{12}d_1(c)d_2(c)(e^{-\Delta_1(t_f-t)-\Delta_2 t} + e^{-\Delta_2 (t_f-t)-\Delta_1 t}) + b_{22} d_2^2(c)e^{-\Delta_2 t_f}]\nonumber\\
&/[ 1+d_1^2(c)e^{-\Delta_1 t_f}+d_2^2(c)e^{-\Delta_2 t_f}] \ ,
\end{align}
where $\Delta_i=m_i-m_0$ denotes the mass difference between excited state $i$ and the ground state. The $b_{ij}$ denotes the matrix element from state $i$ to state $j$, where $b_{00}$ is the ground state matrix element of interest. The values $m_0$, $\Delta_i$, and $b_{ij}$ are independent of $c$, while the coefficients $Z(c)$ and $d_i(c)$ differ between the cases $c=0$ and $c=c^{\rm opt}$. The fitted ground state matrix elements are shown as the grey bands in Figs.~\ref{fig:gTens} and~\ref{fig:gSens}.

\begin{figure}
   \centering
    \includegraphics[width=0.475\linewidth]{figures/ratio/5.C48P14/ratio_tsep_gT.png}
    \includegraphics[width=0.475\linewidth]{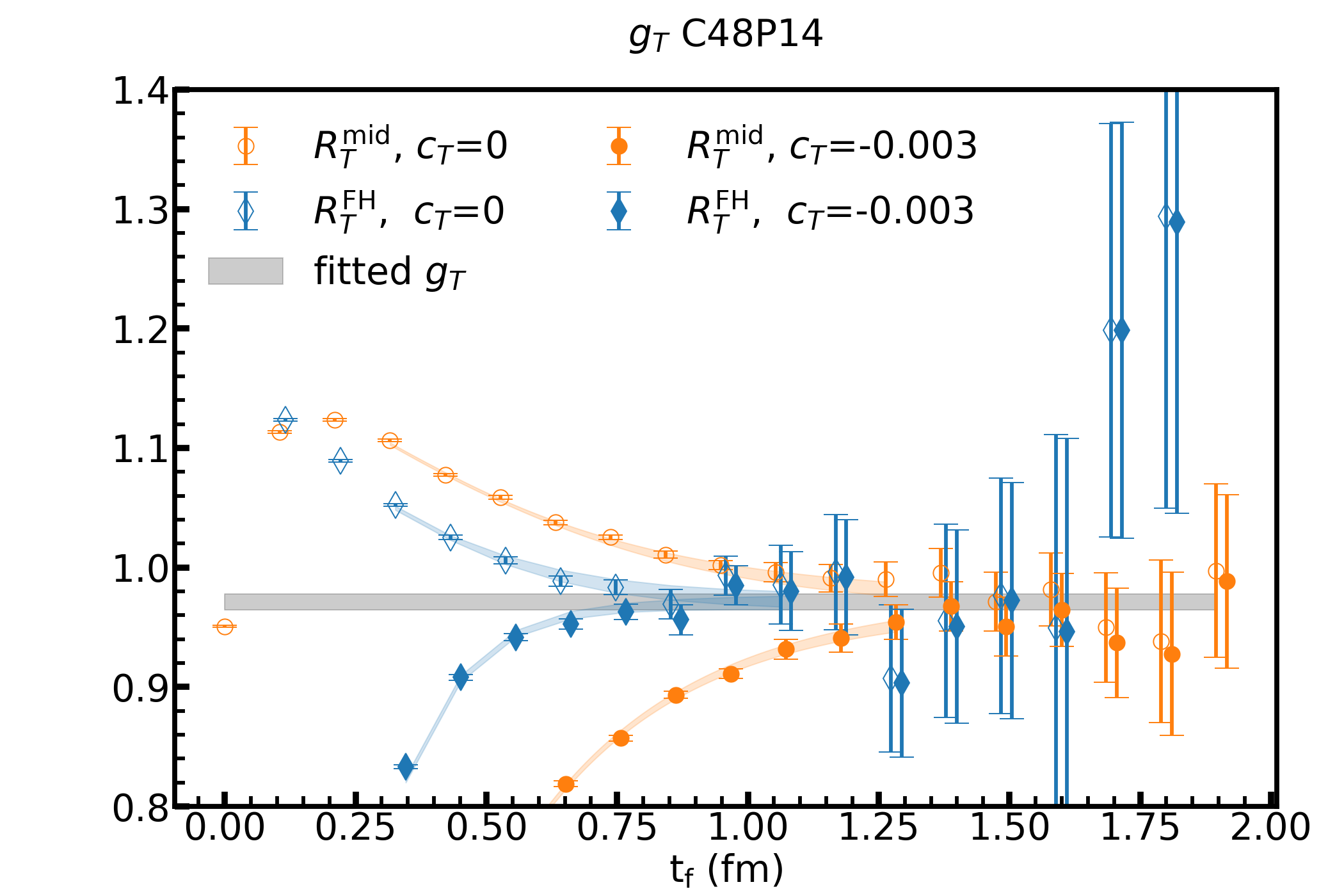}
    \caption{The dependence of the source-sink separation \( t_f \) for the ratios \( \mathcal{R}_X^{\rm mid}(t_f)\equiv \mathcal{R}_X(t_f, t = t_f/2) \) and Feynman-Hellmann functions \( \mathcal{R}_X^{\text{FH}}(t_f) \) for the tensor charges on C48P14, with chosen $c^{\rm opt}$ (left panel) and $2c^{\rm opt}$ (right panels). The ground-state matrix elements from the joint three-state fit (gray bands) agree with each other.}
    \label{fig:fitwithdiffc_t}
\end{figure}

\begin{figure}
   \centering
    \includegraphics[width=0.475\linewidth]{figures/ratio/5.C48P14/ratio_tsep_gS.png}
    \includegraphics[width=0.475\linewidth]{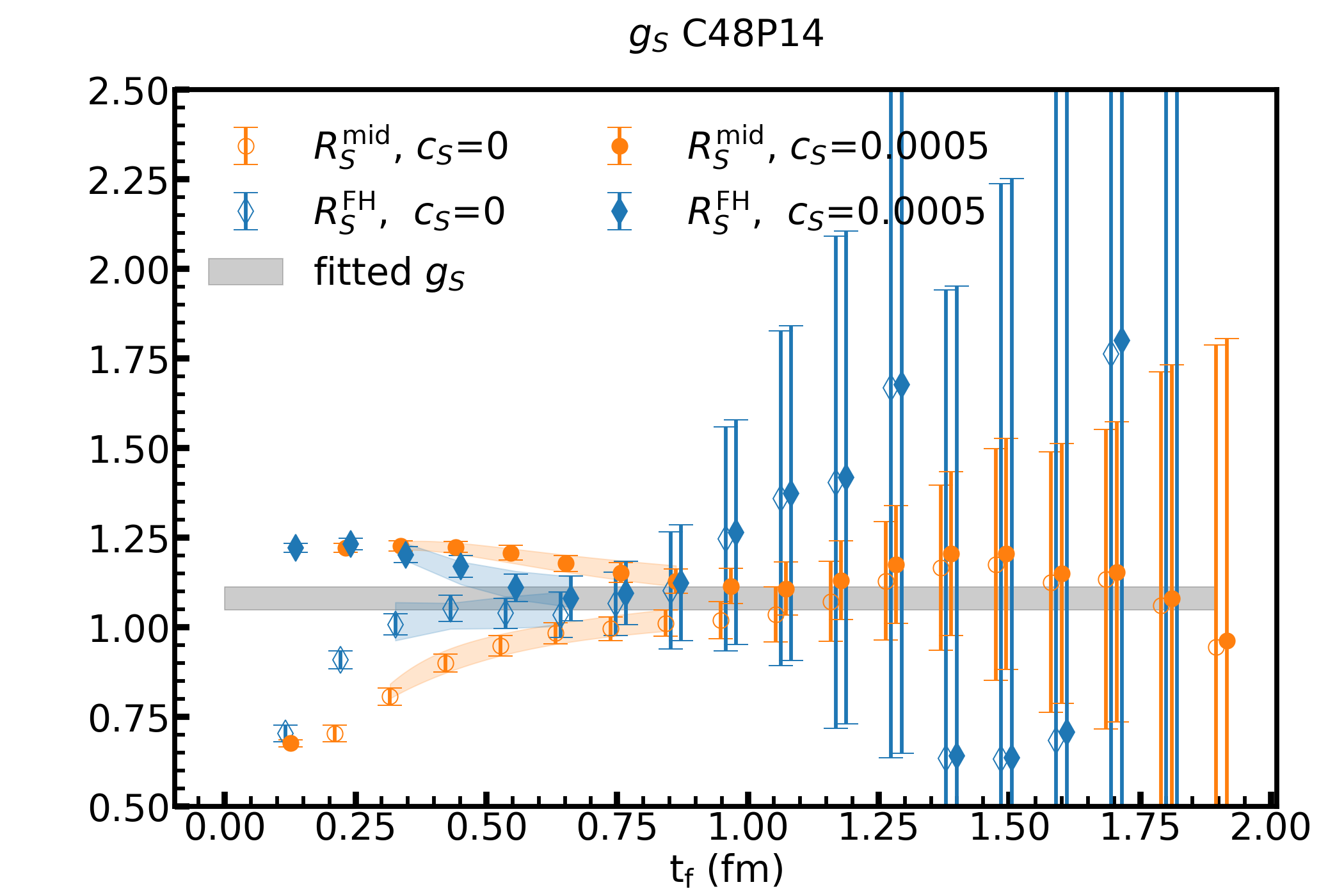}
    \caption{Similar to Fig.~\ref{fig:fitwithdiffc_t} except for the scalar charge.}
    \label{fig:fitwithdiffc_s}
\end{figure}

We expect the fitted ground-state matrix elements to be stable under variations of $c_X$ around its optimized value. The theoretical basis for this is that an interpolator with an arbitrary $c$ can be expressed as a linear combination of the $c=0$ and $c=c^{\rm opt}$ cases, with corrections only at ${\cal O}(c^2)$. This stability is confirmed numerically in Fig.~\ref{fig:fitwithdiffc_t} and \ref{fig:fitwithdiffc_s}, which compares ${\cal R}_X^{\rm mid/FH}$ for $c=c^{\rm opt}$ and $c=2c^{\rm opt}$. The joint fits yield consistent ground-state matrix elements in both cases, validating the robustness of our extraction.

\clearpage

\subsubsection{Influence from $N\pi$}\label{Npi_inf}

\begin{figure}[!h]
    \centering
    \includegraphics[width=0.475\linewidth]{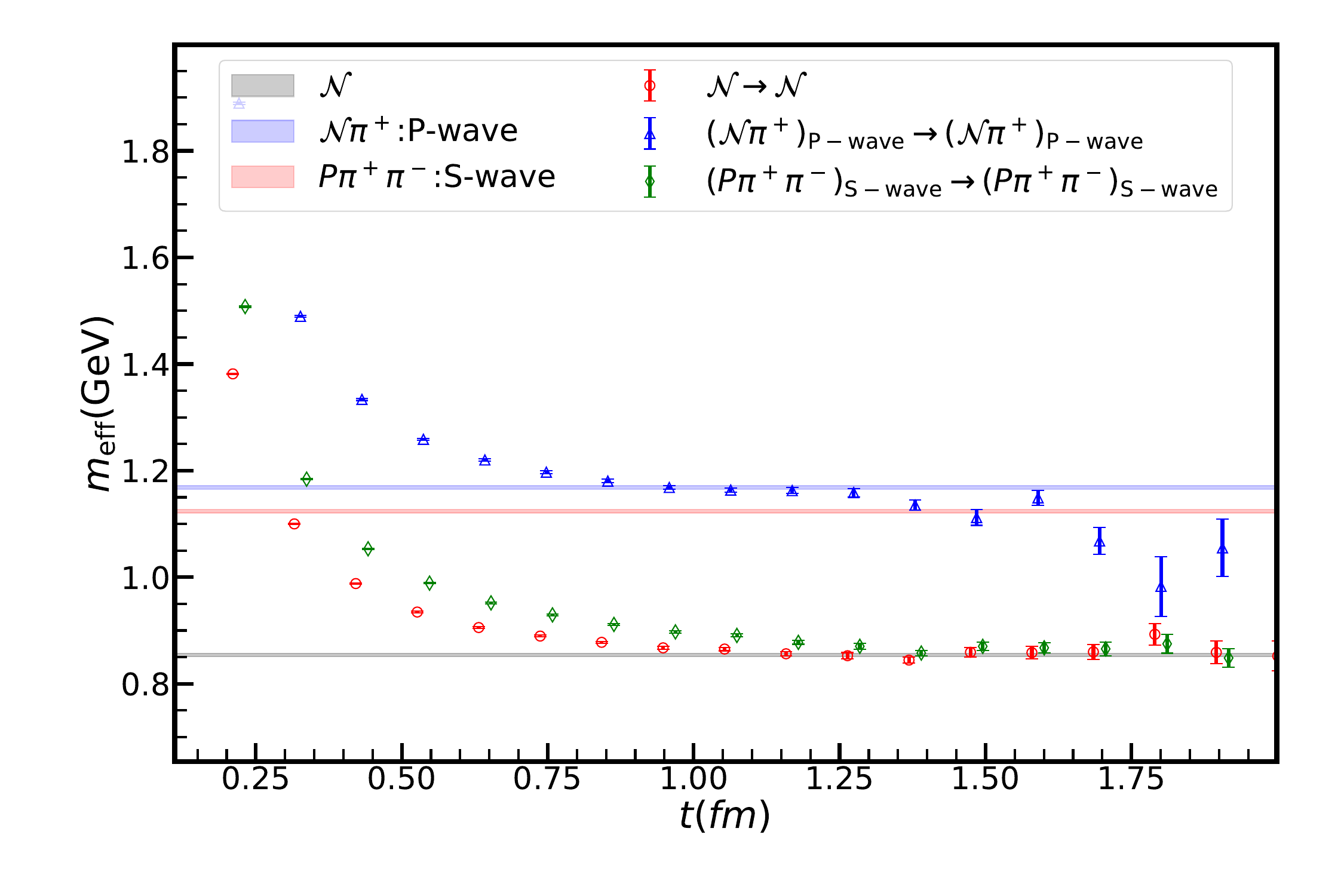}
    \includegraphics[width=0.475\linewidth]{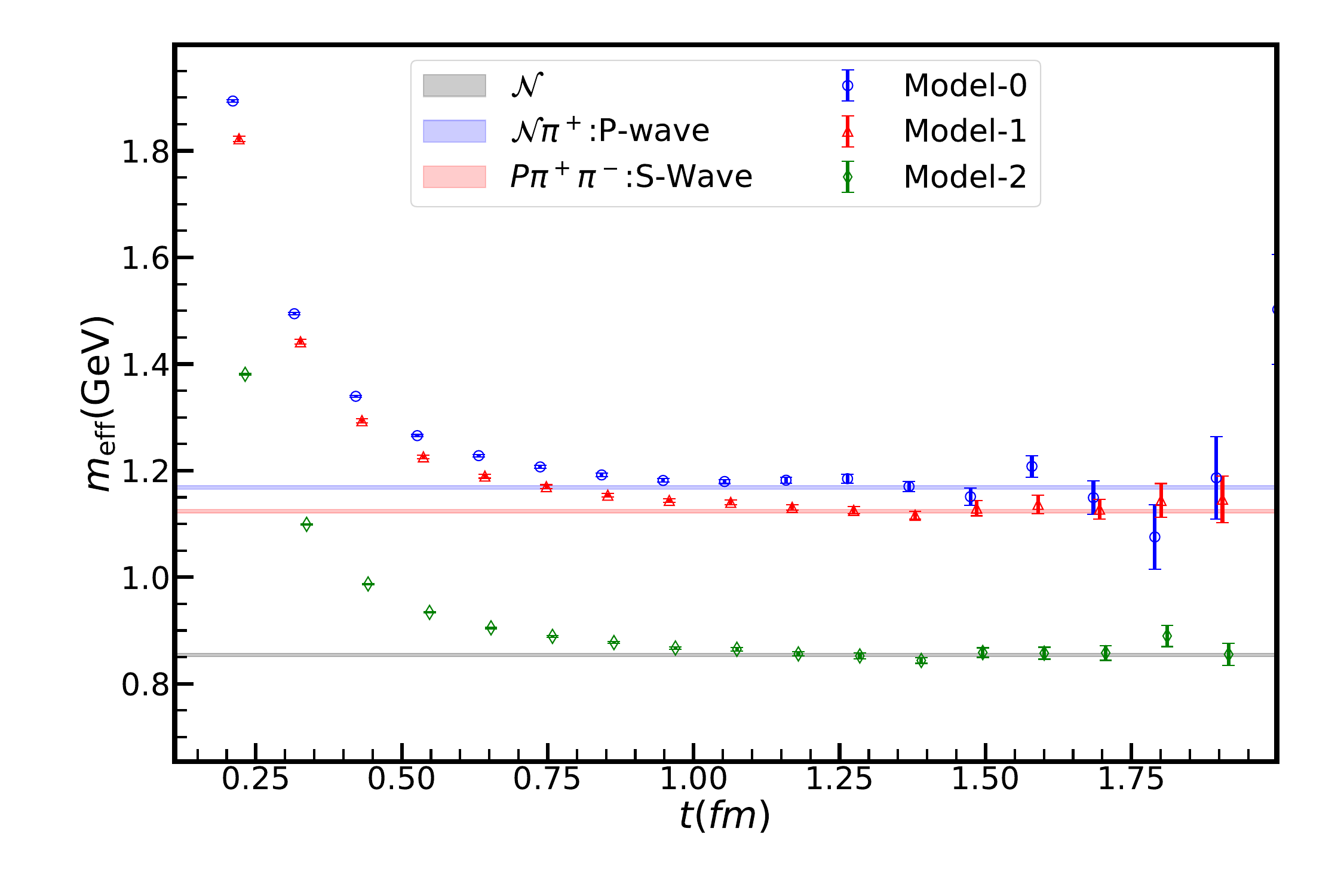}
    \caption{
    Effective masses on the C48P14 ensemble with physical pion mass, before (left panel) and after (right panel) solving the GEVP in Eq.~(\ref{eq:GEVP}) with the three-operator basis $\{{\cal N}, ({\cal N} \pi)_{\rm P-wave}, ({\cal N}\pi\pi)_{\rm S-wave}\}$. The grey, blue and red band shows the energy of nucleon ground state, non-interacting nucleon-pion threshold and nucleon-pion-pion threshold, respectively.
    }
    \label{fig:energNpi}
\end{figure}

{To assess the potential influence of nucleon-pion scattering states, we perform a GEVP analysis using a three-operator basis \(\{{\cal N}, ({\cal N} \pi)_{\rm P-wave}, ({\cal N}\pi\pi)_{\rm S-wave}\}\) on the physical-pion-mass ensemble C48P14. Figure~\ref{fig:energNpi} compares the effective masses before (left panel) and after (right panel) solving the GEVP. The gray, blue, and red bands indicate the energies of the nucleon ground state, the non-interacting \(N\pi\) threshold, and the \(N\pi\pi\) threshold, respectively. After applying the GEVP, the resulting eigenstates align well with these thresholds, confirming that the method successfully isolates the expected low-lying states.}

For the influence on $g_{S,T}$, we just using a two-operator GEVP basis, $\{{\cal N}, ({\cal N} \pi)_{\rm P-wave}\}$ since that involving ${\cal N}\pi\pi)_{\rm S-wave}$ is extremely complicated. The optimized linear combination is obtained by solving Eq.~(\ref{eq:GEVP}), where the stabilized solutions can be obtained at relatively large $t'$ and $t_0$. As tested on C48P14 and shown in Fig.~\ref{fig:gT_GEVPlE}, the results with and without GEVP have similar time dependencies in moderate and large $t_f$, indicating that the nucleon-pion scattering state has a smaller influence on the ESC for both $g_T$ and $g_S$, compared to the current-involved excited states. Therefore, ${\cal N} \pi$ is not considered in extracting the ground state matrix elements in our calculations.

\begin{figure}[!h]
    \centering
    \includegraphics[width=0.475\linewidth]{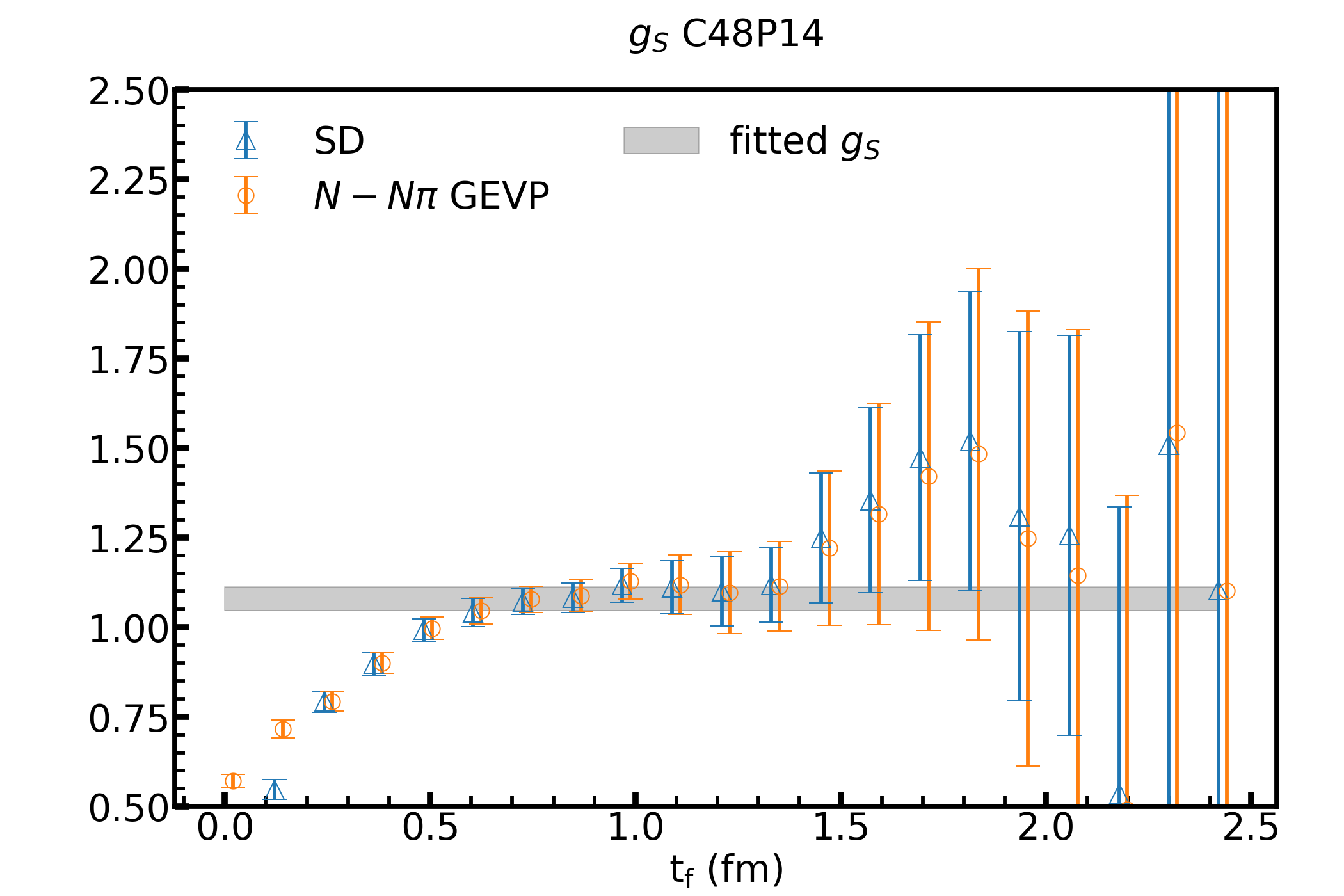}
    \includegraphics[width=0.475\linewidth]{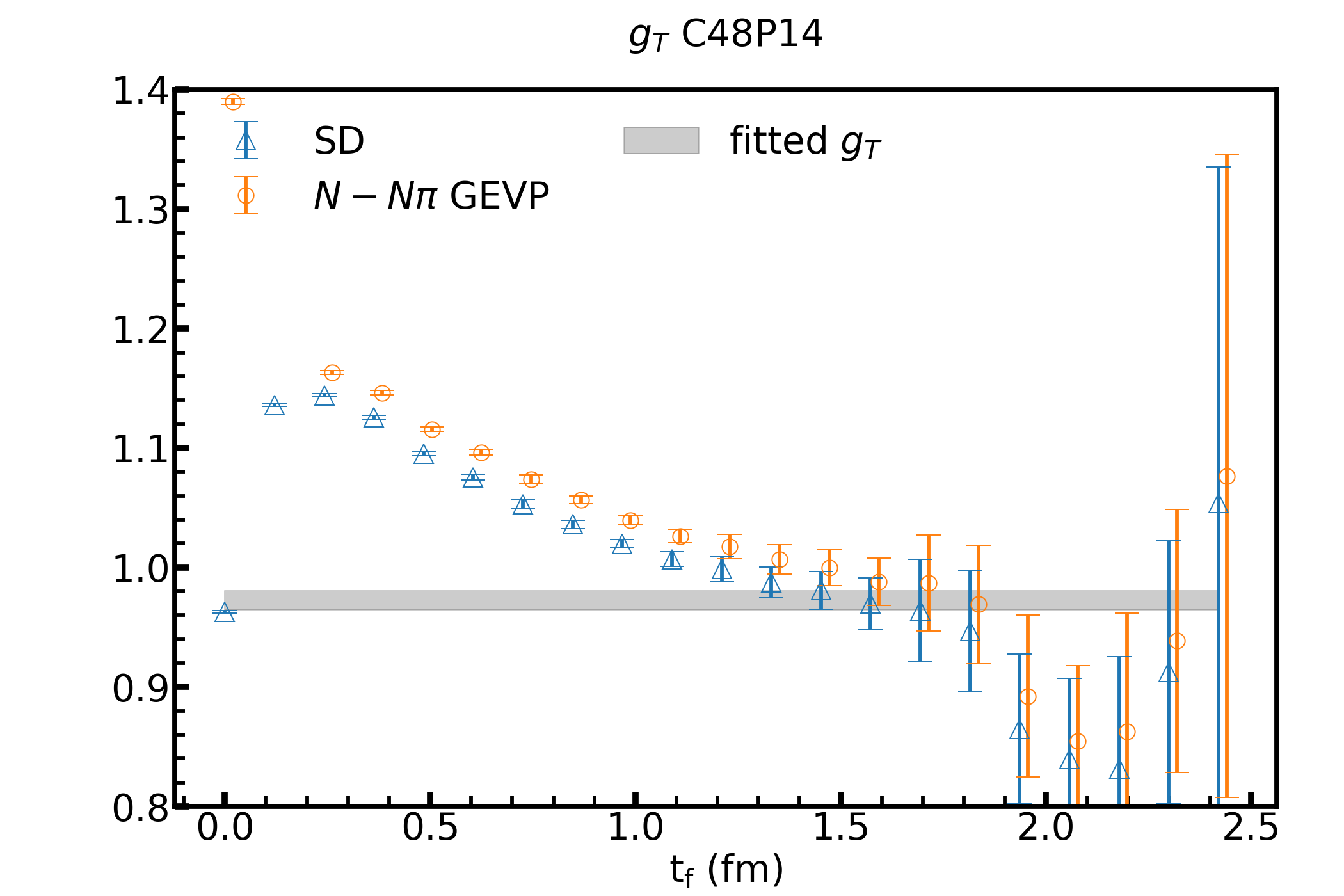}
    \caption{The comparison between the ${\cal N}-{\cal N} \pi$ GEVP (orange circles) and standard (SD, blue triangles) ${\cal N}$ interpolation operator on \( \mathcal{R}_X^{\rm mid}(t_f)\equiv \mathcal{R}_X(t_f, t = t_f/2) \). The left panel displays the scalar charge as a function of the sink source separation $t_f$. The right panel is for tensor charge. The lattice data are generated on C48P14 (physical pion mass).}
    \label{fig:gT_GEVPlE}
\end{figure}

\clearpage

\subsubsection{Contribution of $\langle \mathcal{N}_X (\vec{x}, t_f) O_X(\vec{y},t)  \mathcal{N}^{\dagger}_X(\vec{z},0) \rangle$}\label{sec:ContriNON}

\begin{figure}
    \centering
    \includegraphics[width=0.475\linewidth]{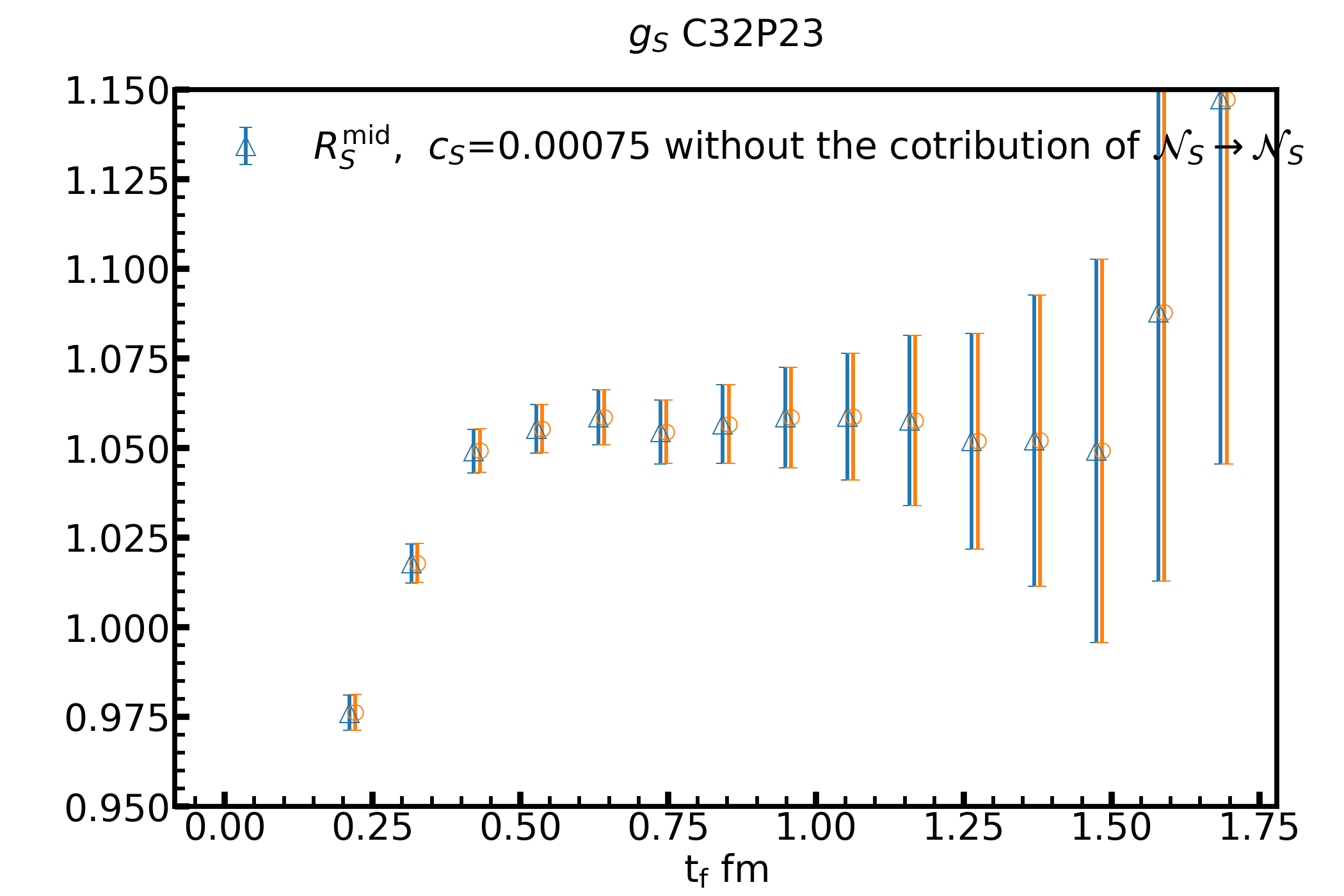}
    \includegraphics[width=0.475\linewidth]{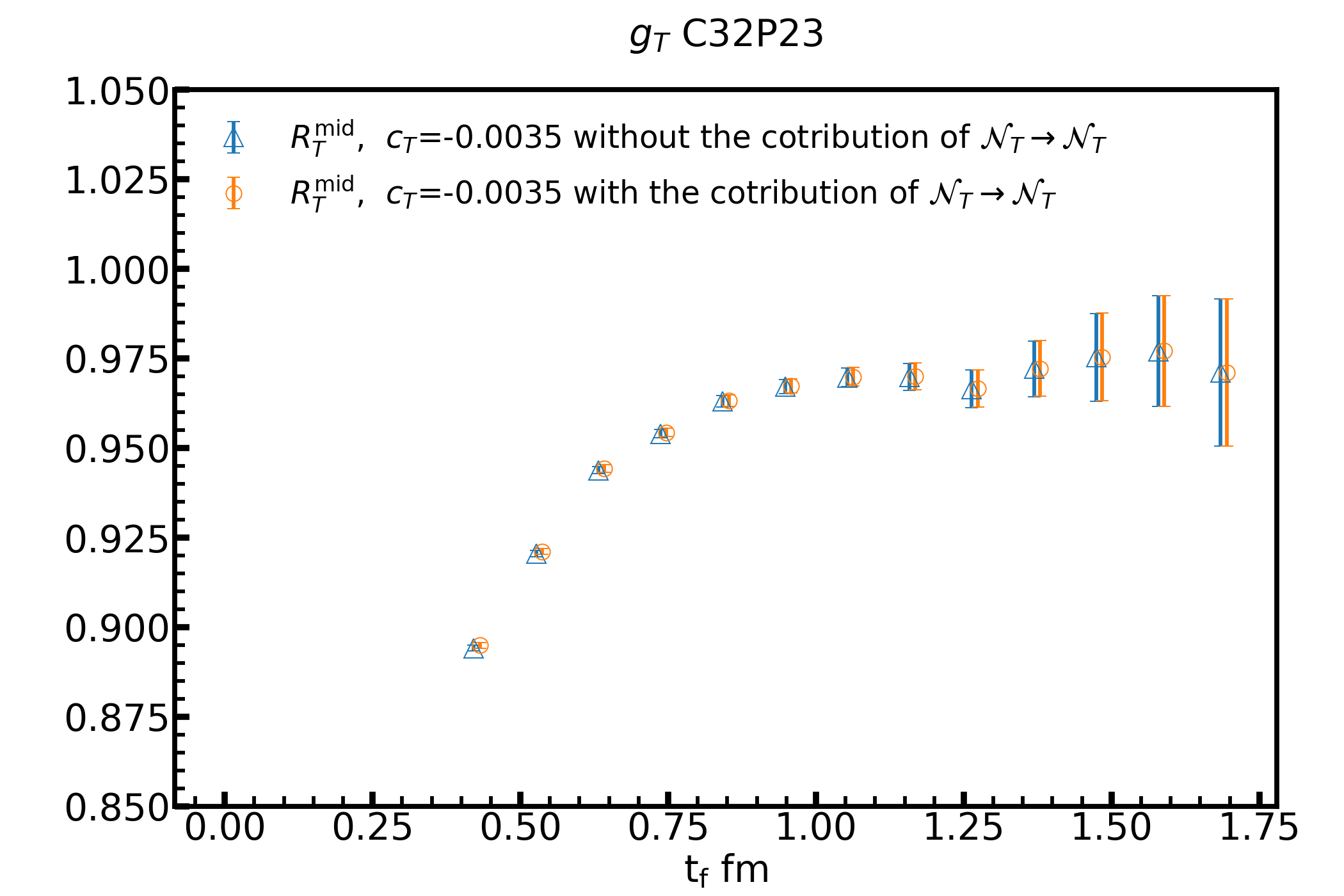}
    \caption{
    Comparison of \( \mathcal{R}_X^{\rm mid}(t_f) \equiv \mathcal{R}_X(t_f, t = t_f/2) \) with and without the contribution from \(\mathcal{N}_X \rightarrow \mathcal{N}_X\), for the scalar (left panel) and tensor (right panel) cases.
    The orange circles indicate results where \(\mathcal{R}_X^{\rm mid}(t_f)\) includes all contributions: \(\mathcal{N} \rightarrow \mathcal{N}\), \(\mathcal{N}_X \rightarrow \mathcal{N}\), \(\mathcal{N} \rightarrow \mathcal{N}_X\), and \(\mathcal{N}_X \rightarrow \mathcal{N}_X\).
    The blue triangles indicate results where \(\mathcal{R}_X^{\rm mid}(t_f)\) includes only the contributions from \(\mathcal{N} \rightarrow \mathcal{N}\), \(\mathcal{N}_X \rightarrow \mathcal{N}\), and \(\mathcal{N} \rightarrow \mathcal{N}_X\).
    The lattice data are generated on the C32P23 ensemble.
    }
    \label{fig:NOtoNO}
\end{figure}

The three-point function for the interpolating field combination $\mathcal{N} + c_X\mathcal{N}_X$ comprises four distinct types of contributions:
\begin{align}
& \langle \mathcal{N} (\vec{x}, t_f) O_X(\vec{y}, t) \mathcal{N}^{\dagger}(\vec{z}, 0) \rangle, 
c_X \langle \mathcal{N}_X (\vec{x}, t_f) O_X(\vec{y}, t) \mathcal{N}^{\dagger}(\vec{z}, 0) \rangle, \nonumber \\ 
& c_X \langle \mathcal{N} (\vec{x}, t_f) O_X(\vec{y}, t) \mathcal{N}_X^{\dagger}(\vec{z}, 0) \rangle, \
c_X^2 \langle \mathcal{N}_X (\vec{x}, t_f) O_X(\vec{y}, t) \mathcal{N}_X^{\dagger}(\vec{z}, 0) \rangle,
\end{align}
which we denote as $\mathcal{N} \rightarrow \mathcal{N}$, $\mathcal{N} \rightarrow \mathcal{N}_X$, $\mathcal{N}_X \rightarrow \mathcal{N}$, and $\mathcal{N}_X \rightarrow \mathcal{N}_X$, respectively.

Previous studies employing similar five-quark interpolating operators (e.g., \cite{Alexandrou:2024tps, Wang:2023omf, Barca:2024hrl, Barca:2025det}) included only the first three contributions, omitting the $\mathcal{N}_X \rightarrow \mathcal{N}_X$ term due to its exceptionally complex contraction structure. As summarized in Table~\ref{tab:quark_diagram}, the number of required quark diagrams increases substantially: from just 2 for the two-point function of the standard three-quark interpolation operator ${\cal N}$, to 3 times more for its three-point function with $O_X=\bar{u}\Gamma_X u-\bar{d}\Gamma_X d$, and similarly for the $\mathcal{N}_X \rightarrow \mathcal{N}$ two-point function. The $\mathcal{N}_X \rightarrow \mathcal{N}_X$ two-point function requires 4.3 times more diagrams than the standard case, with an additional 4.3-fold increase for its three-point function, making the computation of the $\mathcal{N}_X \rightarrow \mathcal{N}_X$ contribution extremely challenging even at the propagator contraction stage.

We have developed an automated framework for constructing correlation functions with arbitrary interpolating and current operators, incorporating optimizations that improve performance by an order of magnitude compared to the initial implementation. Figure~\ref{fig:NOtoNO} shows the ratios $R_{S}^{\rm mid}$ (left panel) and $R_{T}^{\rm mid}$ (right panel) computed with (orange circles) and without (blue triangles) the $\mathcal{N}_X \rightarrow \mathcal{N}_X$ contribution. The results demonstrate that this contribution is negligible.

This observation can be understood from volume scaling arguments: the correlators $\langle \mathcal{N}_X (t_f) O_X(t) \mathcal{N}^{\dagger}(0) \rangle$, $\langle \mathcal{N} (t_f) O_X(t) \mathcal{N}_X^{\dagger}(0) \rangle$, and $\langle \mathcal{N}_X (t_f) O_X(t) \mathcal{N}_X^{\dagger}(0) \rangle$ are each enhanced by a factor of the spatial volume $V$ compared to $\langle \mathcal{N} (t_f) O_X(t) \mathcal{N}^{\dagger}(0) \rangle$. Consequently, the optimal parameter $c_X$ scales as $\mathcal{O}(1/V)$, making $c_X^2$ of order $\mathcal{O}(1/V^2)$. Although $\langle \mathcal{N}_X (t_f) O_X(t) \mathcal{N}_X^{\dagger}(0) \rangle$ itself is volume-enhanced, its overall contribution to the three-point function is suppressed by $1/V$.

Given the negligible impact of the $\mathcal{N}_X \rightarrow \mathcal{N}_X$ term, we calculate $\langle \mathcal{N}_X (\vec{x}, t_f) O_X(\vec{y},t) \mathcal{N}_X^{\dagger}(\vec{z},0) \rangle$ only for ensembles with $N_{\rm e} < 100$ to optimize computational efficiency.

   \begin{table}[htbp]
     \centering
        \caption{Counts of quark diagrams in the correlation functions $c_2 = \langle \mathcal{O}_A \mathcal{O}^{\dagger}_B\rangle$ and $c_3 = \langle \mathcal{O}_A  O_X \mathcal{O}^{\dagger}_B  \rangle$, for the iso-vector current $O_X=\bar{u}\Gamma_X u-\bar{d}\Gamma_X d$.}
       \begin{tabular}{c|cc|cc}
       \hline
       \hline
           \multirow{2}[4]{*}{$\mathcal{O}_A/\mathcal{O}_B$} & \multicolumn{2}{c|}{three-quark} & \multicolumn{2}{c}{five-quark} \\
\cline{2-5}            & $c_2$ & $c_3$ & $c_2$ & $c_3$ \\
       \hline
       three-quark  & 2    & 6    & 6    & 26 \\
       five-quark & 6    & 26   & 26   & 138 \\
       \hline
       \hline
       \end{tabular}%
     \label{tab:quark_diagram}%
   \end{table}%

\clearpage

\subsection{Simulation setup and auto-correlation of $g_{S,T}$}\label{autocorr}

\begin{table}[htbp]
  \centering
  \caption{The $g_V$ , $g_S$ and $g_T$ values from 2-state fit and 3-state fit for comparison.}
\resizebox{1.00\columnwidth}{!}{
    \begin{tabular}{c|cccccccc|cc|c|cc|cc}
    \hline
    \hline
           & $a$(fm) & $n_L^3\times n_T$ & $m_\pi$(MeV) & $m_\pi L$ & $m_{\eta_s}(\rm MeV)$ & $n_{\rm cfg}$ & interval & $\tau$ & $N_{\rm e}$ & $N_{\rm st}$ &  $g_V^{\rm 2-state}$  & $g_S^{\rm 2-state}$ & $g_S^{\rm 3-state}$  & $g_T^{\rm 2-state}$ & $g_T^{\rm 3-state}$ \\
    \hline
    C24P34    & \multirow{8}[2]{*}{0.1053} & 24$\times$64 & 341.1(1.8)  & 4.38 &  748.61(75)  & 48   & 200 & 1.000   & 50     & 250        &     0.9927(62)           & 1.036(18)      & 1.038(23)       & 1.0182(58)      & 1.0167(67) \\
    C24P29    & & 24$\times$72 & 292.7(1.2) & 3.75 & 657.83(64)  &     190 & 200 & 0.707 & 50     & 50         &     0.9943(26)          & 1.027(29)      & 1.027(37)       & 0.9933(37)      & 0.9904(45) \\
    C32P29    & & 32$\times$64 & 292.4(1.1) & 5.01  & 658.80(43)  &   196  & 200 & 0.700 & 60     & 60         &     0.9961(23)          & 1.126(13)      & 1.091(13)       & 1.0099(20)      & 1.0007(55) \\
    C24P23    & & 24$\times$64    & 229.5(3.0) & 2.93  &  645.67(99  &     169  &  20 & 1.000  & 50     & 50         &      0.9903(35)         & 0.880(24)      & 0.942(33)       & 0.9661(40)      & 0.9670(39) \\
    C32P23    & & 32$\times$64    & 228.0(1.2) & 3.91  &  643.93(45) &     333   & 50 &  0.700  & 60     & 60         &     0.9978(38)          & 1.058(29)      & 1.039(19)       & 0.9873(37)      & 0.9796(38) \\
    C48P23    & & 48$\times$96    & 225.6(0.9)  & 5.79 & 644.08(62) &    54 & 50 & 0.700 & 100    & 200        &     0.9950(17)          & 1.115(19)      & 1.107(19)       & 1.0008(36)      & 0.9881(59) \\
    C48P14     & & 48$\times$96     & 135.5(1.6)  & 3.47  & 706.55(39)  &      56    & 100 & 1.000   & 150    & 200        &     0.9961(55)          & 1.029(30)      & 1.050(34)       & 0.9777(69)      & 0.9743(81) \\
    C64P14     & & 64$\times$128     & 134.5(1.6)  & 4.63  &  706.55(39) &     38        & 20 & 1.000 & 150    & 50         &     0.9967(24)          & 1.099(25)      & 1.111(27)       & 0.9866(85)      & 0.9885(75) \\
    \hline                                                                       
    E32P29    & 0.08973 & 32$\times$64 & 286.7(1.8)      &   4.19     & 701.37(92)   &   99   & 20 & 1.000 & 50     & 50         &     0.9980(52)          & 1.046(30)      & 1.012(29)       & 1.0121(45)      & 1.0103(64) \\
    \hline                                                                                          
    F32P30    & \multirow{5}[2]{*}{0.07753} &32$\times$96 & 303.2(1.3)  & 3.56  &  675.98(97)  &    91  & 200 & 0.500 & 100    & 200        &     0.9964(10)          & 0.954(26)      & 1.004(40)       & 1.0054(28)      & 1.0059(37) \\
    F48P30   & & 48$\times$96 & 303.4(0.9)  & 5.72  &  674.76(58)   &    40     & 100 & 0.500 & 100    & 200        &     0.9987(40)          & 1.124(16)      & 1.091(17)       & 1.0276(24)      & 1.0217(35) \\
    F32P21    & & 32$\times$64 & 210.9(2.2) & 2.67  &  658.79(94)   &  369 & 50 & 0.500 & 50     & 50         &     0.9992(57)          & 0.904(34)      & 0.910(42)       & 0.9683(56)      & 0.9685(38) \\
    F48P21   & & 48$\times$96 & 207.2(1.1)  & 3.91  &  661.94(64)   &     150       & 20 & 0.500 & 100    & 60         &     0.9974(23)          & 1.084(26)      & 1.049(26)       & 0.9951(26)      & 0.9897(33) \\
    F64P13    & & 64$\times$128& 134.1(1.5)   & 3.37 &  681.48(59)  &   46  & 20 & 1.000 & 140    & 60         &    1.0010(26)           & 0.990(23)      & 0.997(29)       & 0.9927(64)      & 0.9982(39) \\
    \hline                                                                             
    G36P29    & 0.06887 &36$\times$108 & 297.2(0.9) & 3.73 &  693.05(46)   &    43    & 40 & 1.000 & 60     & 140        &    0.9995(38)           & 0.970(14)      & 1.005(17)       & 1.0142(61)      & 1.0120(61) \\
    \hline                                                                      
    H48P32    & 0.05199 &48$\times$144 & 317.2(0.9) &   4.00   & 691.88(65)  &   46  & 50 & 1.000 & 100    & 200        &    0.9981(09)           & 0.919(19)      & 0.982(36)       & 1.0243(47)      & 1.0260(53) \\
    \hline    
    \hline
    $g_{V/S/T}^{\rm physical}$   & & & &  &  &  &  & &  &            &    0.9998(12)           & 1.098(27)      &  1.106(31)      &1.0236(52)   &  1.0264(53)    \\
    \hline
    \hline
    \end{tabular}%
    }
  \label{tab:two-state}%
\end{table}%

{In this section, we present a comprehensive overview of our simulation setup as well as the methodology adopted for autocorrelation analysis. The Hybrid Monte Carlo (HMC) algorithm is employed for generating gauge field configurations. The integration step size $\tau$ and the interval between configurations for each ensemble are summarized in Table~\ref{tab:two-state}. To further enhance the efficiency of our simulations, we implement the Hasenbusch mass preconditioning technique~\cite{Hasenbusch:2001ne}, which effectively accelerates the HMC algorithm by improving the conditioning of the fermion matrix and thus increasing the acceptance rate.

The thermal equilibrium of our Markov chain are carefully checked. In Fig.~\ref{fig:HMC_history}, we display the evolution of the plaquette value $\langle P\rangle =u_0^4$ and topological charge along the HMC trajectories for the F48P21, C48P14, and H48P32 ensembles. As illustrated, both the plaquette and the topological charge fluctuate randomly around their respective mean values, demonstrating the absence of long-term drifts. This behavior provides a clear indication that the generated configurations are well equilibrated and that our ensembles have reached thermal equilibrium.

A notable advantage of the blending method lies in its ability to extract significantly more information from a single configuration by approximating the all-to-all propagator. Consequently, far fewer configurations are required compared to traditional techniques. However, when using fewer configurations, it becomes essential to properly account for auto-correlations present along the Markov chain trajectory. To quantitatively evaluate the impact of auto-correlation in this context, we perform a thorough statistical analysis, focusing on the variance of binned means as outlined in Ref.~\cite{Chang:2018uxx}.

Building upon the auto-correlation studies of basic observables such as $u_0$, $Q$, $m_{\pi}$, and $m_{\eta_c}$ in Ref.~\cite{CLQCD:2024yyn}, we conduct a dedicated binning analysis on our data. Specifically, we compute $R_{S,T}^{\rm mid}(t_f)$ using different bin sizes $n = \{1, 2, 3, 4\}$ for both the $c=0$ and $c=c_{\text{opt}}$ interpolators across three representative ensembles:
\begin{itemize}
    \item F48P21: $a=0.077\,\mathrm{fm}$, $m_\pi \approx 210\,\mathrm{MeV}$, $m_{\pi}L = 3.91$,
    \item F64P13: $a=0.077\,\mathrm{fm}$, $m_\pi \approx 130\,\mathrm{MeV}$, $m_{\pi}L = 3.37$,
    \item H48P32: $a=0.052\,\mathrm{fm}$, $m_\pi \approx 320\,\mathrm{MeV}$, $m_{\pi}L = 4.00$.
\end{itemize}
For each case, the mean value and standard deviation are estimated using a sufficiently large number of bootstrap resamples. In the absence of significant autocorrelations, the standard deviation should remain nearly constant as the bin size $n$ varies. The dependence of both the mean value and standard deviation on bin size $n$ is depicted in Fig.~\ref{fig:autocorr_all}. The results show that the standard deviation remains stable for all tested ensembles, indicating that autocorrelation effects are negligible in our data. As a result, the statistical errors reported in this study can be considered robust and unaffected by autocorrelation.}

\begin{figure}
    \centering
    \includegraphics[width=0.3\linewidth]{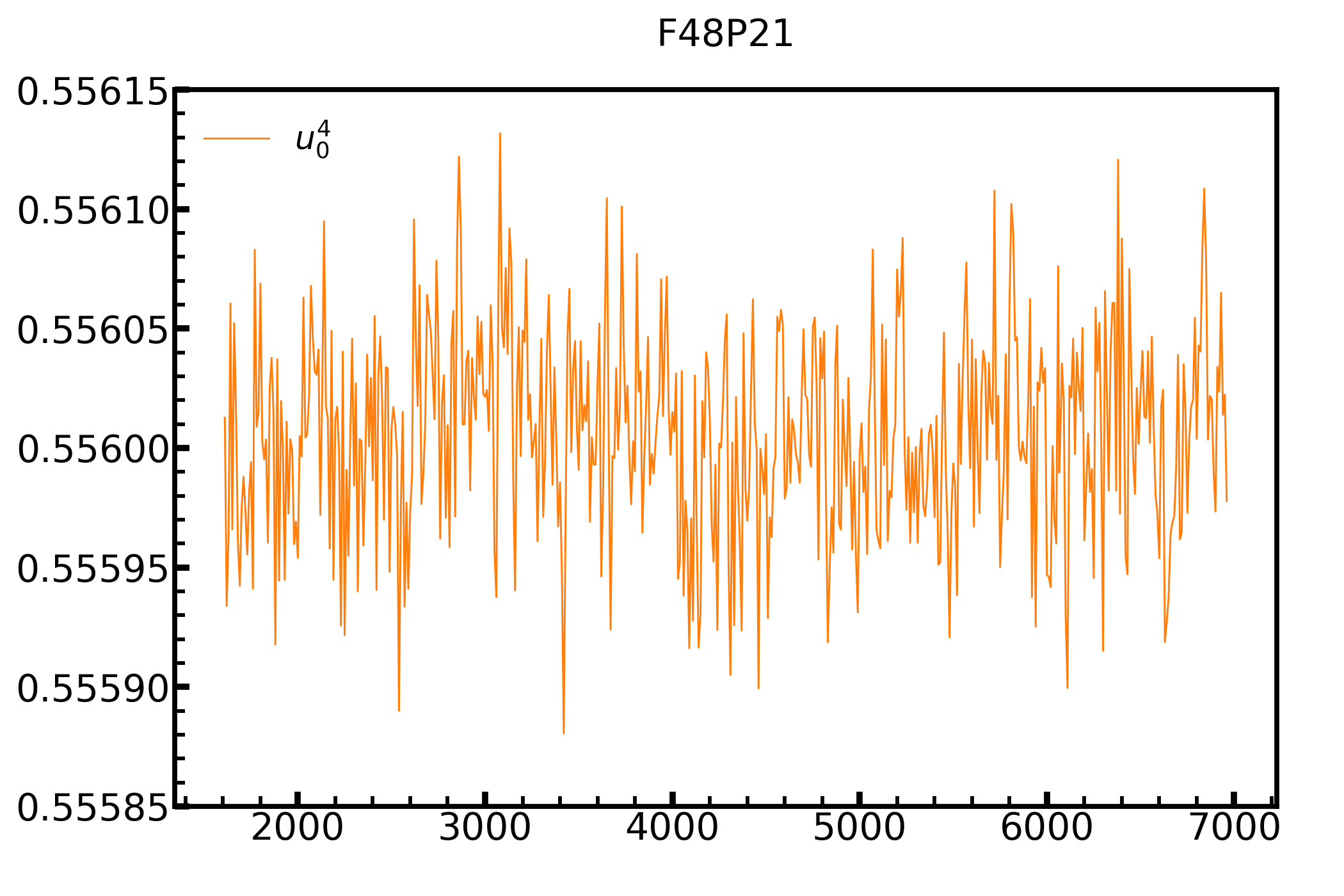}
    \includegraphics[width=0.3\linewidth]{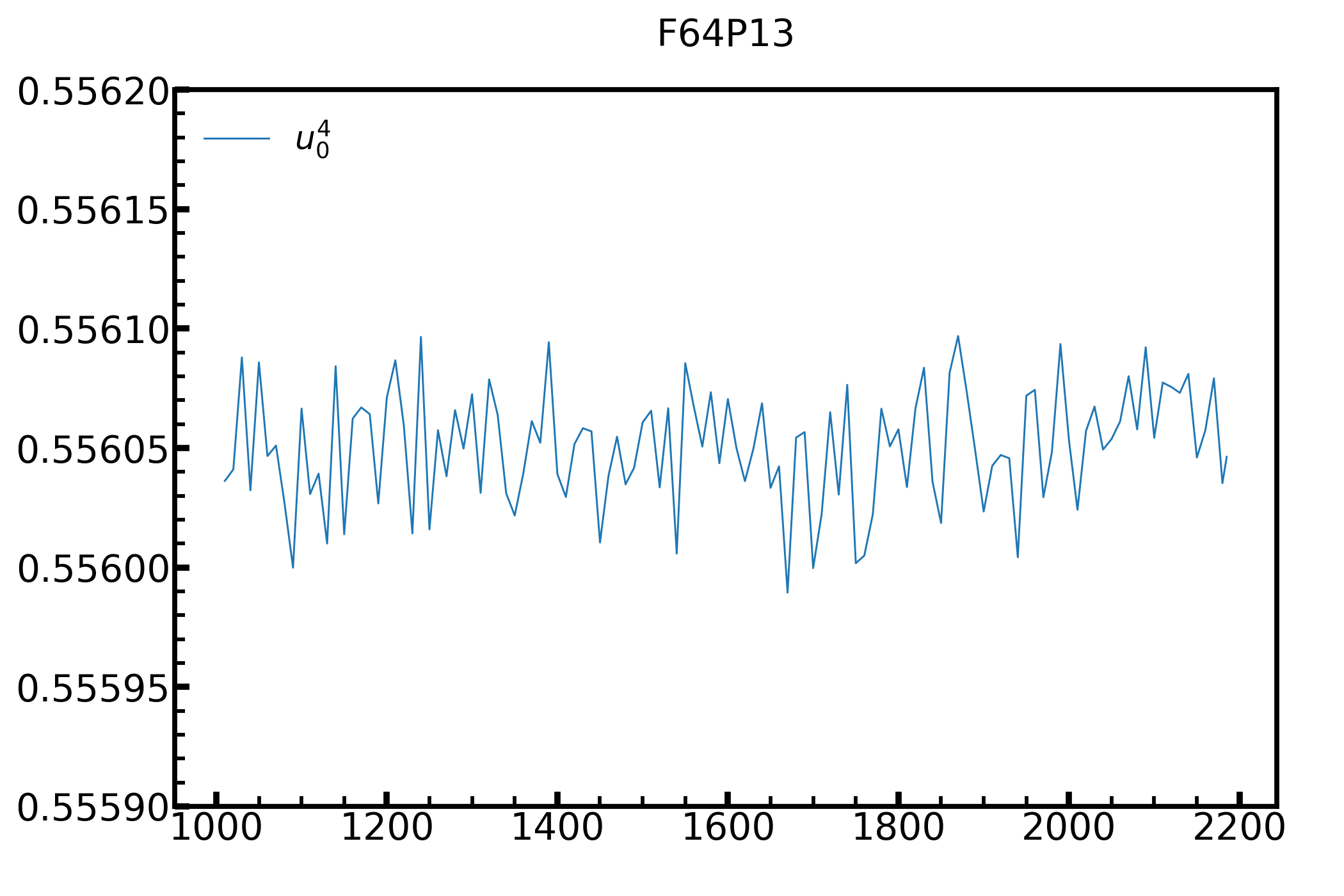}
    \includegraphics[width=0.3\linewidth]{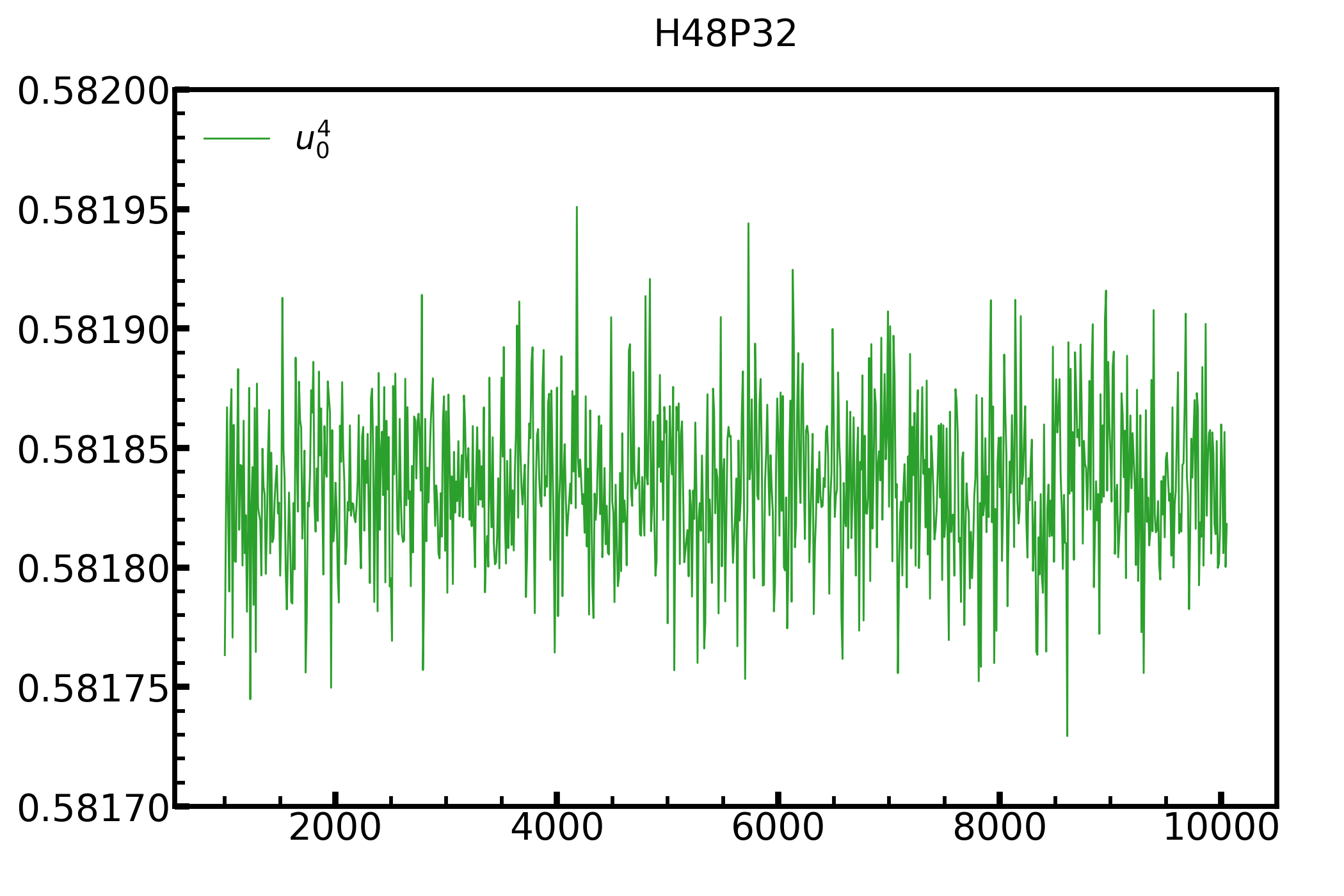}
    \includegraphics[width=0.3\linewidth]{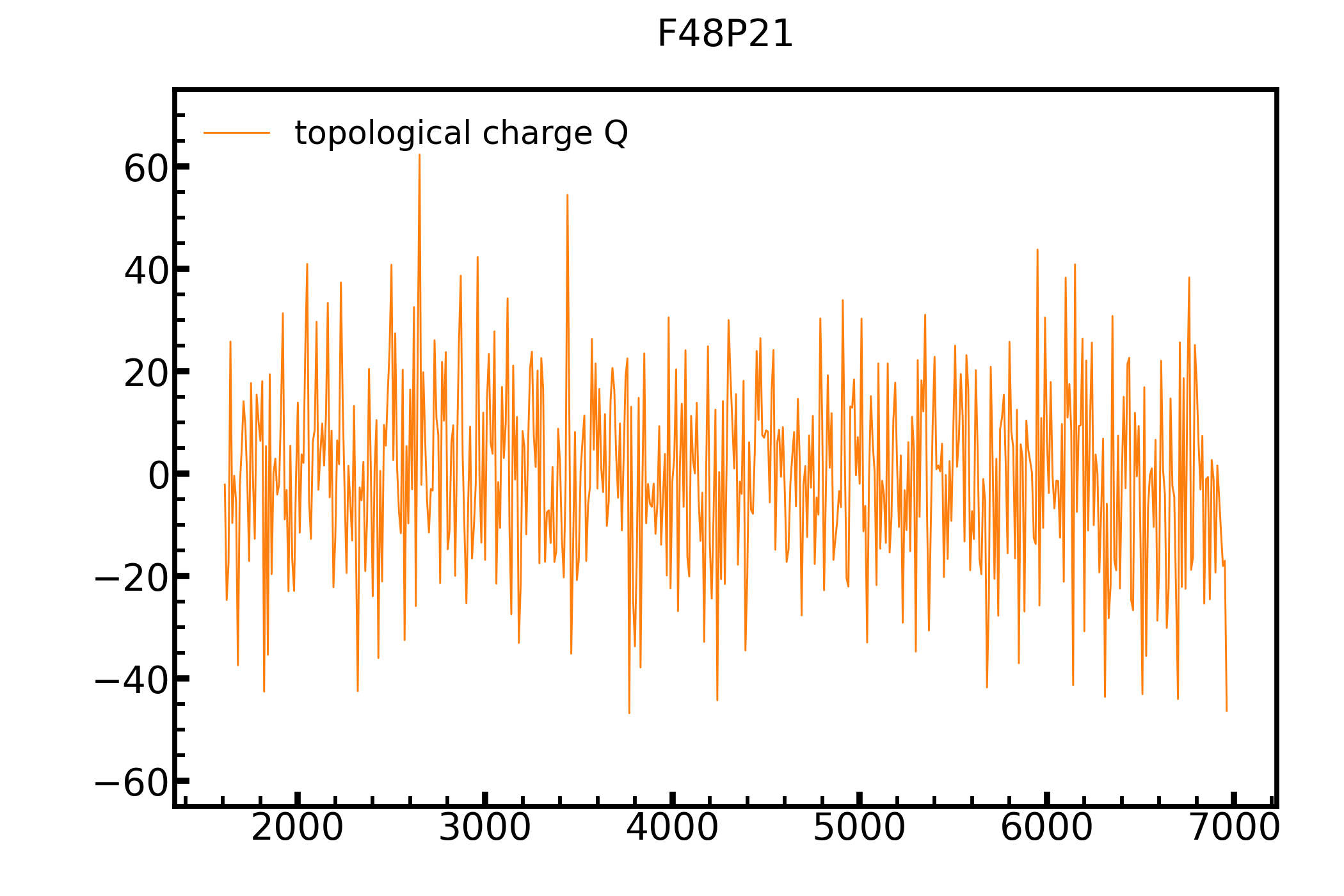}
    \includegraphics[width=0.3\linewidth]{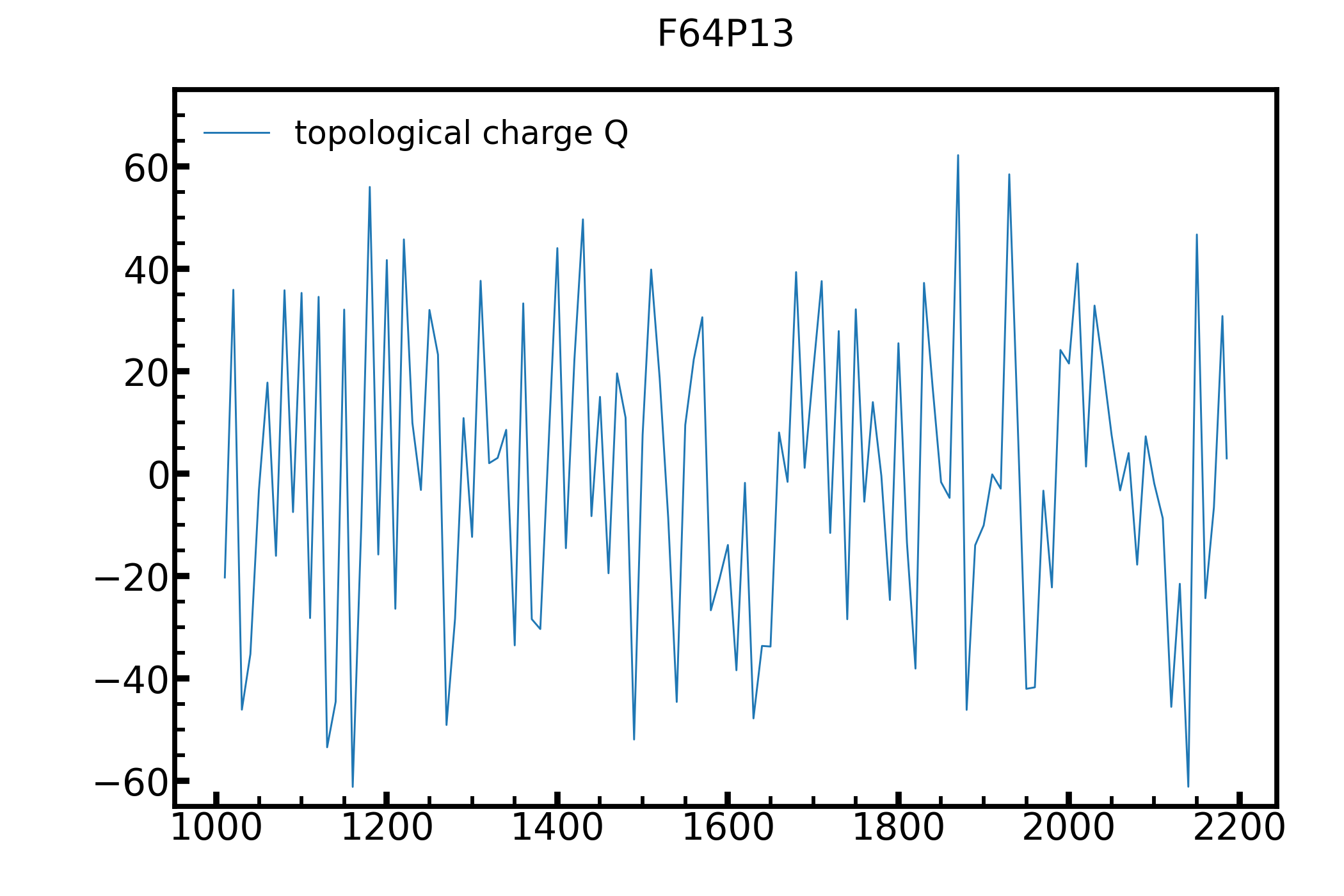}
    \includegraphics[width=0.3\linewidth]{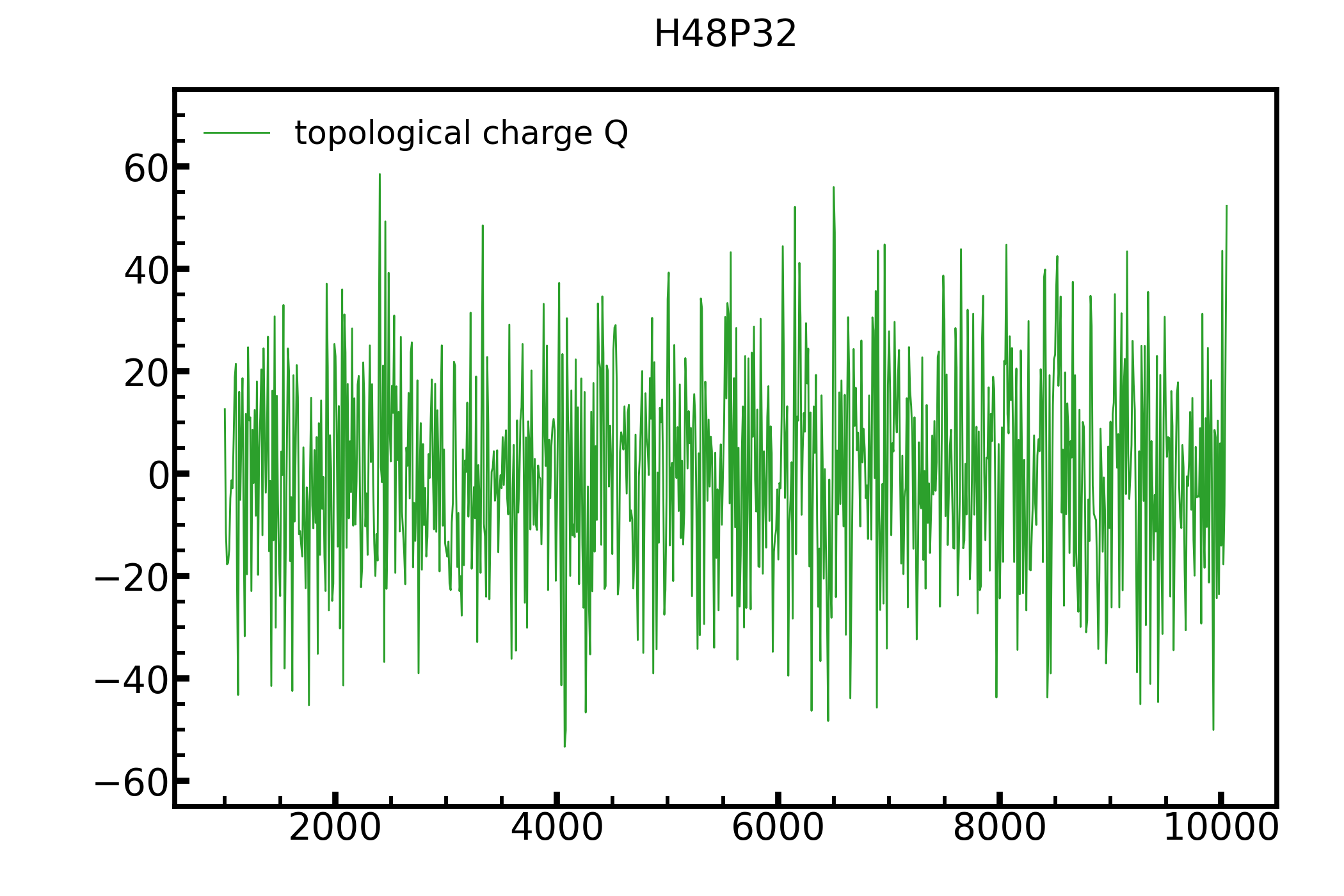}    
    \caption{{The evolution of the plaquette and topological charge along the HMC trajectory. present the plaquette(upon three figures) and topological charge(down three figures) as the function of the configuration number.}}
    \label{fig:HMC_history}
\end{figure}

\begin{figure}
    \centering
    \includegraphics[width=0.3\linewidth]{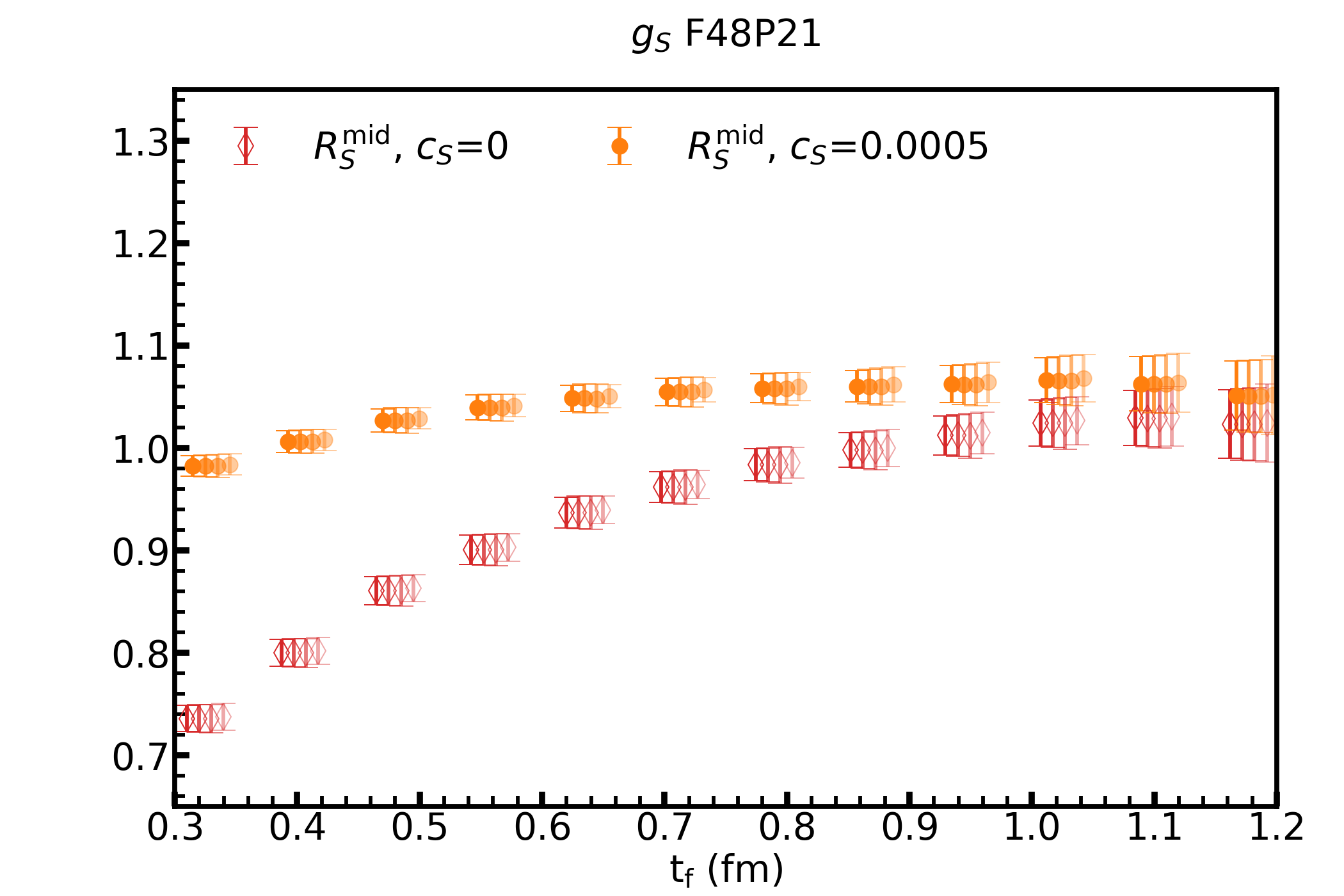}
    \includegraphics[width=0.3\linewidth]{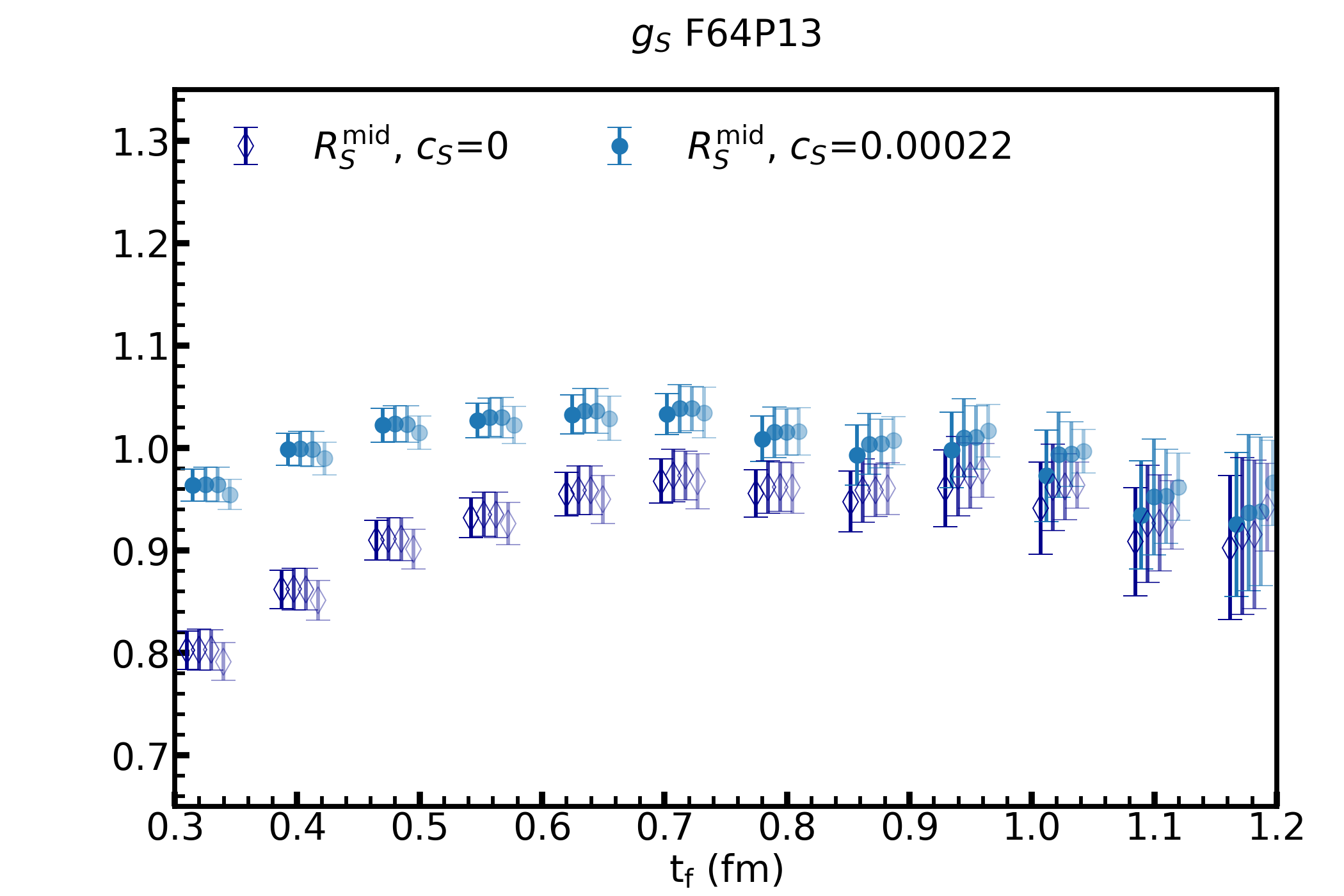}
    \includegraphics[width=0.3\linewidth]{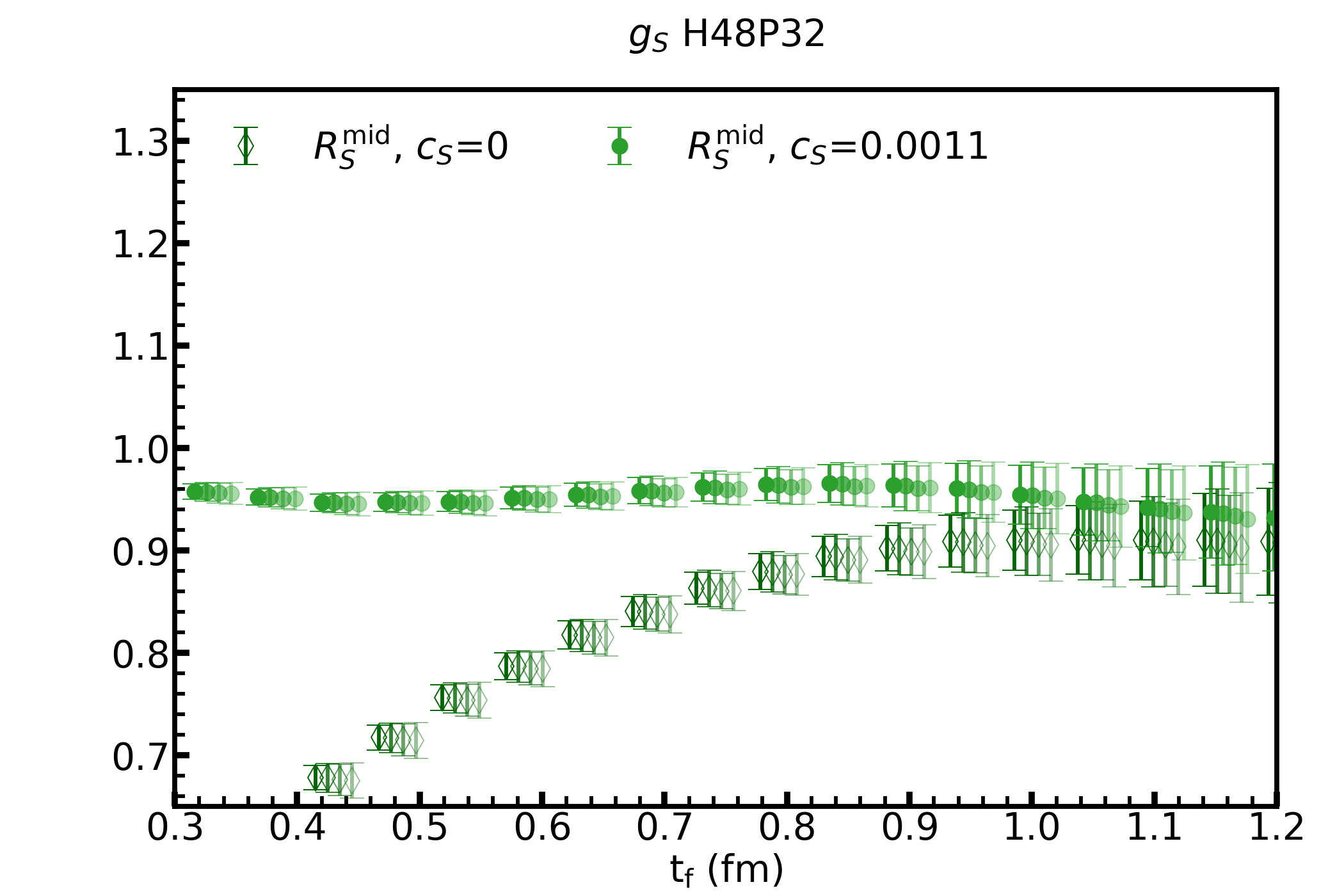}
    \includegraphics[width=0.3\linewidth]{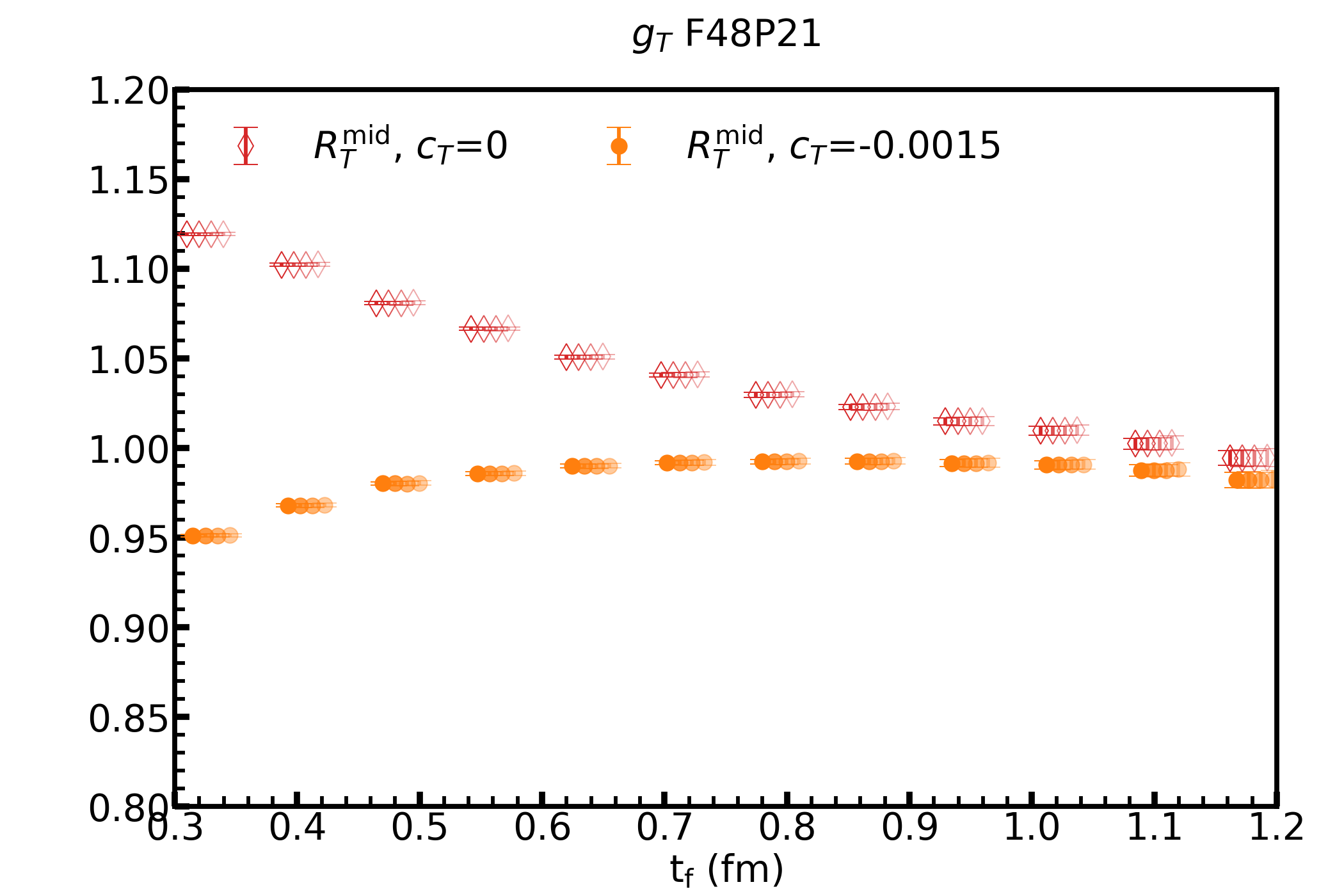}
    \includegraphics[width=0.3\linewidth]{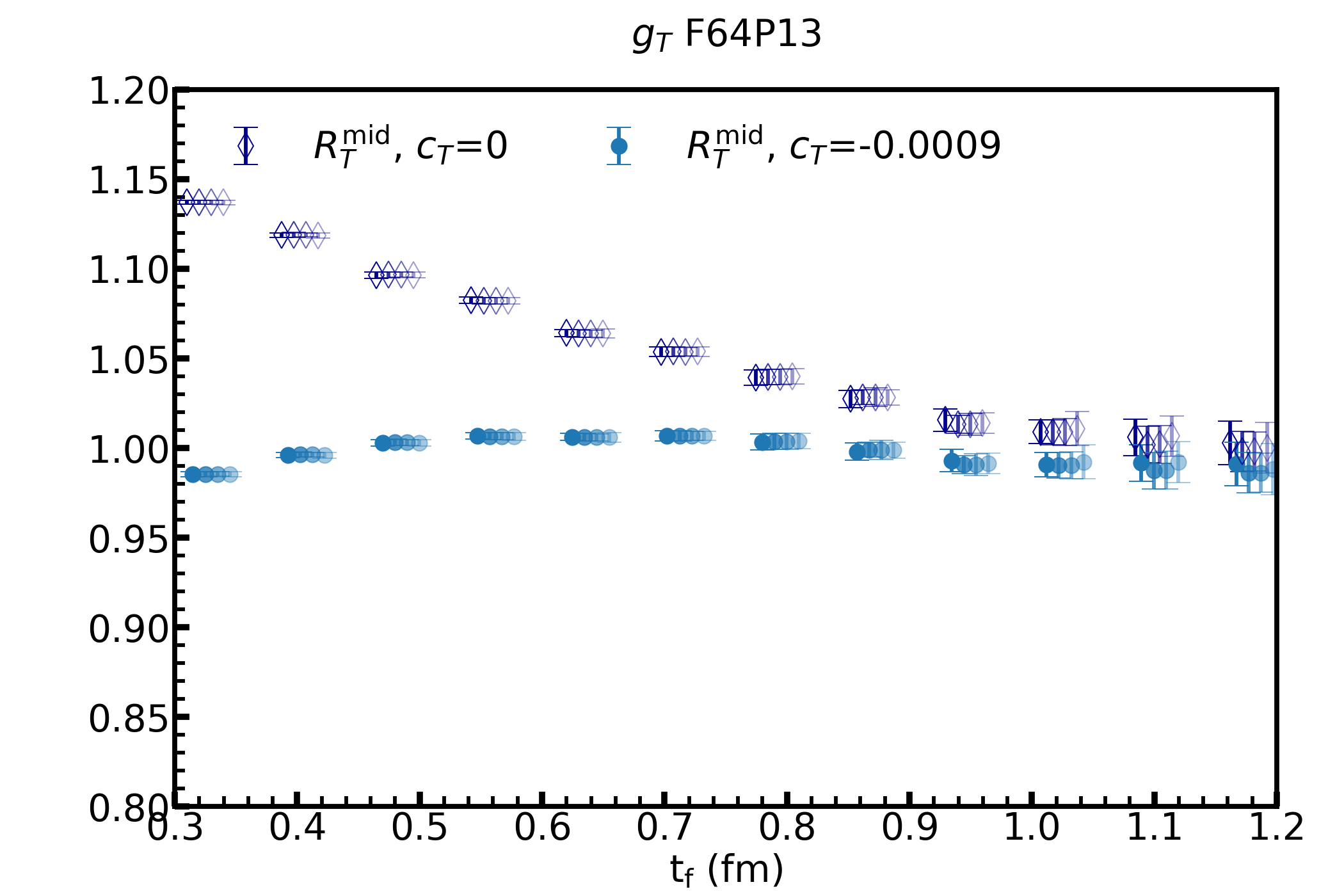}
    \includegraphics[width=0.3\linewidth]{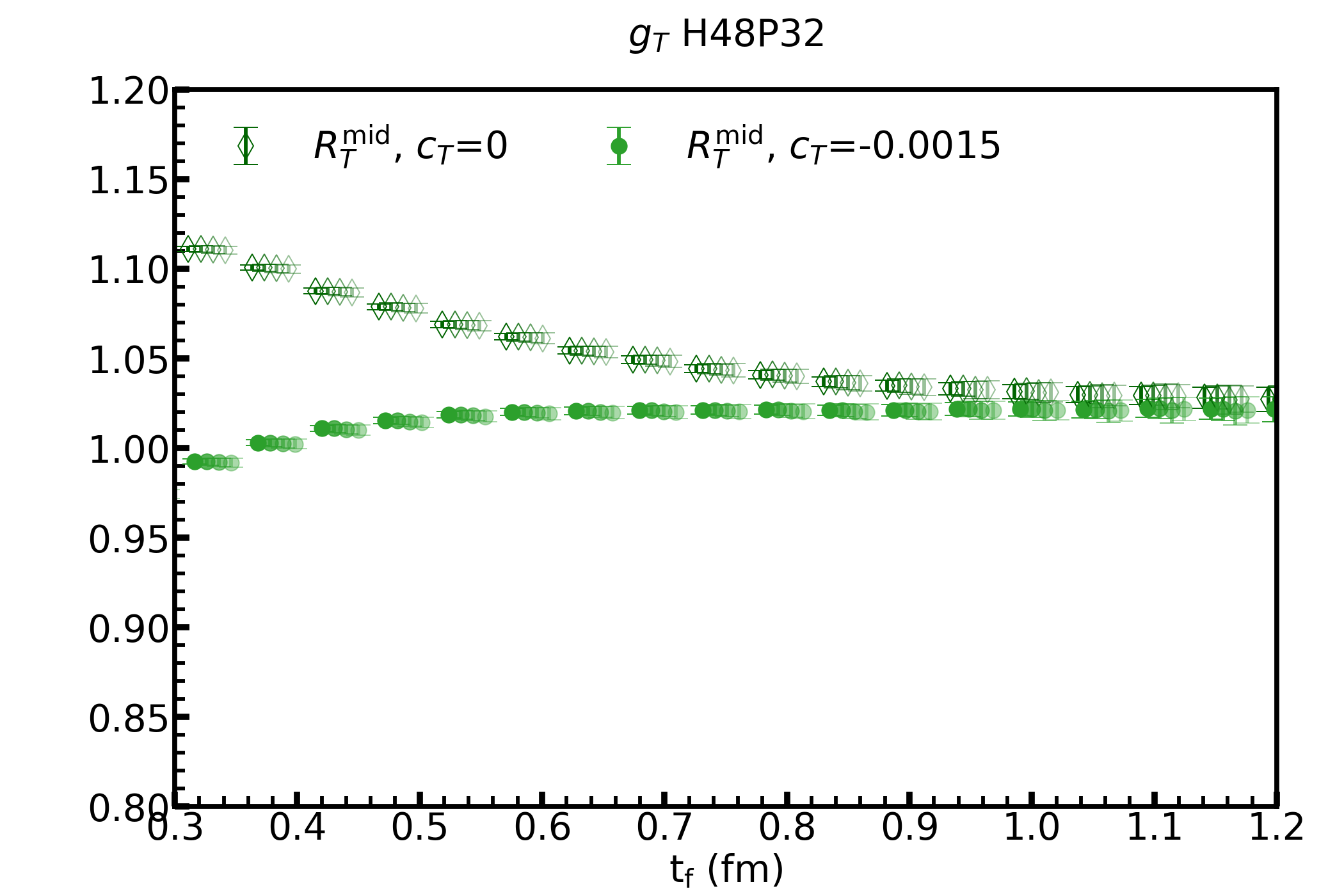}    
    \caption{{The autocorrelation analysis of $g_{\rm S,T}$ is shown by illustrating $R_{S,T}^{\rm min}(t_f)$ with $c=0$ and $c=c_{\text{opt}}$ with bin size is $\{1,2,3,4\}$(from left to right, color from dark to shallow) on the F48P21 (left panels), F64P13 (middle panels) and H48P32 (right panels) ensembles, for the tensor (upper panels) and scalar (lower panels) cases. The crosses represent the results for $R^{\rm min}$ with $c=0$, while the circles represent the results with $c=c_{\rm opt}$.}}
    \label{fig:autocorr_all}
\end{figure}

\clearpage

\subsection{The sensitivity analysis of fit window}

{In this Appendix, we present the fit windows used to extract $g_{\rm S,T}$ on each of the three ensembles, along with the corresponding $\chi^2/\mathrm{d.o.f.}$ values in Table~\ref{tab:fitting_info}. To systematically assess the stability of our results, we performed a dedicated sensitivity analysis by varying the start ($t_{\rm min}$) and end ($t_{\rm max}$) of the fit window by $\pm$1 lattice spacing relative to the values listed in Table~\ref{tab:fitting_info}. This test was conducted on three representative ensembles, F48P21, F64P13 and H48P32, with the results compiled in Table~\ref{tab:fit_window}.

Our analysis demonstrates the robustness of the extraction: on all the three ensembles, the fitted values of $g_S$ and $g_T$ remain consistent across all tested window variations. As expected, statistical uncertainties moderately increase (decrease) when $t_{\rm min}$ is increased ($t_{\rm max}$ is decreased), confirming that our chosen window represents a sound balance between statistical precision and systematic control.

Therefore, our chosen fit windows (documented in Table~\ref{tab:fitting_info}) were specifically selected to satisfy two key criteria: (1) ensure the stability and good constraint of the three-state fit parameters, and (2) achieve an optimal balance between statistical precision and systematic control over excited-states. This rigorous sensitivity analysis confirms that our final results are robust against reasonable variations in the fit window.}

\begin{figure}[!h]
    \centering
    \includegraphics[width=0.45\linewidth]{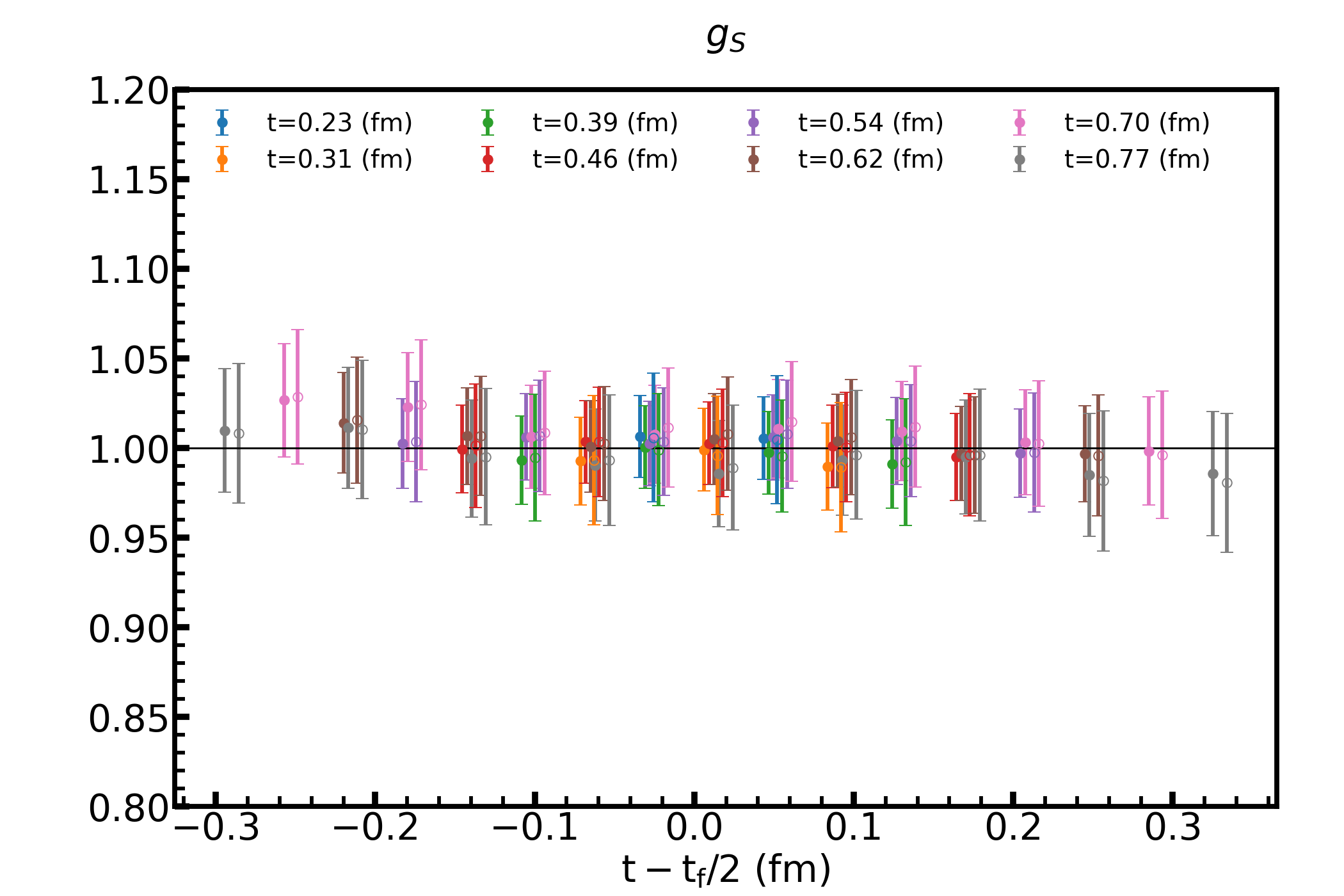}
    \includegraphics[width=0.45\linewidth]{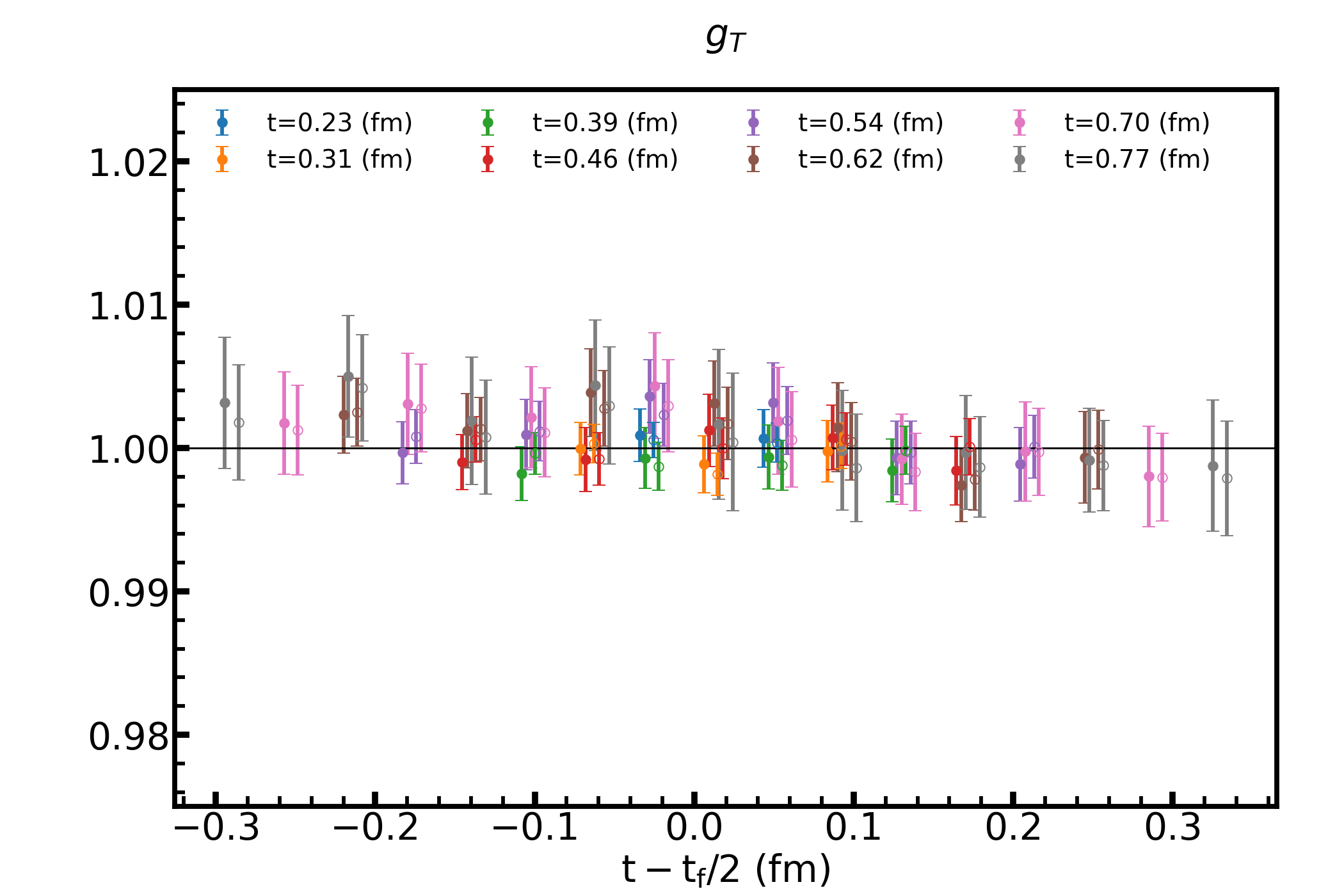}
    \caption{The ratio of the original data to the fit results for $R_S(t,t_f)$ and $R_T(t,t_f)$ as a function of $t - t_f/2$ on the F64P13 ensemble. The hollow and solid data points correspond to $c = 0$ and $c = c_{\rm opt}$, respectively.}
    \label{fig:residual}
\end{figure}

{Fig.~\ref{fig:residual} displays the normalized fit residuals—the original data with statistical uncertainty divided by the central value of the fit function, for both the two-point correlators and the three-point ratios \( R_S(t, t_f) \) and \( R_T(t, t_f) \) on the F64P13 ensemble, plotted as a function of the time separation.}

{The ratios are consistently scattered around unity across the entire fitted range and remain well within the uncertainty. This demonstrates visually that our multi-state fit model provides an excellent, unbiased description of the correlator data, with no discernible systematic trends.}

\begin{table}[htbp]
     \centering
    \caption{
    {The fitting information for $g_S$ and $g_T$ from the two-state and three-state fits.}
    }
    \resizebox{0.75\columnwidth}{!}{
    \begin{tabular}{c|c|cccc|cccc}
    \hline
    \hline
    \multirow{2}{*}{Ensemble} & \multirow{2}{*}{$n_{\text{state}}$} & \multicolumn{4}{c|}{$g_S$} & \multicolumn{4}{c}{$g_T$}\\
    \cline{3-10}            
    & & $t_{\text{min}}(a)$ & $t_{\text{max}}(a)$ & $g_S$ & $\chi^2$/d.o.f & $t_{\text{min}}(a)$  & $t_{\text{max}}(a)$ & $g_T$ & $\chi^2$/d.o.f \\
     \hline
    \multirow{2}{*}{C24P34}  & 2 & 2  & 8  & 1.036(18) &0.57 & 6  & 10  & 1.0182(58) &0.02\\
    \cline{2-10}
    & 3 & 2  & 8  & 1.038(23) &0.04 & 3  & 12  & 1.0167(67) &0.04\\
    \hline
    \multirow{2}{*}{C24P29}  & 2 & 3  & 10  & 1.027(29) &0.63 & 6  & 10  & 0.9933(37) &0.10\\
    \cline{2-10}
    & 3 & 3  & 12  & 1.027(37) &0.06 & 3  & 12  & 0.9904(45) &0.23\\
    \hline
    \multirow{2}{*}{C32P29}  & 2 & 3  & 10  & 1.126(13) &0.98 & 6  & 10  & 1.0099(20) &0.17\\
    \cline{2-10}
    & 3 & 3  & 10  & 1.091(13) &0.19 & 3  & 12  & 1.0007(55) &0.34\\
    \hline
    \multirow{2}{*}{C24P23}  & 2 & 2  & 8  & 0.880(24) &0.84 & 6  & 10  & 0.9661(40) &0.09\\
    \cline{2-10}
    & 3 & 3  & 8  & 0.942(33) &0.03 & 3  & 10  & 0.9670(39) &0.09\\
    \hline
    \multirow{2}{*}{C32P23}  & 2 & 6  & 12  & 1.058(29) &0.09 & 6  & 10  & 0.9873(37) &0.19\\
    \cline{2-10}
    & 3 & 3  & 12  & 1.039(19) &0.07 & 3  & 12  & 0.9796(38) &0.28\\
    \hline
    \multirow{2}{*}{C48P23}  & 2 & 3  & 12  & 1.115(19) &1.30 & 6  & 10  & 1.0008(36) &0.33\\
    \cline{2-10}
    & 3 & 3  & 12  & 1.107(19) &0.01 & 3  & 12  & 0.9881(59) &0.27\\
    \hline
    \multirow{2}{*}{C48P14}  & 2 & 3  & 8  & 1.029(30) &0.58 & 6  & 10  & 0.9777(69) &0.16\\
    \cline{2-10}
    & 3 & 3  & 8  & 1.050(34) &0.02 & 3  & 12  & 0.9743(81) &0.33\\
    \hline
    \multirow{2}{*}{C64P14}  & 2 & 4  & 10  & 1.099(25) &1.52 & 6  & 10  & 0.9866(85) &0.29\\
    \cline{2-10}
    & 3 & 4  & 10  & 1.111(27) &0.15 & 6  & 12  & 0.9885(75) &0.66\\
    \hline
    \multirow{2}{*}{E32P29}  & 2 & 4  & 14  & 1.046(30) &0.70 & 7  & 12  & 1.0121(45) &0.06\\
    \cline{2-10}
    & 3 & 4  & 14  & 1.012(29) &0.11 & 4  & 14  & 1.0103(64) &0.14\\
    \hline
    \multirow{2}{*}{F32P30}  & 2 & 6  & 16  & 0.954(26) &1.21 & 8  & 14  & 1.0054(28) &0.30\\
    \cline{2-10}
    & 3 & 6  & 16  & 1.004(40) &0.12 & 4  & 17  & 1.0059(37) &0.01\\
    \hline
    \multirow{2}{*}{F48P30}  & 2 & 6  & 16  & 1.124(16) &0.69 & 8  & 14  & 1.0276(24) &0.45\\
    \cline{2-10}
    & 3 & 6  & 16  & 1.091(17) &0.06 & 4  & 13  & 1.0217(35) &0.31\\
    \hline
    \multirow{2}{*}{F32P21}  & 2 & 3  & 11  & 0.904(34) &1.26 & 8  & 14  & 0.9683(56) &0.16\\
    \cline{2-10}
    & 3 & 3  & 11  & 0.910(42) &0.12 & 4  & 13  & 0.9685(38) &0.21\\
    \hline
    \multirow{2}{*}{F48P21}  & 2 & 4  & 16  & 1.084(26) &0.55 & 8  & 14  & 0.9951(26) &0.42\\
    \cline{2-10}
    & 3 & 4  & 16  & 1.049(26) &0.04 & 4  & 13  & 0.9897(33) &0.28\\
    \hline
    \multirow{2}{*}{F64P13}  & 2 & 6  & 11  & 0.990(23) &0.42 & 10  & 17  & 0.9927(64) &0.93\\
    \cline{2-10}
    & 3 & 3  & 12  & 0.997(29) &0.16 & 3  & 15  & 0.9982(39) &0.83\\
    \hline
    \multirow{2}{*}{G36P29}  & 2 & 5  & 12  & 0.972(32) &0.43 & 9  & 16  & 1.0142(61) &0.08\\
    \cline{2-10}
    & 3 & 5  & 12  & 0.956(35) &0.01 & 5  & 19  & 1.0120(61) &0.16\\
    \hline
    \multirow{2}{*}{H48P32}  & 2 & 6  & 18  & 0.919(19) &0.55 & 12  & 21  & 1.0243(47) &0.11\\
    \cline{2-10}
    & 3 & 6  & 18  & 0.982(36) &0.03 & 6  & 23  & 1.0260(53) &0.43\\
    \hline
    \hline
    \end{tabular}%
    }
    \label{tab:fitting_info}%
    \end{table}%
\clearpage
\begin{table}[htbp]
     \centering
    \caption{{Analysis of the sensitivity of $g_S$ and $g_T$ to the choice of fit window.}}
    \resizebox{0.75\columnwidth}{!}{
    \begin{tabular}{c|c|cccc|cccc}
    \hline
    \hline
    \multirow{2}{*}{Ensemble} & \multirow{2}{*}{$n_{\text{state}}$} & \multicolumn{4}{c|}{$g_S$} & \multicolumn{4}{c}{$g_T$}\\
    \cline{3-10}            
    & & $t_{\text{min}}(a)$ & $t_{\text{max}}(a)$ & $g_S$ & $\chi^2/d.o.f$ & $t_{\text{min}}(a)$  & $t_{\text{max}}(a)$ & $g_T$ & $\chi^2/d.o.f$ \\
    \hline
    \multirow{5}{*}{F48P21}& 3 & 4  & 16  & 1.049(26) &0.04 & 4  & 13  & 0.9897(33) &0.28\\
    \cline{2-10}
    & 3 & 3  & 16  & 1.048(24) &0.05 & 3  & 13  & 0.9900(33) &0.43\\
    \cline{2-10}
    & 3 & 5  & 16  & 1.050(26) &0.04 & 5  & 13  & 0.9890(35) &0.23\\
    \cline{2-10}
    & 3 & 4  & 15  & 1.051(23) &0.02 & 4  & 12  & 0.9913(31) &0.25\\
    \cline{2-10}
    & 3 & 4  & 17  & 1.047(27) &0.09 & 4  & 14  & 0.9876(39) &0.31\\
    \hline
    \multirow{5}{*}{F64P13}& 3 & 3  & 12  & 0.997(29) &0.16 & 3  & 15  & 0.9982(39) &0.83\\
    \cline{2-10}
    & 3 & 2  & 12  & 1.000(29) &0.18 & 2  & 15  & 1.0003(47) &1.15\\
    \cline{2-10}
    & 3 & 4  & 12  & 0.992(32) &0.15 & 4  & 15  & 0.9974(41) &0.79\\
    \cline{2-10}
    & 3 & 3  & 11  & 1.004(26) &0.12 & 3  & 14  & 0.9990(37) &0.82\\
    \cline{2-10}
    & 3 & 3  & 13  & 0.993(31) &0.16 & 3  & 16  & 0.9968(42) &0.91\\
    \hline
    \multirow{5}{*}{H48P32}& 3 & 6  & 18  & 0.982(36) &0.03 & 6  & 23  & 1.0260(53) &0.43\\
    \cline{2-10}
    & 3 & 5  & 18  & 0.982(42) &0.03 & 5  & 23  & 1.0258(48) &0.46\\
    \cline{2-10}
    & 3 & 7  & 18  & 0.981(36) &0.03 & 7  & 23  & 1.0261(69) &0.41\\
    \cline{2-10}
    & 3 & 6  & 17  & 0.986(33) &0.02 & 6  & 22  & 1.0258(50) &0.43\\
    \cline{2-10}
    & 3 & 6  & 19  & 0.973(38) &0.05 & 6  & 24  & 1.0261(57) &0.42\\
    \hline
    \hline
    \end{tabular}%
    }
    \label{tab:fit_window}%
    \end{table}%

\clearpage

\subsection{Systematic uncertainties}\label{syserr}

Using the current-involved interpolation field combination discussed previously, we conducted a systematic analysis employing both two-state and three-state fits. To satisfy the criterion $\chi^{2}$/d.o.f $\leq 1$, we varied the minimum sink time $t_f^{\rm min}$ and implemented a temporal cutoff $t_{\rm cut}$, restricting the current insertion to the interval $[t_{\rm cut}, t_f - t_{\rm cut}]$. We use smaller $t_f^{\rm min}$ and $t_{\rm cut}$ in the three-state fits than in the corresponding two-state fit, within the constraint of $\chi^{2}$/d.o.f $\leq 1$. The extracted charges $g_{S,T}$ are summarized in Table~\ref{tab:two-state}, with the values of $N_{\rm st}$ used to generate the general-purpose blending data. For the specific case of $g_{S,T}$, we find that $N_{\rm st}=30$ is sufficient to saturate the statistical uncertainty. The three-state fit results are selected as our primary values, and the systematic uncertainty associated with excited-state contamination is estimated from their difference with the two-state fit results.

The statistical uncertainties of the renormalization constants $Z_{S,T}$ are independent across ensembles and are combined in quadrature with the uncertainties of the bare nucleon matrix elements. In contrast, the systematic uncertainties $\delta_{\rm sys}Z_{S,T}$—arising from the truncation of the perturbative matching series, the value of the strong coupling, and scale evolution—are fully correlated. To account for this, we perform the global fit again using the shifted renormalization constant $Z_{S,T}+\delta_{\rm sys}Z_{S,T}$, and take the difference from the central fit as the systematic uncertainty attributable to renormalization.

For the continuum, chiral and infinite-volume extrapolations, we use the Akaike information criterion (AIC) to estimate the systematic error from different extrapolation models, with the AIC weight we use here are~\cite{Borsanyi:2020mff}:
\begin{equation}
    \omega_i = \frac{\exp \left[-\frac{1}{2}(\chi^2_i + 2n_{i,\text{par}} - n_{i,\text{data}}) \right] }{\sum_{j}\exp \left[-\frac{1}{2}(\chi^2_j + 2n_{j,\text{par}} - n_{j,\text{data}})\right]},
    \label{Eq.AIC}
\end{equation}
where the $\chi^2_i$,  $n_{i,\text{par}}$ and $n_{i,\text{data}}$ refer to the $\chi^2$, the number of fit parameters and the number of data points for the i-th fit model. The cumulative distribution function(CDF) for the probability  distribution of the result is defined as:
\begin{equation}
    P(y;\lambda) = \int_{-\infty}^{y} {\rm d} x \sum_i \omega_i N_i(x,m_i,\sigma \sqrt{\lambda}), 
\end{equation}
where $N_i(x,m_i,\sigma \sqrt{\lambda})$ is a Gaussian distribution with mean $m_i$ and standard deviation $\sigma \sqrt{\lambda}$, and $\lambda$ is a parameter to control the width of the distribution for extraction systematic error.
The median of the CDF is our choice for the central value of $y$ and its total error is given by the interval between the 16\% and 84\% percentiles of the CDF.
\begin{equation}
    \sigma^2_{\rm tot}  =  \left[ \frac{1}{2} (y_{84} - y_{16}) \right]^2   
     ~~~~~~~~     {\rm with} ~~~~~~~~  P(y_{16},1) = 0.16, ~~~~~~~~ P(y_{84},1) = 0.84.
\end{equation}
Assuming the systematic error and statistical error are independent, the total error is given by:
\begin{equation}
    \sigma_{\rm tot}^2 =  \lambda \sigma^2_{\rm stat} + \sigma^2_{\rm sys}.
\end{equation}
Then $\sigma^2_{\rm sys}$ can be obtained by repeating the analysis with different $\lambda$.

\begin{table}[htbp]
  \centering
  \caption{The parameters determined by the default joint fit ansatz Eq.~(\ref{eq:global_fit_sm}).}
\resizebox{0.55\columnwidth}{!}{
    \begin{tabular}{c|ccccccc}
    \hline
    \hline
          & $g_{S/T}^{\text{QCD}}$ & $c_{l}^{(2)} (\text{GeV}^{-2})$ &  $c_{l}^{(3)} (\text{GeV}^{-3})$  & $c_{a}^{(2)} (\text{fm}^{-2})$  & $c_V$  & $\chi^2$/d.o.f\\
    \hline
    $g_S$   & 1.106(31)     & -0.6(1.5)        &     0.1(3.9)           & 3.4(2.2)       & -3.19(41) & 0.47 \\
    \hline
    $g_T$   & 1.0264(53) &  -1.53(39)     & 4.8(1.1)        &     -2.71(45)        &  -0.469(66) & 0.49 \\
    \hline
    \hline
    \end{tabular}%
    }
  \label{tab:params}%
\end{table}%

The fit parameters and $\chi^2$/d.o.f. of the default joint fit ansatz,
\begin{align}
\label{eq:global_fit_sm}
g_X (a,m_\pi,L) = & g_{X}^{\rm QCD} \left(1 + \sum_{i=2,3}c_l^{(i)}(m_\pi^i-m_{\pi,\rm phy}^i)) \right) \nonumber \\
& (1+c_Ve^{-m_\pi L}) +c_{a} a^2,
\end{align}
are collected in Table~\ref{tab:params} for both $g_S$ and $g_T$.

\subsubsection{Continuum extrapolation}

{For clover fermions, a general $O(a)$-improved operator takes the form $O^{\rm imp} = (1 + c_0 m_q a) O + c_1 a O^{\rm D}$, where $O^{\rm D}$ is a higher-dimensional operator containing a derivative. Perturbative calculations~\cite{Capitani:2000xi} show that, after accounting for cancellations between quark-field and operator corrections, $c_0 = 1 + \mathcal{O}(\alpha_s)$ and $c_1 = \mathcal{O}(\alpha_s)$. In our analysis, the $O(a)$ effects (specifically the $O(m_q a)$ terms) for the scalar and tensor currents are removed by the renormalization procedure. This is achieved by combining the vector charge renormalization constant $Z_V$ (determined from a hadron matrix element at the unitary light quark mass) with the ratios $Z_{S/T}/Z_V$ taken in the chiral limit.

Consequently, the leading residual discretization errors are of order $O(a \alpha_s)$. As demonstrated even for the charm quark case in Ref.~\cite{CLQCD:2024yyn}, these effects are numerically small. Thus} we include the following models in the AIC for continuum extrapolation:

\begin{itemize}
    \item Simple $a^2$ form for the $\mathcal{O}(a)$-improved clover fermion action, $c_a^{(2)}a^2$;
    \item $a^2+a^3$ form for higher order correction, $c_a^{(2)}a^2+c_a^{(3)}a^3$;
    \item $a^2+a^3+a^4$ form for even higher order correction, $c_a^{(2)}a^2+c_a^{(3)}a^3+c_a^{(4)}a^4$;
    \item Simple $a^2$ form with the data at $a<0.1$ fm only;
    \item $a^2$ form with additional $a\log(u_0)$ term for the residual $a\alpha_s$ effect,  $c_a^{(1)} a\log(u_0) + c_a^{(2)}a^2$;
    \item $a\log(u_0)+a^2+a^3$ form for higher order correction, $c_a^{(1)} a\log (u_0) + c_a^{(2)}a^2 + c_a^{(3)}a^3$
    \item Most conservative $a\log(u_0)+a^2+a^3+a^4$ form, $c_a^{(1)} a\log (u_0) + c_a^{(2)}a^2 + c_a^{(3)}a^3 + c_a^{(4)}a^4$
\end{itemize}

\begin{figure}
    \centering
    \includegraphics[width=0.45\linewidth]{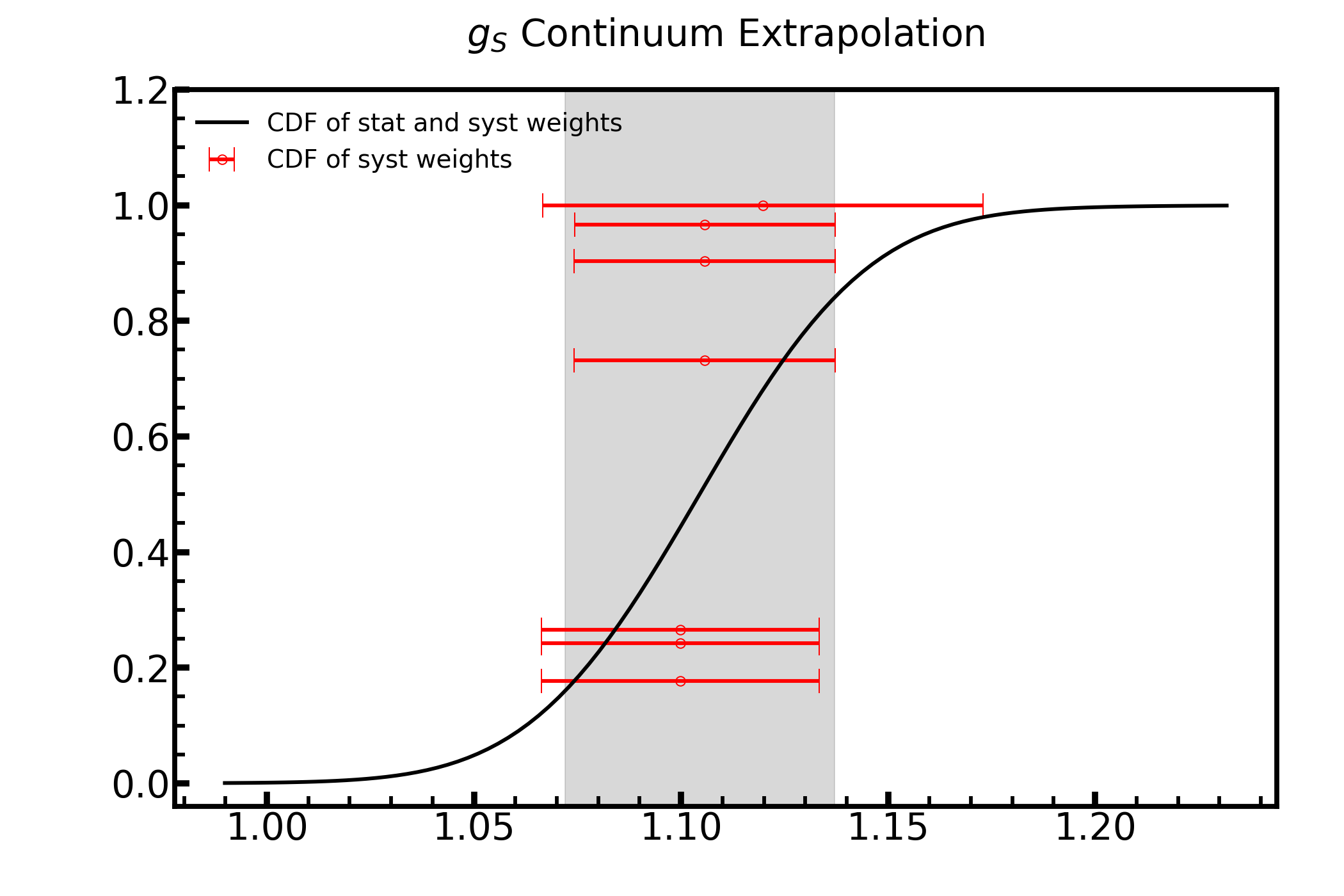}
     \includegraphics[width=0.45\linewidth]{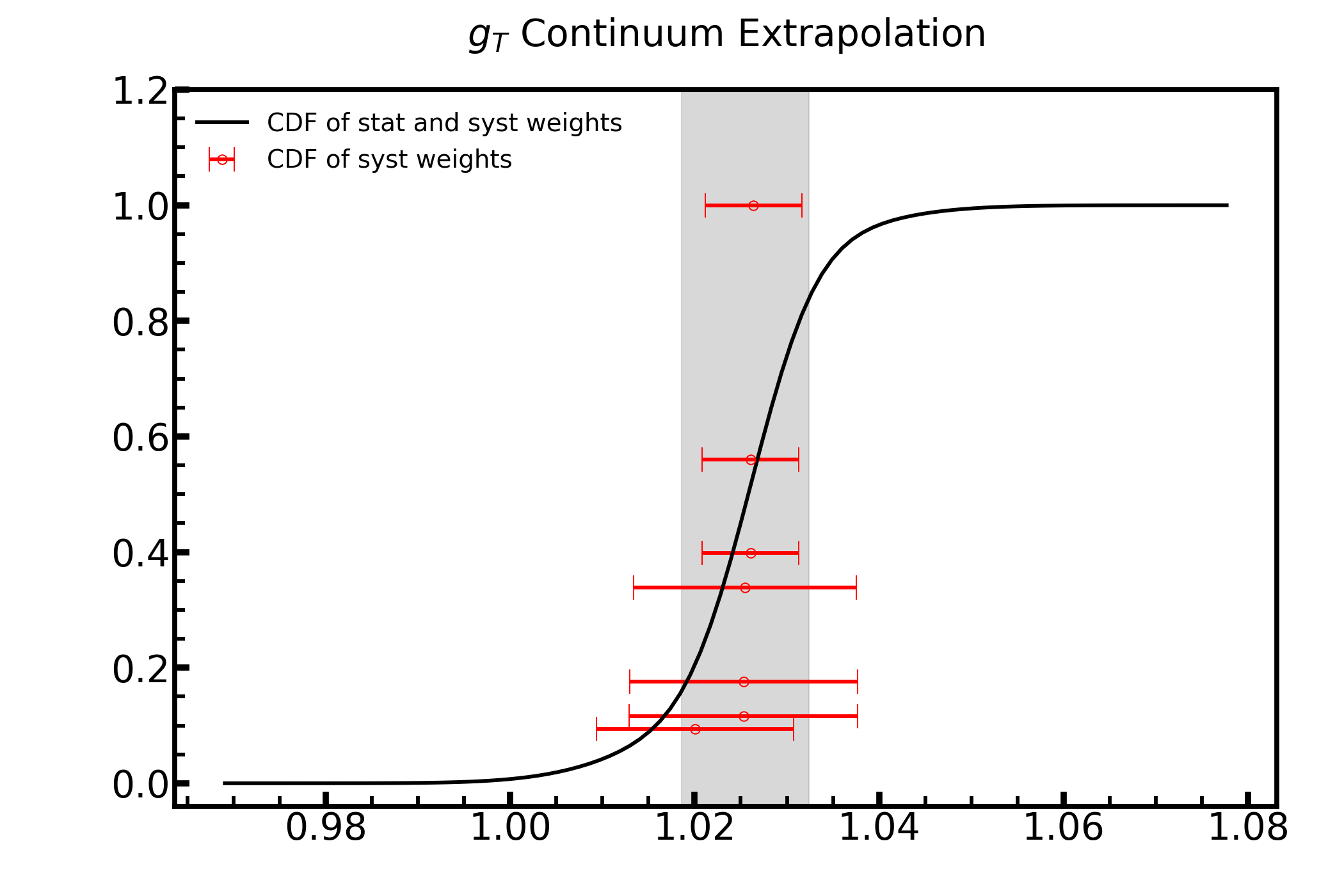}
    \caption{
    The CDF of different fit models of the continuum extrapolation of $g_{S}$ (left panel) and $g_T$ (right panel). 
    }
    \label{fig:continue_AIC}
\end{figure}

As shown in Fig.~\ref{fig:continue_AIC}, the extracted $g_{S,T}$ using different fit models are consistent with each other very well. 

\begin{figure}
    \centering
    \includegraphics[width=0.45\linewidth]{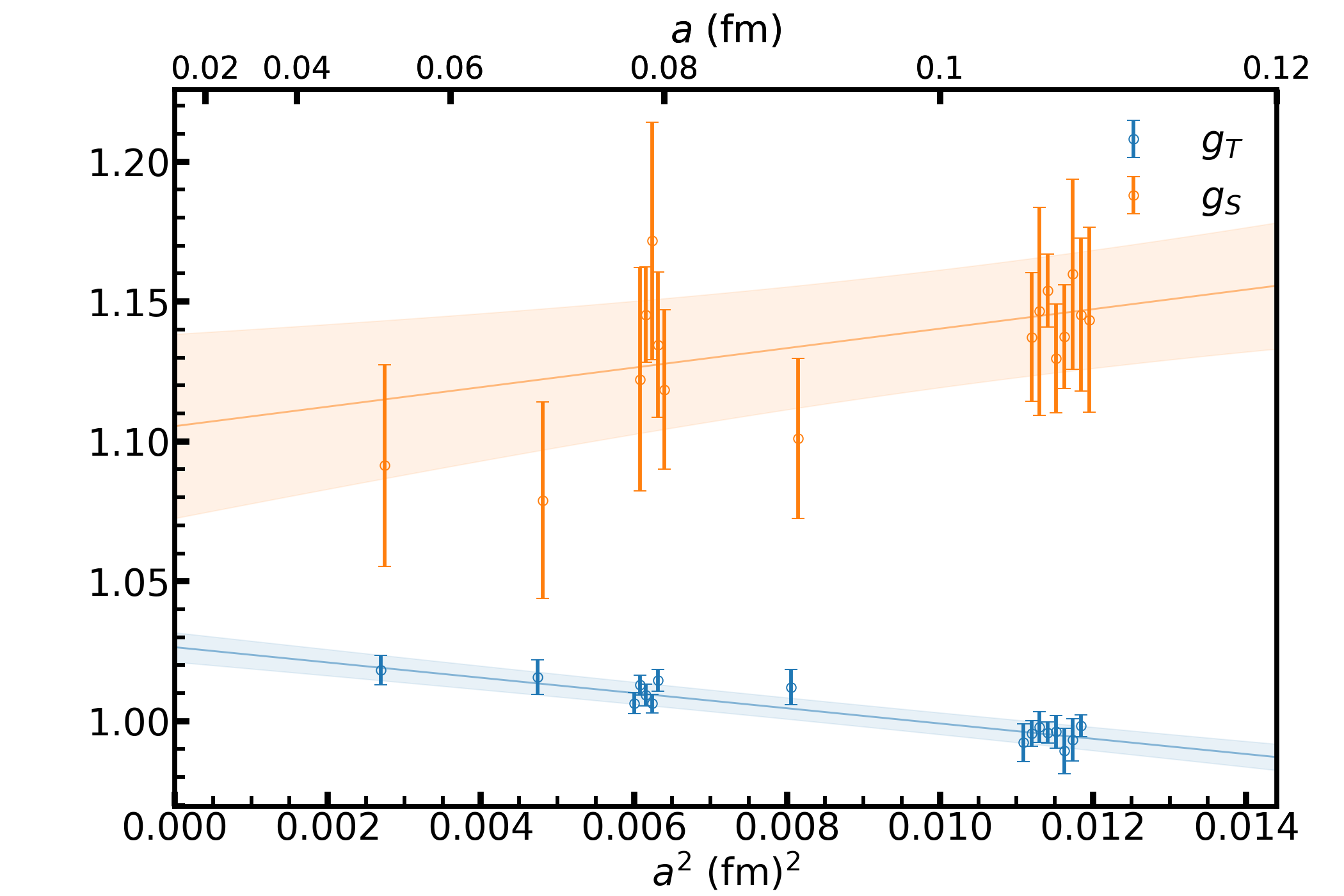}
    \caption{
        The lattice spacing dependence of the \( g_S \) and \(g_T \) with data points corrected to physical pion mass and infinite volume limits {using the default $a^2$ ansatz}. The data points at the same lattice spacing but different pion mass and volume are horizontally shifted to improve readability.
    }
    \label{fig:a}
\end{figure}

Fig.~\ref{fig:a} presents the continuum extrapolation for $g_T$ (blue dots and band) and $g_S$ (orange dots and band), using the default $a^2$ ansatz. The data shown have been corrected for finite-volume effects and unphysical pion masses using the parameters determined from the global fit. As shown in the figure, results at the five lattice spacings display a clear linear $a^2$ dependence with no evident $a^4$ corrections. The slopes for $g_S$ and $g_T$, however, exhibit opposite signs. {Fig.~\ref{fig:a_123} displays the continuum extrapolation using six additional models, with each subfigure containing two fit lines corresponding to two distinct models. As shown in the figures, higher-order terms ($a^3$ or $a^4$) do not significantly affect the extrapolated value. Including the $a\alpha_s$ term or restricting the data to finer lattice spacings increases the uncertainty considerably, while the central value remains largely unchanged. }

\begin{figure}
    \centering
    \includegraphics[width=0.30\linewidth]{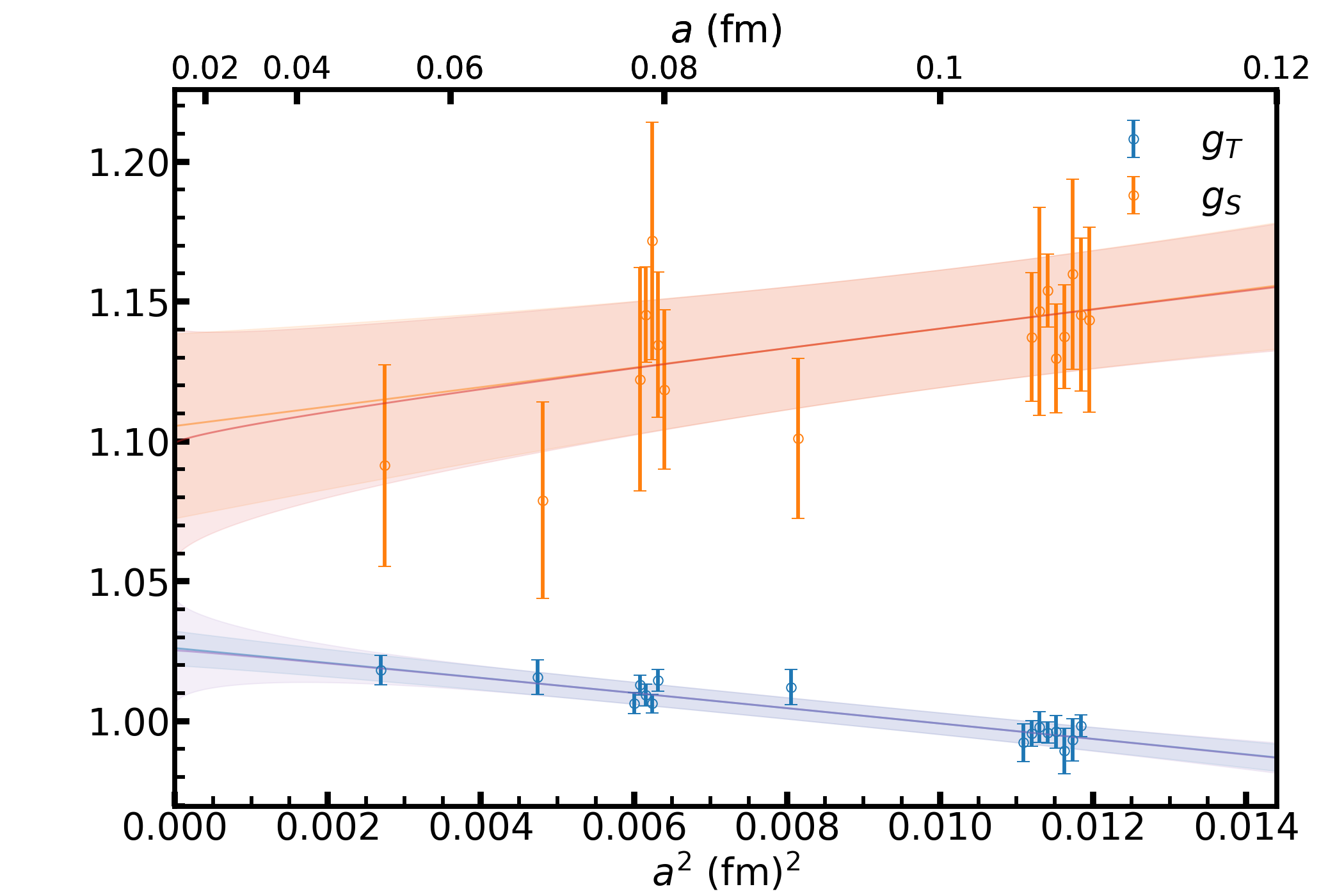}
    \includegraphics[width=0.30\linewidth]{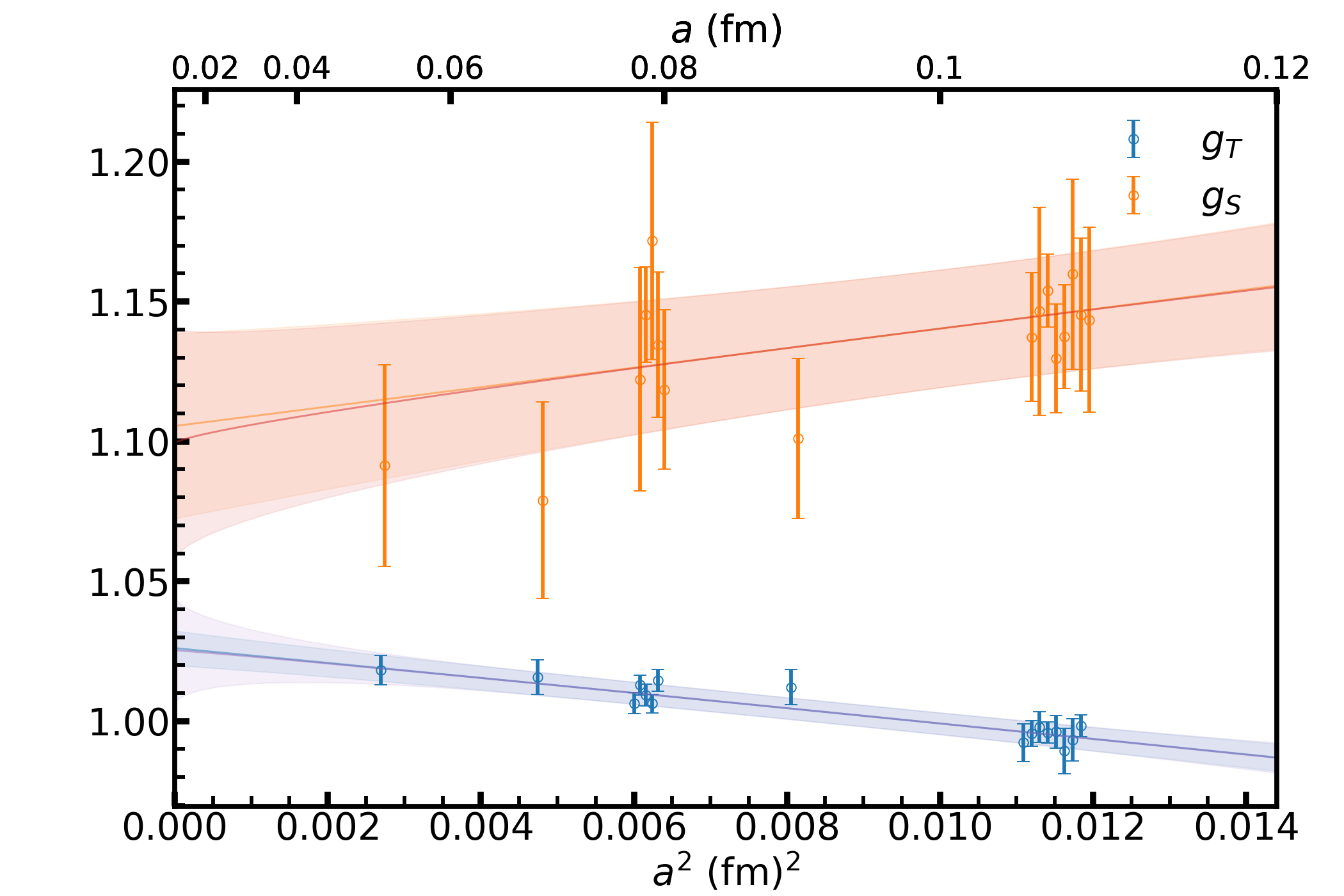}
    \includegraphics[width=0.30\linewidth]{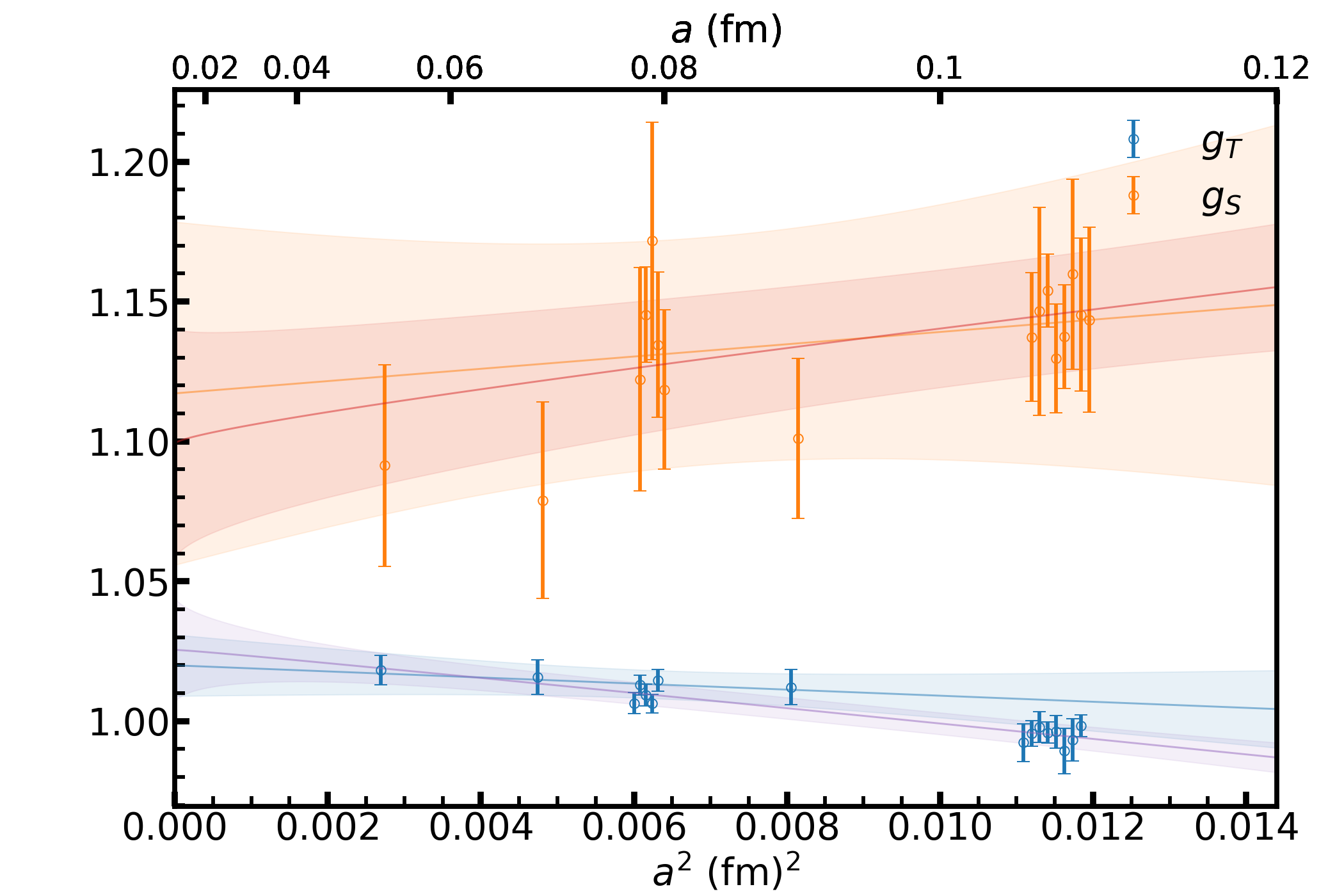}
    \caption{
        {The lattice spacing dependence of the \( g_S \) and \(g_T \) with data points corrected to physical pion mass and infinite volume limits. The data points at the same lattice spacing but different pion mass and volume are horizontally shifted to improve readability. For the three figures, the left panel presents the extrapolation models $a^2+a^3$ and $a\log(u_0)+a^2+a^3$; the middle panel shows the extrapolation models $a^2+a^3+a^4$ and $a\log(u_0)+a^2+a^3+a^4$; and the right panel displays the extrapolation model $a^2$ using only ensembles with $a<0.1$ fm, along with $c_a^{(1)} a\log(u_0) + c_a^{(2)}a^2$. The darker fit line corresponds to the model that includes the $a\log(u_0)$ term.}
    }
    \label{fig:a_123}
\end{figure}

\subsubsection{Chiral extrapolation}

For continuum extrapolation, we include the following 3 models in the AIC:

\begin{itemize}
    \item Default model with linear light quark mass $m_q\propto m_{\pi}^2$ dependence and also non-linear correction for heavier pion mass as suggested by $\chi$PT, $c_l^{(2)}m_\pi^2 + c_l^{(3)}m_\pi^3$;
    \item Simplied model with the linear term only, $ac_l^{(2)}m_\pi^2$;
    \item Most conservative one with even higher order term, $c_l^{(2)}m_\pi^2 + c_l^{(3)}m_\pi^3 + c_l^{(4)}m_\pi^4$
\end{itemize}

As shown in Fig.~\ref{fig:chiral_AIC}, the extracted $g_{S,T}$ using different fit models are consistent with each other very well, since our $g_{S,T}$ at the physical pion mass have been quite precise. Thus, different fit models are found to produce nearly identical results at the physical pion mass but can exhibit pronounced differences at heavier pion masses.

\begin{figure}
    \centering
    \includegraphics[width=0.45\linewidth]{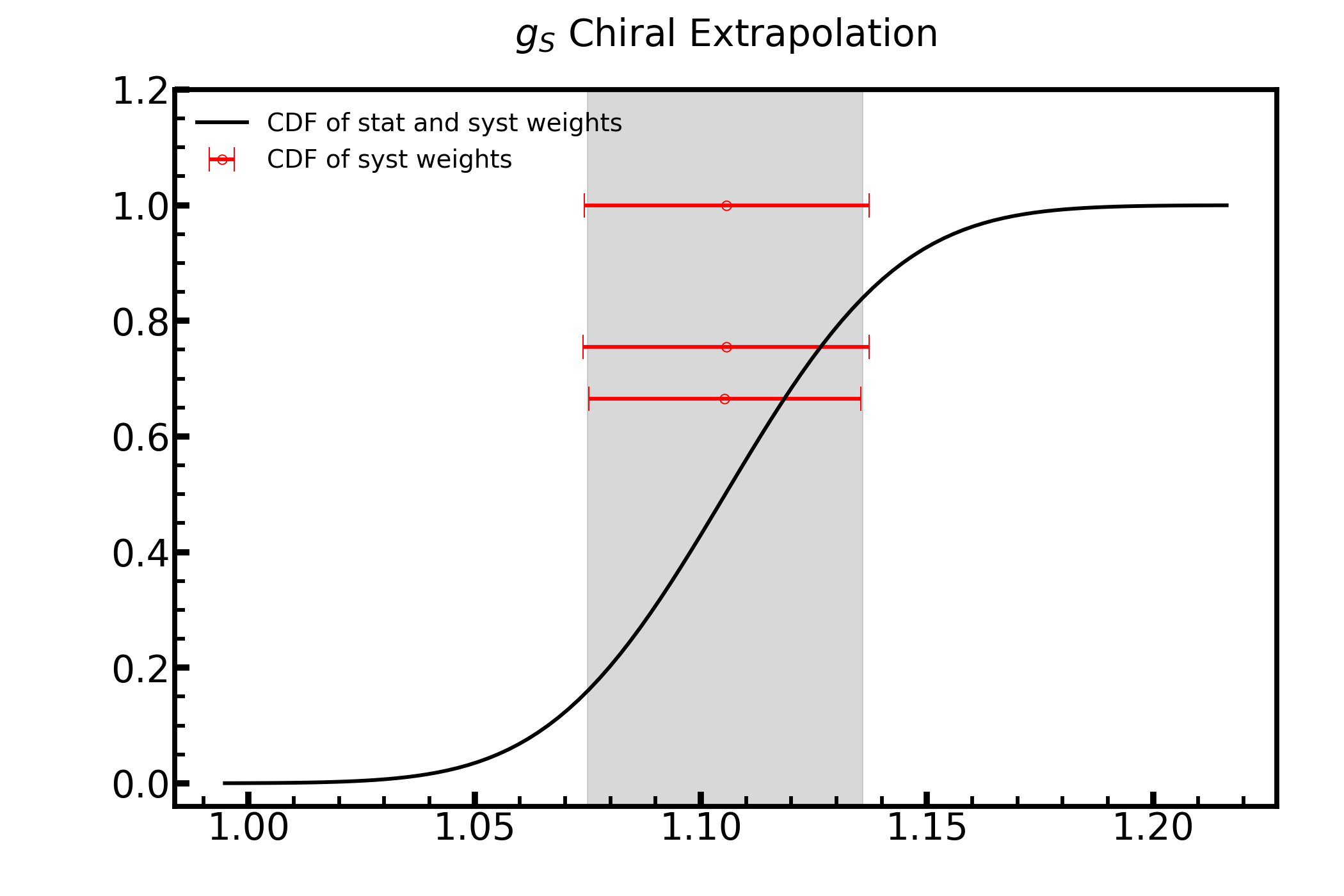}
     \includegraphics[width=0.45\linewidth]{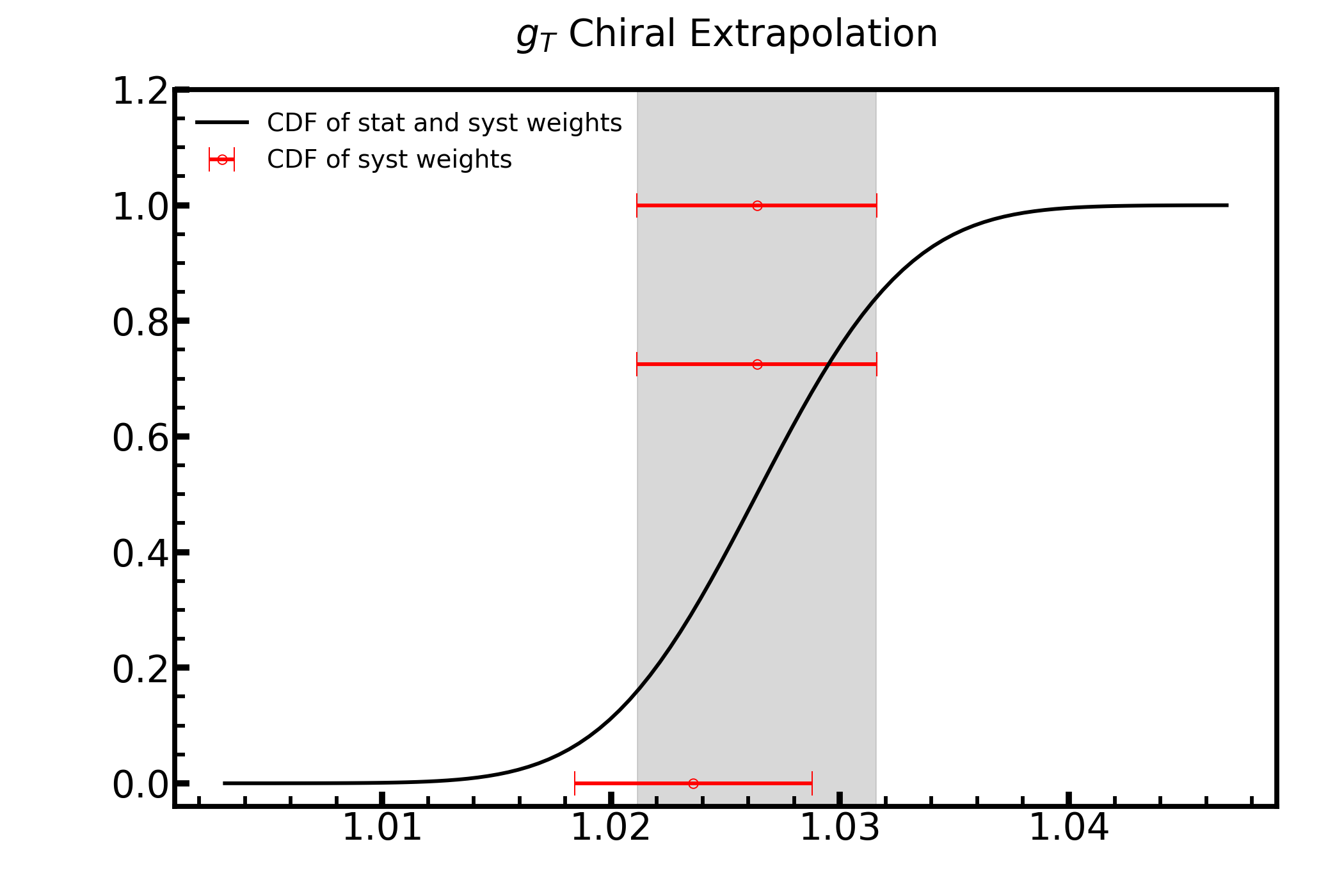}
    \caption{
    The CDF of different fit models of the chiral extrapolation of $g_{S}$ (left panel) and $g_T$ (right panel). 
    }
    \label{fig:chiral_AIC}
\end{figure}

\subsubsection{Infinite volume extrapolation}

Since the HB$\chi$PT FVE ansatz $m_{\pi}^{2}(m_\pi L )^{-1/2} e^{-m_\pi L}$ is disfavored by the $g_S$ data, we consider a general FVE model $m_{\pi}^{i}(m_\pi L )^{j/2} e^{-m_\pi L})$  with $i\in\{0,\pm 1,\pm2 \}$ and $j\in\{0,\pm 1,\pm2 \}$ in the AIC average. Those models cover both the HB$\chi$PT FVE ansatz and also naive $e^{-m_\pi L})$ one which is consistent with that of nucleon mass. 

\begin{figure}
    \centering
    \includegraphics[width=0.45\linewidth]{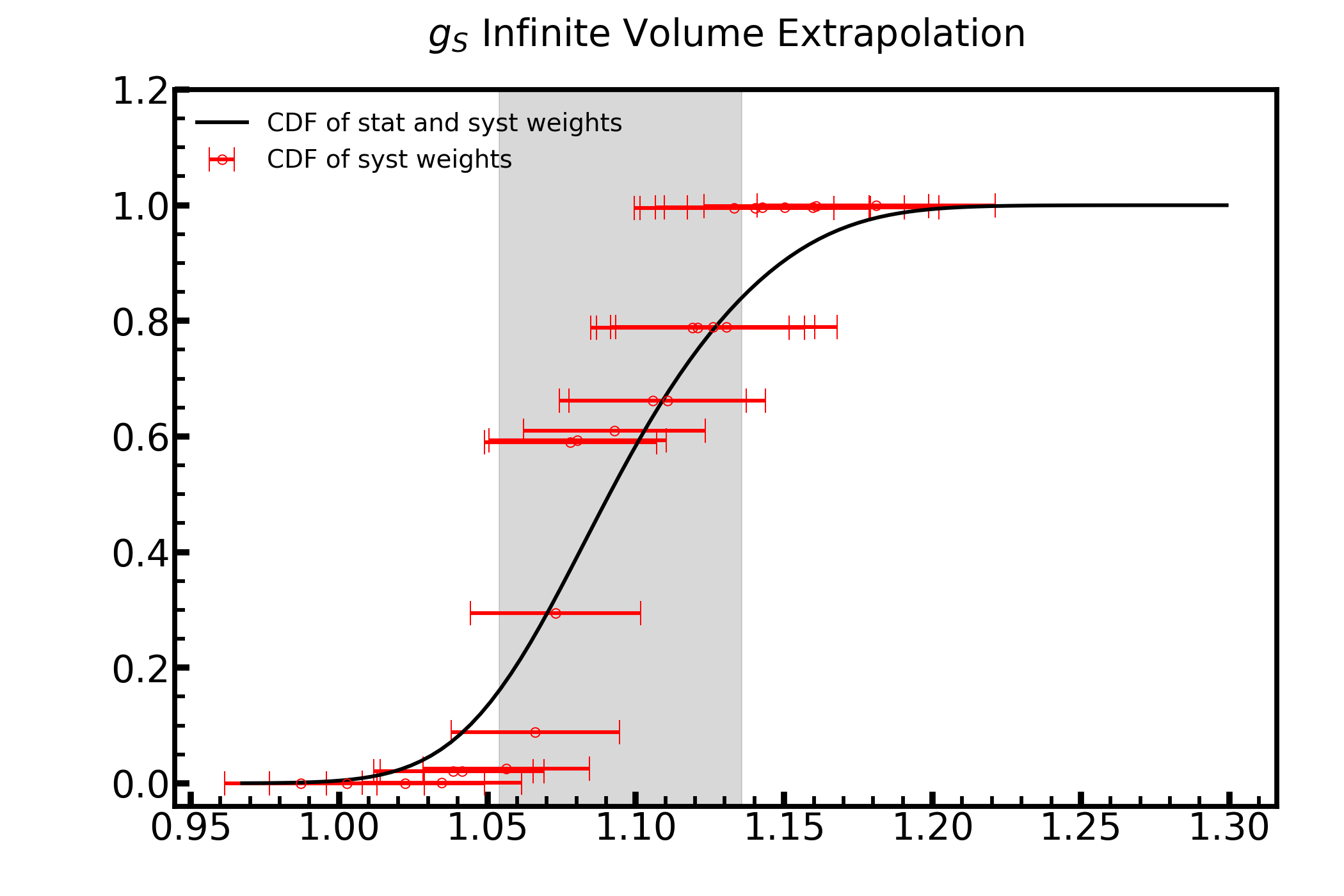}
     \includegraphics[width=0.45\linewidth]{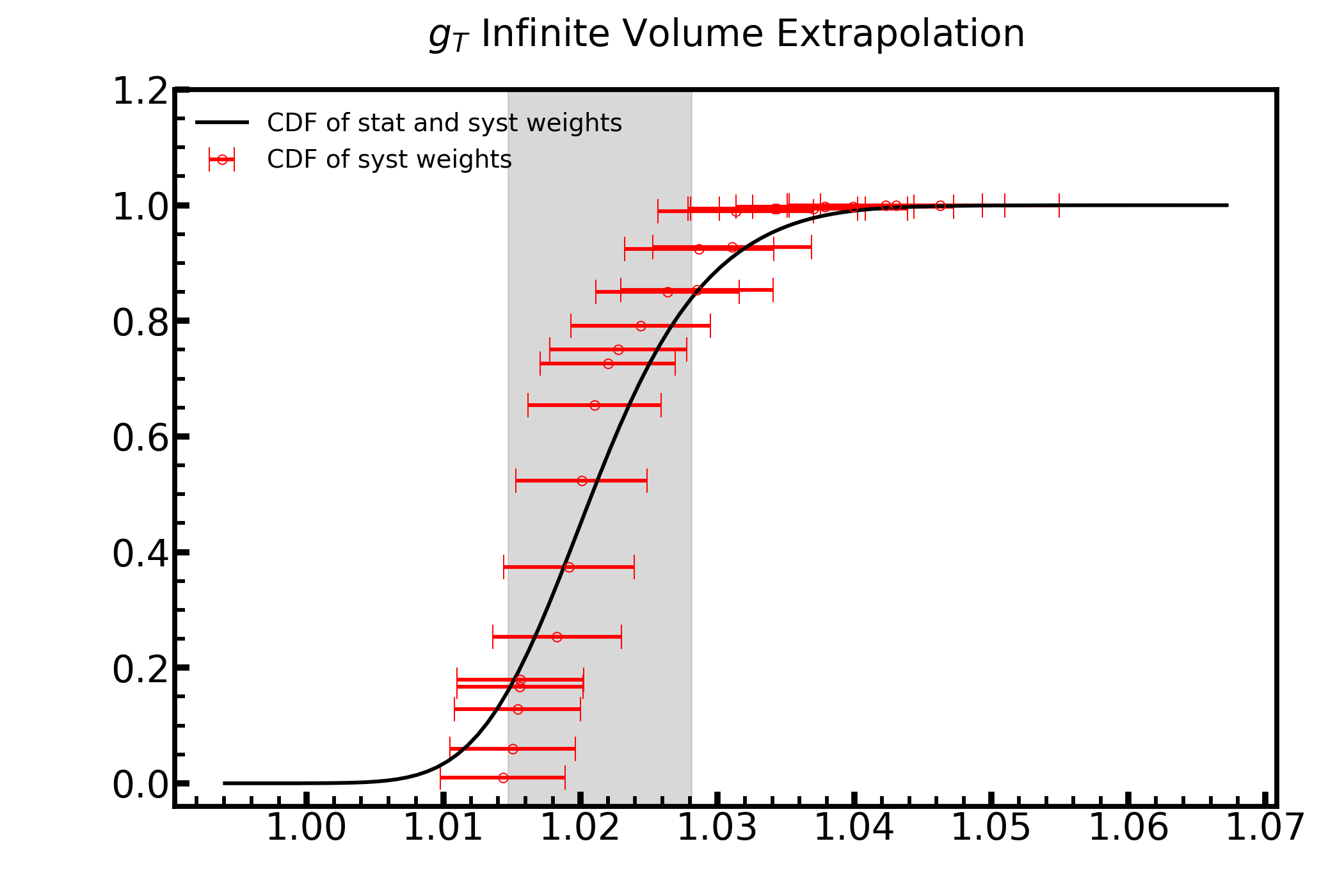}
    \caption{
    The CDF of different fit models of the infinite volume extrapolation of $g_{S}$ (left panel) and $g_T$ (right panel). 
    }
    \label{fig:volume_AIC}
\end{figure}

Fig.~\ref{fig:volume_AIC} shows a relatively strong dependence of the extracted $g_S$ and $g_T$ on different finite-volume ansatz, highlighting a sizable systematic uncertainty, especially for $g_S$. The fit quality favors the naive ansatz $e^{-m_\pi L}$ ($\chi^2/\text{d.o.f.} = 0.46$) over the HB$\chi$PT form $m_{\pi}^{2}(m_\pi L )^{-1/2} e^{-m_\pi L}$ ($\chi^2/\text{d.o.f.} = 2.5$) for $g_S$, but the  $\chi^2/\text{d.o.f.}$ using either HB$\chi$PT or naive ansatz is smaller than 1 for $g_T$. The ambiguous result for $g_T$ is expected, as its small finite-volume effects are comparable to the current statistical uncertainties, making the ansatz indistinguishable.

\begin{figure}[!h]
    \centering
    \includegraphics[width=0.45\linewidth]{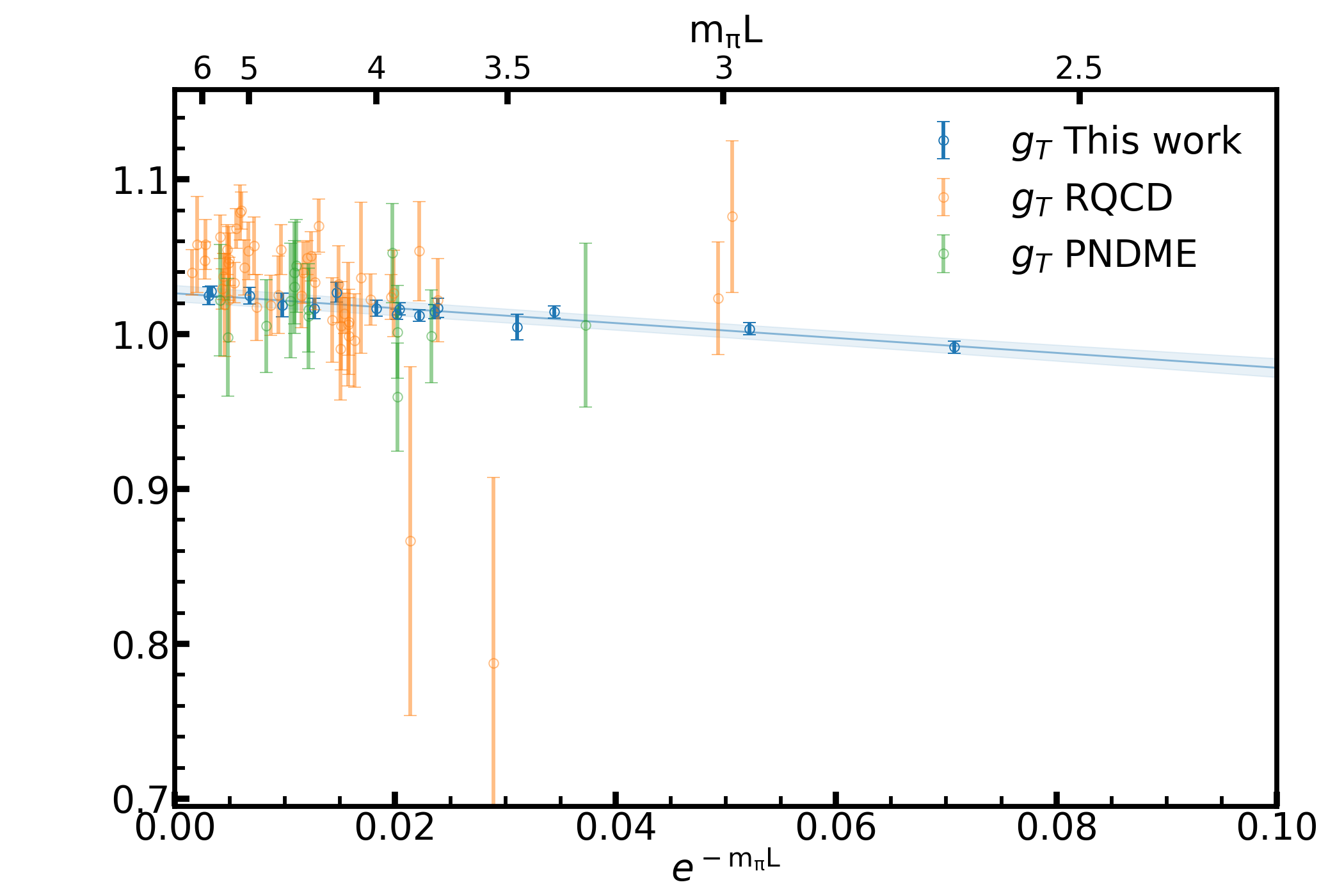}
    \includegraphics[width=0.45\linewidth]{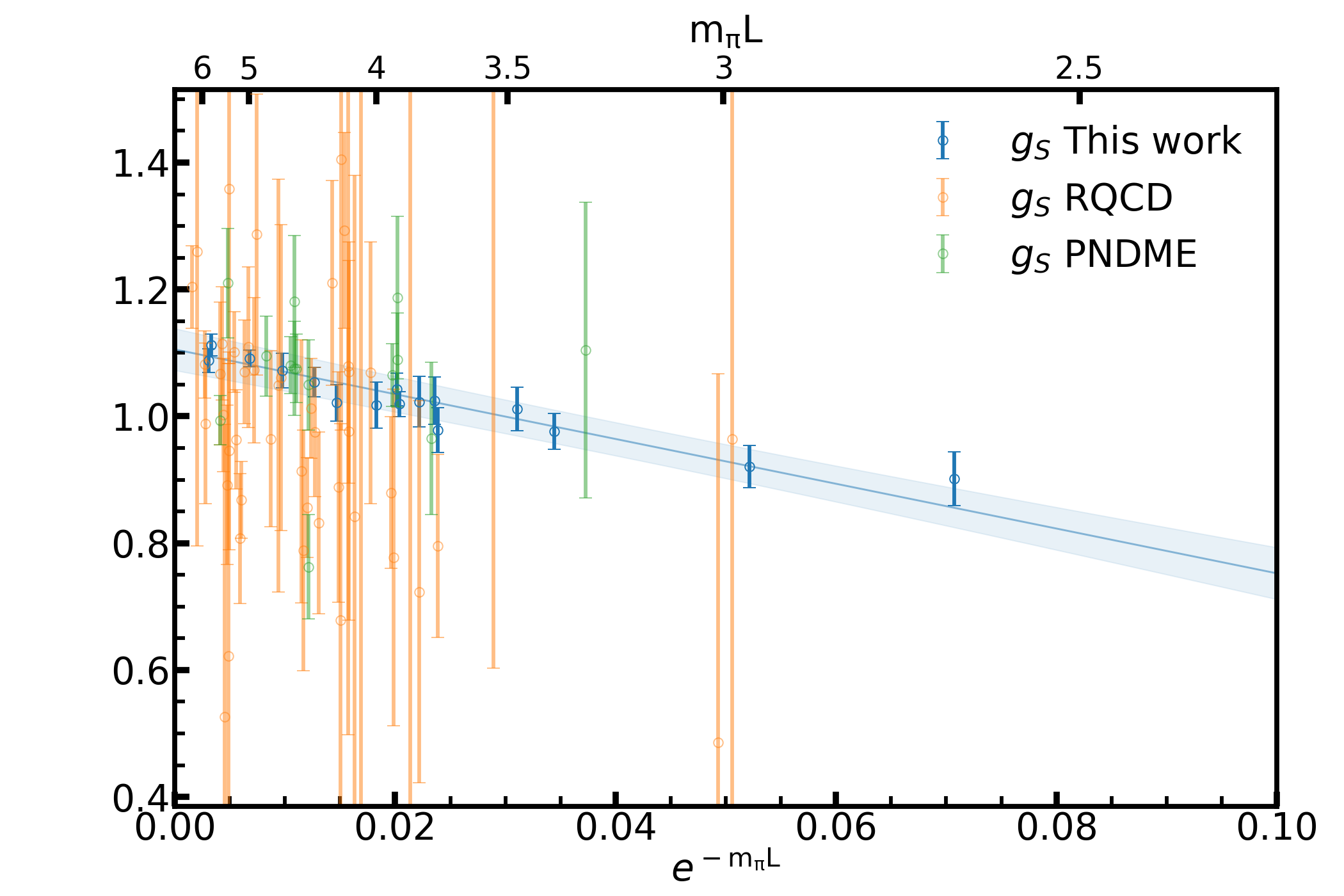}
    \caption{
    Comparison of this work versus other collaborations at finite volume, with all data extrapolated to the continuum limit.
    }
    \label{fig:volume_compare}
\end{figure}

{Fig.~\ref{fig:volume_compare} further plots $g_T$ and $g_S$, corrected for discretization and chiral extrapolation effects, against $e^{-m_\pi L}$. It shows that our high-precision data (blue triangles) are fully consistent within $2\sigma$ with the results from RQCD~\cite{Bali:2023sdi} (orange circles) and PNDME~\cite{Jang:2023zts} (green crosses) at comparable volumes. The statistical uncertainties of the earlier results are substantially larger, which prevented those studies from definitively identifying the FV trend or discriminating between different model ansatzes.}

\end{widetext}


%

\end{document}